\documentclass[a4paper, openright, 11pt, oneside]{book}

\usepackage{bbm}
\usepackage{mathrsfs}
\usepackage[british]{babel}

\usepackage{newcent}

\usepackage{fancyhdr}
\usepackage{lscape}
\usepackage{float}
\usepackage{graphicx}
\usepackage{epsfig}

\usepackage{pstricks}
\usepackage{amsmath}
\usepackage{rotating}

\usepackage{latexsym}

\newlength{\parindentdummy}
\newlength{\baselinedummy}
\numberwithin{equation}{section}
\numberwithin{table}{section}

	\newcommand{\HRule}{\rule{\linewidth}{1mm}}

	\renewcommand{\chaptername}{Chapter}
	\newcommand{\sectionname}{Section}

\newcommand{\cc}[2]{c\genfrac{[}{]}{0pt}{}{#1}{#2}}
\newcommand{\oo}[2]{\left(#1\left|#2\right.\right)}
\newcommand{\bartheta}[2]{\bar{\theta}\begin{bmatrix} #1 \\ #2 \end{bmatrix}}

\def\chir{\text{Ch}}
\newcommand{\nn}[0]{\nonumber}
\newcommand{\ba}{\begin{eqnarray}}
\newcommand{\ea}{\end{eqnarray}}

\newcommand{\ket}[1]{|#1\rangle}
\newcommand{\nahe}{{\sc nahe}}
\newcommand{\cption}[1]{Inequivalent liftable $S^3$ models with a #1 gauge
group. The chiral content of each model is listed per plane and
numbered, $+$ lists all the positive chiral states per plane
while $-$ lists all the negative states per plane. The total
sum of all the planes is then listed and subsequently the net
total number of chiral states. The list is ordered by the total
net number of chiral states.}
\newcommand{\instate}{\emph{in} state}
\newcommand{\outstate}{\emph{out} state}
\newcommand{\mssm}{{\sc mssm}}

	\def\titletext{Classif\mbox{}ication of the Chiral $\mathbbm{Z}_2 \times \mathbbm{Z}_2$ Heterotic String Models}
	\title{\titletext}
	\author{Sander E.M. Nooij}
	\date{June 2004}

\begin{document}
\null
\pagestyle{empty}
\def\background{The Standard Model has been shown to be consistent with observations up to the electroweak scale. The $SO(10)$ grand unification group provides an elegant way to unify the gauge interactions and the particle content of the Standard Model. The replication of the three generations of the Standard Model cannot be explained in this framework. At present the best candidate for providing a unified theory for matter and its interactions is string theory.
Among the most advanced realistic string models that have three generations with a $SO(10)$ embedding are the heterotic free fermionic $\mathbbm{Z}_2 \times \mathbbm{Z}_2$ orbifold models. This thesis provides a classification of the chiral content of the heterotic $\mathbbm{Z}_2 \times \mathbbm{Z}_2$ orbifold models. We show that the chiral content of the heterotic $\mathbbm{Z}_2 \times \mathbbm{Z}_2$ orbifold models at any point in the moduli space can be described by a free fermionic model. We present a direct translation between the orbifold formulation and the free fermionic construction. We use the free fermionic description for the classification wherein we consider orbifolds with symmetric shifts.
We show that perturbative three generation models are not obtained in the case of $\mathbbm{Z}_2 \times \mathbbm{Z}_2$ orbifolds with symmetric shifts on complex tori, and that the perturbative three generation models in this class necessarily employ an asymmetric shift. We show that the freedom in the modular invariant phases in the $N = 1$ vacua that control the chiral content, can be interpreted as vacuum expectation values of background fields of the underlying $N = 4$ theory, whose dynamical components are projected out by the $\mathbbm{Z}_2$ fermionic projections. In this class of vacua the chiral content of the models is determined by the underlying $N = 4$ mother theory.}

\setlength{\baselineskip}{2\baselinedummy}
\begin{center}
{\Huge  Classification of the}\\
{\Huge Chiral $\mathbbm{Z}_2 \times \mathbbm{Z}_2$ Heterotic String Models}\\
\end{center}
\setlength\parskip{0cm}

\setlength{\baselineskip}{\baselinedummy}\HRule
\newcount\scratch \scratch=7
\setbox0=\vbox{\hsize=8.5in \parindent=0pt 
        \baselineskip=6.5pt \gray \footnotesize \loop \background\ 
        \advance\scratch by-1 \ifnum\scratch>0 \repeat \black}
\psclip{\psframe*[linecolor=white, fillstyle=solid, fillcolor=lightgray](0in,0in)(\textwidth,-11.5cm)}
        \rput[t]{45}(-3cm,7cm){\reflectbox{\box0}}
\endpsclip
\setlength\parskip{\baselineskip}

\vspace{\stretch{5}}

\HRule
\vspace{\baselineskip}
{\large\sc Sander E.M. Nooij\ \hspace{\fill}\ Trinity 2004}
\vspace{\stretch{1}}
\newpage
		\null\cleardoublepage

\frontmatter
\thispagestyle{empty}\setcounter{page}{1}
\vspace{\stretch{1}}

\setlength{\baselineskip}{1.5\baselinedummy}
\begin{center}
{\Huge \titletext}
\end{center}
\setlength{\baselineskip}{\baselinedummy}

\vspace{\baselineskip}

\begin{center}
{\Large\sc Sander E.M. Nooij\\
Oriel College\\
Trinity 2004}
\end{center}

\begin{center}
{\large \sc Rudolf Peierls Centre for Theoretical Physics\\
University of Oxford}
\end{center}

\vspace{\stretch{1}}

\begin{center}
\begin{tabular}{c @{\hspace{2cm}} c}
\includegraphics[width=0.205\textwidth]{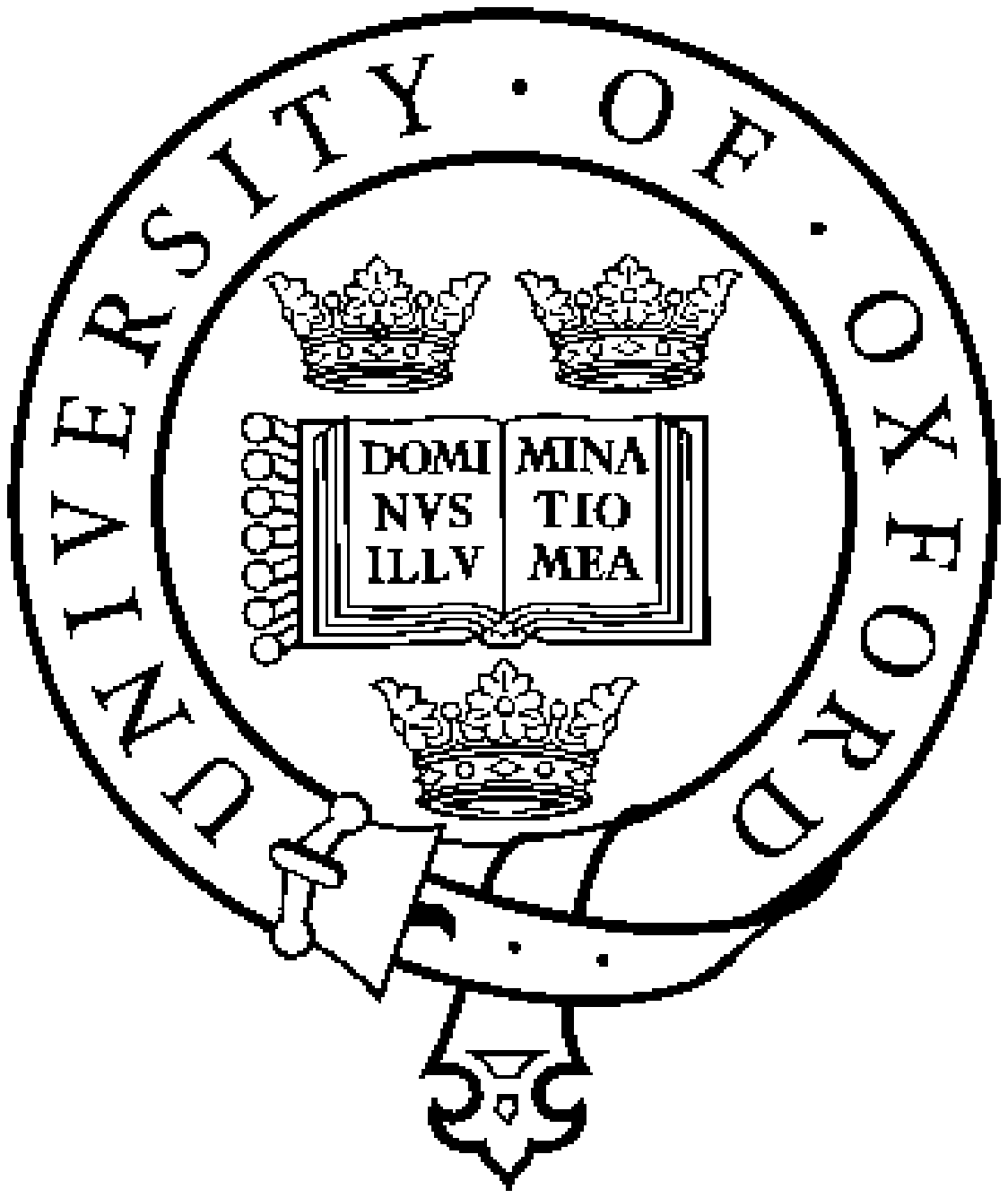} & \includegraphics[width=0.2\textwidth]{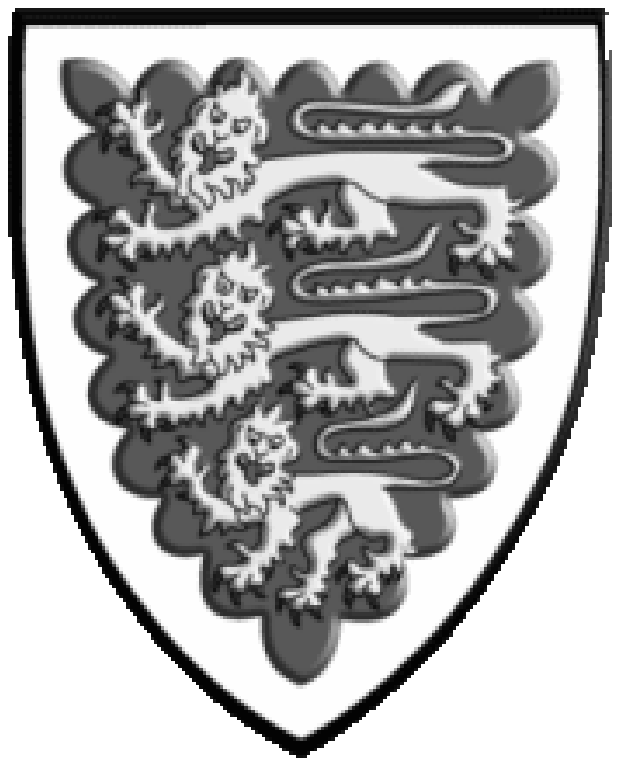}
\end{tabular}
\end{center}

\vspace{\stretch{1}}

\begin{center}
{\Large Thesis submitted in partial fulf\mbox{}illment of the requirements\\ for the Degree of Doctor of Philosophy\\ at the University of Oxford}
\end{center}

\vspace{\stretch{1}}

\newpage
		\null\cleardoublepage
\cleardoublepage
\null
\setlength{\parindent}{\parindentdummy}
\pagestyle{fancy}
\vspace{\stretch{1}}
\begin{center}Abstract\\
\rule{0.5\textwidth}{0.1mm}\end{center}
\begin{quote}
The Standard Model has been shown to be consistent with observations up to the electroweak scale. The $SO(10)$ grand unification group provides an elegant way to unify the gauge interactions and the particle content of the Standard Model. The replication of the three generations of the Standard Model cannot be explained in this framework. At present the best candidate for providing a unified theory for matter and its interactions is string theory.

Among the most advanced realistic string models that have three generations with a $SO(10)$ embedding are the heterotic free fermionic $\mathbbm{Z}_2 \times \mathbbm{Z}_2$ orbifold models. This thesis provides a classification of the chiral content of the heterotic $\mathbbm{Z}_2 \times \mathbbm{Z}_2$ orbifold models. We show that the chiral content of the heterotic $\mathbbm{Z}_2 \times \mathbbm{Z}_2$ orbifold models at any point in the moduli space can be described by a free fermionic model. We present a direct translation between the orbifold formulation and the free fermionic construction. We use the free fermionic description for the classification wherein we consider orbifolds with symmetric shifts.

We show that perturbative three generation models are not obtained in the case of $\mathbbm{Z}_2 \times \mathbbm{Z}_2$ orbifolds with symmetric shifts on complex tori, and that the perturbative three generation models in this class necessarily employ an asymmetric shift. We show that the freedom in the modular invariant phases in the $N = 1$ vacua that control the chiral content, can be interpreted as vacuum expectation values of background fields of the underlying $N = 4$ theory, whose dynamical components are projected out by the $\mathbbm{Z}_2$ fermionic projections. In this class of vacua the chiral content of the models is determined by the underlying $N = 4$ mother theory.
\end{quote}
\begin{center}\rule{0.5\textwidth}{0.1mm}\end{center}
\vspace{\stretch{3}}
\setlength{\parindent}{0pt}
\newpage

		\null\cleardoublepage
\cleardoublepage
\null
\pagestyle{fancy}
\vspace{\stretch{1}}
\begin{center}Publications\\
\rule{0.5\textwidth}{0.1mm}\end{center}
\begin{quote}
This thesis is the culmination of my years at the University of Oxford. It is based on several papers that have been written during that time. Chapter \ref{chap:realistic} is based on a paper published in Nuclear Physics B \cite{Cleaver:2002ps}. The chapters \ref{chap:orbifold}, \ref{chap:fermform} and the results in chapters \ref{chap:models} and \ref{chap:results} are based on three papers. The first paper reports on a talk given at the second String Phenomenology Conference in Durham, UK \cite{Faraggi:2003yd}. The second paper is published in Nuclear Physics B \cite{Faraggi:2004rq}. The last paper on which parts of this thesis are based, is currently in preparation \cite{Faraggi:2004xnew}.  
\end{quote}
\begin{center}\rule{0.5\textwidth}{0.1mm}\end{center}
\vspace{\stretch{3}}
\newpage

		\null\cleardoublepage
\chapter*{Acknowledgements}

During my time at the University of Oxford I have enjoyed the support of many people and institutions, allowing me to complete this work. I would like to thank my supervisor Alon Faraggi for his guidance and many useful discussions. The enlightening conversations and meetings with both Costas and Ioannis have proven essential for the completion of this work. I would like to thank Richard and Dave for many interesting discussions as well as Filipe for helping me out even while he was in Portugal.

This work could not have been completed without the financial support I have recieved from many institutions as well as my father Hans. I would like to thank him especially for his fabulous support while I was working on my research. Furthermore, this work would not have been possible without the support of the Pieter Langerhuizen Lambertszoon Fund of the Royal Holland Society of Sciences and Humanities, the \mbox{VSBfonds}, the Fundatie van de Vrijvrouwe van Renswoude te  \mbox{'s Gravenhage} and the Noorthey Academy. I would also like to thank Oriel College for their support and generosity. I am thankful that the Rudolf Peierls Centre for Theoretical Physics at the University of Oxford made it possible for me to strengthen my knowledge outside of Oxford during summer.

I would like to thank my beloved Carolijn for supporting me while I was in Oxford. It was a delight having her around in Oxford during my second year. While she was in the Netherlands, I truly enjoyed our many great conversations on the phone as well as her presence during her frequent visits. The support from my mother Kiek and our many conversations I enjoyed hugely. The interest and support from my sister Claire were remarkable, our conversations good fun. Oxford has been a very interesting and stimulating place, notably because of its highly international environment. The many conversations I had with Josef were very intriguing and I would like to thank him for introducing me to the world of politics. I have had many wonderfull evenings with both chaps Geoff and Mark. I would like to thank Geoff for his insistence on rationality and silly dances, and Mark for his skills in the English language as well as sometimes being a simulacrum of the cerebral and the abstract. When I was not working on my research, I had many great times on the hockey pitch. I would like to thank both Tony and Oli of New College Hockey Club for their enthousiasm. During the year I played at Rover Hockey Club, I had some many good games and evenings with Rich and Tom. I would also like to thank Bernhard and Kudrat for our many conversations and their generosity at the end of my time in Oxford.

Having completed my Doctor of Philosophy at the Universty of Oxford, I will leave behind a place that has given me many invaluable experiences. I will treasure these for the rest of my life.

	\tableofcontents

\mainmatter

\chapter{Introduction}

One of the most remarkable achievements in physics has been the development of the Standard Model that describes the strong, weak and electromagnetic forces. These three forces together with the matter content of the Standard Model have been shown to be consistent with experimental observations up to the electroweak scale.

In the Standard Model the interactions are invariant under the gauge group $SU(3)_C \times SU(2)_L \times U(1)_Y$ above the electroweak scale. In the Standard Model there are three families of chiral leptons and quarks. Each family has two $SU(2)_L$ weak doublets, of which one weak doublet transforms as a strong $SU(3)_C$ triplet, and three $SU(2)_L$ singlets, two of which transform as strong $SU(3)_C$ triplets. The Higgs weak doublet gives mass to particles below the electroweak scale. The forces are mediated by the gauge bosons of the Standard Model gauge group. The chiral matter content for each family is summarized in table \ref{tab:sm}.

\begin{table}[t]
\begin{center}
\begin{tabular}{c | c c c }
			& $SU(3)_C$ 	& $SU(2)_L$ 	& $U(1)_Y$ \\\hline
$L = \binom{\nu_e}{e}_L$& $1$ 		& $2$ 		& $-\frac{1}{2}$\\
$Q = \binom{u}{d}_L$	& $3$		& $2$		& $\frac{1}{6}$\\
$\bar{e}_L$		& $1$		& $1$		& $1$\\
$\bar{u}_L$		& $\bar{3}$	& $1$		& $-\frac{2}{3}$\\
$\bar{d}_L$		& $\bar{3}$	& $1$		& $\frac{1}{3}$
\end{tabular}
\caption{The chiral matter content of the Standard Model decomposed under the gauge group $SU(3)_C \times SU(2)_L \times U(1)_Y$ for each family.}\label{tab:sm}
\end{center}
\end{table}

Below the electroweak scale the $SU(2)_L \times U(1)_Y$ symmetry is broken to $U(1)_{QED}$ thereby giving mass to the $W^\pm$ and $Z$ gauge bosons of the broken symmetry through the Higgs mechanism. The Higgs mechanism furthermore ensures that matter in the Standard Model acquires a mass. In this process the $SU(2)_L$ gauge bosons mix with the $U(1)_Y$ gauge boson giving rise to the Weinberg or weak mixing angle $\theta_W$.

To fully determine the Standard Model we need to fix its parameters. Three gauge couplings determine the strength of the gauge interactions. Nine Yukawa couplings describe the mass of three electron type chiral matter, three down type chiral matter and three up type chiral matter. The Standard Model requires three mixing angles and one weak CP violating phase for the CKM matrix. The $\theta_{QCD}$ phase parameterizes possible CP violation in the strong sector. Additionally, there are two parameters in the Higgs potential setting the electroweak symmetry breaking scale. All these parameters are measured by experiments and one would like to have a theoretical explanation for their origin. The results of the experiments done at {\sc lep} show a repetition of $2.994 \pm 0.012$ light neutrinos in the Standard Model adding strong experimental support to the existence of just three families of quarks and leptons.\cite{Hagiwara:2002fs} This thesis focuses on the replication of three generations in the Standard Model.

An elegant way to explain the gauge couplings and the chiral matter content of the Standard Model is by introducing a $SO(10)$ grand unified theory ({\sc gut}). The chiral matter content of a three family $SO(10)$ model is described by three chiral $16$ representations of $SO(10)$. Each chiral $16$ $SO(10)$ representation contains the chiral matter of the Standard Model plus a right handed neutrino. Breaking each $SO(10)$ to the Standard Model gauge group $SU(3)_C \times SU(2)_L \times U(1)_Y$ decomposes each $16$ of $SO(10)$ multiplet into the Standard Model multiplet structure listed in table \ref{tab:sm} plus the right handed neutrino which is a singlet under each Standard Model gauge group. The right handed neutrino allows for massive neutrinos. A nonzero mass for the neutrino has been supported by the observation of neutrinos oscillations.\cite{Fukuda:1998mi, Cleveland:1998nv} However, the $SO(10)$ grand unified theory cannot explain the number of generations observed in nature.

Ever since the first {\sc gut} was proposed, physicists have been troubled by the existence of two mass scales, the electroweak scale and the Planck scale, that differ by approximately thirteen orders of magnitude. This is known as the hierarchy problem. In terms of perturbative field theory, this implies an almost exact cancellation between the bare mass of the Higgs boson, determining the electroweak scale, and its radiative corrections from the {\sc gut} scale. In order to stop the mass of the Higgs boson from becoming too large, an intermediate scale between the Planck and the electroweak scale should be introduced. At this intermediate scale a symmetry between bosons and fermions, known as supersymmetry ({\sc susy}), is present. If supersymmetry is exact then diagrams with boson loops exactly cancel diagrams with fermion loops. In the case of broken global supersymmetry the boson and fermion loops will cancel up to corrections from the broken {\sc susy} scale. The introduction of supersymmetry therefore solves the hierarchy problem by suppressing radiative corrections to the Higgs mass terms above the {\sc susy} breaking scale. A promising feature of $SO(10)$ unification, with supersymmetry included, is that by giving the Standard Model couplings in terms of the $SO(10)$ coupling, it predicts the weak mixing angle $\sin^2\theta_W$ that agrees with observations to a very high precision.\cite{Ramond:1999vh, Altarelli:2004fq} However, supersymmetry cannot explain the number of generations.

Although this scheme of unification and the introduction of supersymmetry at an intermediate scale gives very good results, we have left out the force of gravity in our description of elementary matter and its interactions. Unifying the four forces using the highly successful method of field quantization as used in the Standard Model has been shown to give a nonrenormalizable theory.\cite{Bern:2002kj} At present the best candidate for unifying the four forces is string theory. In string theory the fundamental objects are one dimensional particles, known as strings. The Lagrangian describing string theory has only one parameter, known as the string tension. However, a consistent string theory has enough degrees of freedom to describe a ten dimensional target space. One way to reduce the ten dimensional target space to the observed four dimensional target space is by compactifying the additional six dimensions. Another way to describe a string in four space time dimensions is by treating the additional degrees of freedom in the Lagrangian as fields on the world sheet not related to any spacetime coordinate a priori. The two descriptions have been shown to be equivalent in a wide class of models.\cite{Lopez:1994ej, Faraggi:2004rq} There are many different ways of reducing the number of dimensions. Consequently we have introduced even more degrees of freedom than we started out with in the Standard Model. The problem of finding the correct vacuum of string theory is one of the major problems in string theory today. 

Among the most realistic phenomenological string models to date are the three generation heterotic string models, constructed in the free-fermion formulation. These models have been the subject of detailed studies, showing that they can, at least in principle, account for desirable physical features including the small neutrino masses, the consistency of gauge-coupling unification with the experimental data from {\sc lep} and elsewhere.  An important property of the fermionic construction is the standard SO(10) embedding of the Standard Model spectrum, which ensures natural consistency with the experimental values for $\sin^2 \theta_W$ at the electroweak scale and allows neutrinos to have a nonzero mass below the electroweak scale.\cite{Ramond:1999vh}

A key property of the realistic free fermionic models is their underlying $\mathbbm{Z}_2 \times \mathbbm{Z}_2$ orbifold structure. The emergence of the three chiral generations in a large class of fermionic constructions is correlated with the existence of three twisted sectors in the $\mathbbm{Z}_2 \times \mathbbm{Z}_2$ orbifold of the six dimensional internal space. Each twisted sector produces exactly one of the light chiral generations and there is no additional chiral matter. Thus, the fermionic construction offers a plausible and compelling explanation for the existence of three generations in nature. However, 
the geometrical structures that underlies the realistic free fermionic models are not fully understood. This thesis tries to shed light on this aspect and on the geometrical mechanism that reduces the number of generations in the $\mathbbm{Z}_2 \times \mathbbm{Z}_2$ orbifold models. 

We are interested in the net number of chiral generations. In the $\mathbbm{Z}_2 \times \mathbbm{Z}_2$ orbifold models, the net number of generations comes from the twisted sectors. The untwisted sector has an equal number of generations and anti-generations. By compactifying the heterotic string on a $\mathbbm{Z}_2 \times \mathbbm{Z}_2$ orbifold we are left with the moduli of the three planes $(T_i, U_i)$, where $i$ runs over the three twisted planes. We argue in section \ref{sec:orbifoldsandshifts} that the matter spectrum of the twisted sectors does not depend on the moduli. Consequently we can choose any value for the moduli space to investigate the chiral content of the $\mathbbm{Z}_2 \times \mathbbm{Z}_2$ orbifold. We choose the maximal symmetry point in the moduli space as this point can be described by a free fermionic model. By selecting a specific free fermionic model and describing its chiral content we have therefore described the chiral content of all models that are related to the selected free fermionic model by a change of the moduli $(T_i, U_i)$ of the internal space. 

In this thesis we classify the chiral content of the heterotic $\mathbbm{Z}_2 \times \mathbbm{Z}_2$ $N=1$ supersymmetric orbifolds. This classification is possible since we can choose any point in the moduli space to describe the chiral content as explained above. This choice gives us the opportunity to describe the chiral content of the heterotic $\mathbbm{Z}_2 \times \mathbbm{Z}_2$ orbifolds in the free fermionic construction. We argue in section \ref{sec:general} that in the realistic free fermionic models the number of generations is reduced by symmetric shifts. In the orbifold description symmetric shifts on the internal space are well known. We show in chapter \ref{chap:orbifold} the correspondence between symmetric shifts on the internal space and symmetric  shifts in the free fermionic description. As the hidden $\mathbbm{E}_8$ gauge group is broken to $SO(16)$ and then further broken to $SO(8) \times SO(8)$ in the realistic free fermionic models we include these possibilities of symmetry breaking in our classification. The classification of the chiral content therefore includes all $N=1$ $\mathbbm{Z}_2 \times \mathbbm{Z}_2$ models where symmetric shifts are realized on the internal space and where the hidden gauge group is at most broken to $SO(8) \times SO(8)$ at \emph{any} point in the moduli space. 

In the classification we define four subclasses of $\mathbbm{Z}_2 \times \mathbbm{Z}_2$ orbifolds. The first subclass of models has spinorial representations on each plane. The second subclass has spinorial representations on two of the three planes. The third subclass has spinorial representations on only one of the three planes. The fourth subclass does not have spinorial representations on any of the three planes. In chapter \ref{chap:fermionic} we explain that a free fermionic model is defined by a set of basis vectors and a set of generalized GSO coefficients. We show in chapter \ref{chap:orbifold} that in each subclass of models, the basis vectors that induce the $\mathbbm{Z}_2 \times \mathbbm{Z}_2$ twists are \emph{completely} fixed. The basis vectors describing the symmetric shifts on the internal space and the Wilson lines in the hidden sector are \emph{completely} fixed as well. The only free parameters in the classification are therefore the generalized GSO coefficients in the free fermionic formulation. 
We classify the chiral content of the heterotic $\mathbbm{Z}_2 \times \mathbbm{Z}_2$ orbifolds by looking at all possible values for these GSO coefficients. This is done using a computer program written in {\sc fortran}. We show that a subclass of $\mathbbm{Z}_2 \times \mathbbm{Z}_2$ orbifold models have a geometrical interpretation at the $N=4$ level. In these models the symmetric shifts on the internal space and the Wilson lines that break the hidden $\mathbbm{E}_8 \to SO(8) \times SO(8)$ are completely separated from the twists on the internal space. We show that the freedom in the modular invariant phases in the $N = 1$ vacua that control the chiral content, can be interpreted as vacuum expectation values of background fields of the underlying $N = 4$ theory, whose dynamical components are projected out by the $\mathbbm{Z}_2$ fermionic projections. In this class of vacua the chiral content of the models is predetermined at the $N = 4$ level. We restrict the numerical classification to those models that have a geometrical interpretation at the $N=4$ level due to computational limitations.

The results of the classification show that three generation models can be realized using symmetric shifts. The observable gauge group in these models cannot be broken perturbatively to the Standard Model gauge group while preserving the matter content. Additionally, the complex structure of the internal space is necessarily broken, thereby splitting the internal $\Gamma_{6,6}$ lattice into six $\Gamma_{1,1}$ lattices. The classification shows that in the context of the realistic free fermionic $\mathbbm{Z}_2 \times \mathbbm{Z}_2$ models the reduction of the number of generations to the observed three, \emph{necessarily} requires asymmetric shifts on the internal space. This is one of the main results of the analysis and it reveals, at least in the context of the three generation free fermionic models, that the geometrical structures that underly these models may not be simple Calabi--Yau manifolds, but it corresponds to geometries that are yet to be defined. 

This thesis is organized as follows. In chapter \ref{chap:heterotic.setup} we give a short review of the heterotic string. We show that in a conformal field theory bosons can be interchanged by fermions. We give a short overview of the closed string models. In chapter \ref{chap:orbifold.setup} we discuss the orbifold formulation of a string on a compact space. We discuss the construction of the free fermionic models in chapter \ref{chap:fermionic}. Chapter \ref{chap:realistic} contains a typical derivation of the particle content of a free fermionic model. Two models are discussed in detail. In chapter \ref{chap:orbifold} we isolate the free parameters of $\mathbbm{Z}_2 \times \mathbbm{Z}_2$ orbifolds in the heterotic string using the orbifold description. In chapter \ref{chap:fermform} we translate this orbifold description to the fermionic formulation. We then focus on the chiral content of these models. Using the formulas derived in chapter \ref{chap:fermform}, we give in chapter \ref{chap:models} an example of a model that has $48$ families as well as a model that has $3$ generations. Similar examples of models with a particular number of generations are given. In chapter \ref{chap:results} we discuss the results we obtained from our analysis. We end this thesis with chapter \ref{chap:conclusions} giving some suggestions for future work and an overall conclusion. In appendix \ref{app:spectrum} we give the full spectrum of two sample free fermionic models of the heterotic string. We give the explicit results of the classification of the chiral $\mathbbm{Z}_2 \times \mathbbm{Z}_2$ fermionic models of the heterotic super string in appendix \ref{app:tables}.

\renewcommand{\theta}{\vartheta}


	\part{Introduction to the Orbifold and Free Fermionic Construction}
\chapter{The Heterotic String}\label{chap:heterotic.setup}

In this chapter we construct the heterotic string. We describe the bosonic string.  We show how bosons are related to fermions in a conformal field theory description. This will be one of the building blocks for the free fermionic description of the heterotic string. The four different closed strings are discussed. We explain how the different string theories are related by dualities.

\section{The String}\label{sec:string}

We construct the string from basic principles. We introduce the operator product expansion of two operators that will allow us to explain the equivalence between bosons and fermions. We refer to \cite{master2001} for more details.

\begin{figure}[t]
\begin{center}
\includegraphics[width=0.8\textwidth]{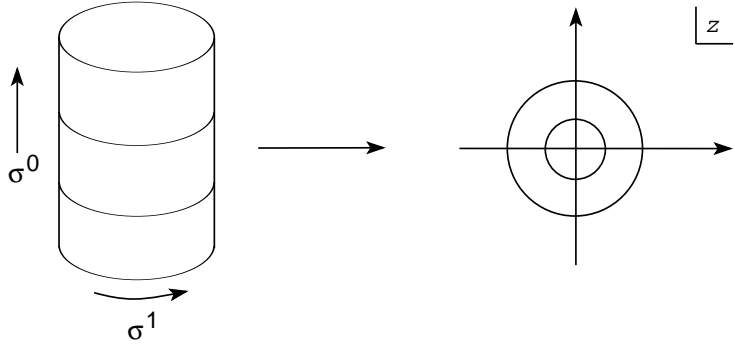}
\caption{The map from the cylinder to the compactified complex plane}\label{fig:complex}
\end{center}
\end{figure}

Considering higher dimensional objects as elementary particles, where the world manifold is scale invariant leads automatically to the notion of strings. These strings have necessarily a $2$ dimensional world sheet and their action is described as
\begin{equation}
S = \frac{1}{2\pi\alpha'} \int_{M} d^2\sigma \left( -\det \partial_{a}X^{\mu}\partial_{b}X_{\mu} \right)^{\frac{1}{2}},
\end{equation}
where $\alpha'$ is the string tension, $M$ is the string world sheet. The bosonic fields $X^\mu$ are maps from the world sheet to the target space and $\mu$ runs over the number of spacetime dimensions and $a$ runs over the number of world sheet dimensions i.e. $2$. $\sigma$ parameterizes the world sheet. This action is known as the Nambu-Goto action\index{Nambu-Goto action}. The $\alpha'$ is the string tension and $M$ is the string world sheet. This action is rewritten as the Polyakov action\index{Polyakov action}
\begin{equation}\label{eq:polyakov1}
S_{P}= \frac{1}{4\pi\alpha'} \int_{M} d^{2} \sigma\ {(-\gamma)}^{1/2}\gamma^{ab}\partial_aX^{\mu}\partial_bX_{\mu},
\end{equation}
where $\gamma^{ab}$ is the metric on the world sheet. From now on we consider only closed strings. This means that the world sheet becomes a cylinder. In this case, the Polyakov action can be rewritten as an integral over the complex plane by first defining light cone coordinates following a Wick rotation. We then map the world sheet to the complex plane by mapping the \instate\ to the origin and the \outstate\ to infinity as is done in figure \ref{fig:complex}
\begin{equation}
z = e^{-iw} = z^1 + i z^2,
\end{equation}
where $w = i\sigma^0 + \sigma^1$. This can be realized due to the scale or conformal invariance of the theory. We then complexify the fields on the world sheet. Deriving the equations of motion shows that the fields split into left and right moving sectors. Since the left and right moving sectors are pure holomorphic and anti-holomorphic functions on the world sheet respectively, we can expand these fields into a Laurent series.

An interesting feature has arisen when we projected the world sheet to the complex plane. When calculating the integral of equation \eqref{eq:polyakov1} on the complex plane the integral in the radial direction is trivial while the contour integral only picks up residues. Since we have mapped the \instate\  to the origin we have effectively removed the origin from the complex plane and replaced this with a residue, or \instate.  This contour integral therefore picks up the residue at the origin. The \instate\ is therefore realized by the residue. In a similar fashion we can replace each incoming or outgoing state on the world sheet by a residue on the complex plane. The only requirement induced by time like ordering of the QFT is that the poles are ordered radially. The interesting part of the theory therefore is encrypted in the poles of the fields on the complex plane. 

The operator product expansion\index{operator product expansion} is a result of this realization. First consider a field $\phi$. Any change of this field induced by a conformal transformation is identical to the commutator of the field with the conformal Noether charge
\begin{equation}\label{eq:gen}
\partial_\epsilon \phi(z,\bar{z}) = [ Q_\epsilon, \phi(z,\bar{z}) ],
\end{equation}
where $\partial_\epsilon$ is the infinitesimal flow along the conformal direction and $Q_\epsilon$ is the conformal charge which generates the conformal transformations. This can be seen to be identical to the contour integral over the radially ordered product of the stress tensor with the field $\phi$. Since the contour integral only picks up the values at the poles we can write the product of two operators $\mathcal{A}_{1,2}(\sigma_{1,2})$ defined at $\sigma_{1,2}$ on the world sheet as an expansion of the poles plus a regular part
\begin{equation}
\mathcal{A}_1(\sigma_1) \mathcal{A}_2(\sigma_2) = poles\ +\ :\mathcal{A}_1(\sigma_1) \mathcal{A}_2(\sigma_2):,
\end{equation}
where $:\ :$ defines the standard normal ordering of operators in a QFT and is known as the regular part. Usually the regular part is omitted and one only writes down the poles. For more details we refer to \cite{Ketov:1995yd, Schellekens:1996tg}

\section{Bosonization}\index{bosonization|(}\index{fermionization|(}\label{sec:bosonization}

In this section we describe the correspondence of fermions and bosons in a conformal field theory. We derive this equivalence from the operator products of both fermions and bosons. 


We show that in a conformal field theory we can interchange bosonic fields with fermionic fields. This is important for the construction of the free fermionic models. In this section we set $\alpha'=2$. We follow the methods employed in \cite{Polchinski:1998rr}.

As we have explained in the previous section all information of the conformal field theory is embedded in the operator products. Let us now consider the operator product expansion of the bosonic fields $X(z)$, $X(w)$\index{operator product expansion!Z@$X(z)X(w)$}. We find.
\begin{equation}\label{eq:xxope}
X(z) X(0) = - \ln|z|^2 + \mathcal{O}(z).
\end{equation}
Consider now the operators $e^{\pm iX(z)}$. Using the Campbell-Baker-Hausdorff formula
\begin{equation}
e^{ipX}e^{iqX} = e^{ipX(z) + iqX(0) + \frac{1}{2} pq  [X(z), X(0)] + \ldots}
\end{equation}
we find they have the operator product expansions 
\begin{subequations}\label{eqs:opeboson}
\begin{eqnarray}
e^{ iX(z)}e^{ -iX(0)} 	&=& \frac{1}{z} + \mathcal{O}(z),\\
e^{ iX(z)}e^{ iX(0)}	&=& \mathcal{O}(z),\\
e^{ -iX(z)}e^{ -iX(0)}	&=& \mathcal{O}(z).
\end{eqnarray}
\end{subequations}
Similarly, consider now the operator product expansion of two complexified Ma\-jo\-ra\-na-Weyl fermions $\psi^{1,2}(z)$\index{operator product expansion!psi@$\psi(z)\bar{\psi}(0)$}
\begin{align}\label{eq:mayoranaweylfermions}
\psi &= \frac{1}{\sqrt{2}} \left( \psi^1 + i \psi^2 \right), & \bar{\psi} &=  \frac{1}{\sqrt{2}} \left( \psi^1 - i \psi^2 \right).
\end{align}
Their operator product expansions are
\begin{subequations}\label{eqs:opefermion}
\begin{eqnarray}
\psi(z)\bar{\psi}(0) 		&=& \frac{1}{z} + \mathcal{O}(z),\\
\psi(z)\psi(0) 			&=& \mathcal{O}(z),\\
\bar{\psi}(z)\bar{\psi}(0) 	&=& \mathcal{O}(z).
\end{eqnarray}
\end{subequations}
We see that equations \eqref{eqs:opeboson} are equal to equations \eqref{eqs:opefermion}. Therefore, we find the equivalence between bosons and fermions and write
\begin{align}\label{eq:fermionization}
\psi(z) &\cong e^{iX(z)},&\bar{\psi}(z) &\cong e^{-iX(z)},
\end{align}
where the equal sign should be interpreted as being valid primarily as a statement for the expectation values of the two fields. This can be easily extended to the anti-holomorphic or right moving sector. From equation \eqref{eq:fermionization} we see that a shift around half the compactified dimension is realized by
\begin{align}
X(z) &\to X(z)+\pi\ \  \leftrightarrow\ \  \psi(z) \to -\psi(z)
\end{align}
and similarly for the right moving sector. Therefore \emph{any} shift on the bosonic coordinates can be realized by a suitable choice of boundary conditions for the free fermionic degrees of freedom.

This boson-fermion equivalence is crucial for the further development of the construction of the heterotic string in lower than $10$ dimensions.\index{bosonization|)}\index{fermionization|)}

\section{The Closed String Models}

In this section we give a small overview of the four different closed string models. We first discuss the two type II strings. The heterotic strings are then discussed. The Type I string is mentioned briefly.

In section \ref{sec:string} we have discussed that the left and right moving sectors are completely decoupled in the closed string. We can therefore construct several string theories by setting the configuration of the left and right moving sectors. 

The type II strings\index{string!type II}\index{type II string|see{string, type II}} are defined such that both their left and right moving sectors are $N=1$ super symmetric. The different types of the Type II strings arise due to the chirality of the super current on the left and right moving sector. The type IIA strings\index{string!type IIA} have opposite chirality, while the Type IIB strings\index{string!type IIB} have the same chirality of the super current in the two sectors. 

The heterotic strings\index{string!heterotic}\index{heterotic string|see{string, heterotic}} are of a different form. In our formulation the left moving sector is defined to be super symmetric, while the right moving sector is non super symmetric. Since the left moving sector gives a $10$ dimensional target space, the $16$ additional degrees of freedom in the right moving sector are considered to be free. They can either make up an $SO(32)$ gauge group or a $\mathbbm{E}_8 \times \mathbbm{E}_8$. This gives rise to the two different heterotic string theories.

\subsection{The type II string}\index{string!type II|(textbf}

Constructing a super symmetric string action puts constraints on the number of dimensions and as such, constraints on the type of world sheet fermions. It can be shown that the string action can be only super symmetric in $D=3,4,6,10$\cite{Green:1987sp}. We consider only the case when $D=10$. The super coordinates are then necessarily Majorana-Weyl which requires the assignment of chirality. Although chirality is a matter of convention the relative chirality of the two string directions, left and right, is physically different. The Type IIA string\index{string!type IIA} has opposite chirality for the left and right movers.  Type IIB strings\index{string!type IIB} have the same chirality in the left and right moving sector. It can be shown that the type IIA strings are not chiral contrary to the type IIB strings.

In general relativity, space time spans up four dimensions. We therefore need to configure the target space of the super string in such a way that the low energy region of the theory exhibits effectively a four dimensional space. One way to reduce the number of dimensions is by the compactification\index{compactification} of six dimensions. The simplest way to achieve this is to compactify the string on a torus\cite{Green:1987mn}. It has been shown that the maximal number of space time super symmetries for the type II strings is $N=8$\cite{Green:1984ct, Schwarz:1982jn}. 

The Minimal Super Symmetric Standard model (\mssm)\index{MSSM} is realized with $N=1$ spacetime super symmetry. There are several ways to reduce the number of super symmetries from $N=8$ down to $N=1$. In early works lattice constructions, orbifold methods, sigma models techniques, exactly solvable conformal blocks and notably the free fermionic formulation were used for this reduction\cite{Ferrara:1989nm}. More recently the connection between the free fermionic formulation\index{free fermionic formulation} and the orbifold construction\index{orbifold construction} has lead to the classification of the type II strings\cite{Gregori:1999ny}. 

It is believed that all string models are related by duality
. One duality which is easy to see is known as T-duality\index{duality!T-duality}\index{T-duality|see{duality, T-duality}}. This equivalence of two string theories is realized when the closed string is compactified on a circle. In this configuration not only the momenta contribute to the mass but also the winding modes of the string, i.e. how many times the string is wound on the circle. The spectrum of the string is then invariant under winding and momenta mode exchange by inverting the radius of the circle $R \to R'=\frac{\alpha'}{R}$. The result is that physically the $R\to \infty$ limit is equivalent to the $R'\to 0$ limit. This is purely due to the extended nature of the string.

Having constructed a theory with fields on a two dimensional world sheet, we would like to see what the low energy effective field theory description is. We can realise such a theory by describing the variation of the couplings as a function of the conformal invariance. Because we impose conformal invariance at the quantum level we find that the couplings in the $2D$ theory cannot run. This leads to constraints. Since constraints in general can be realised as equations of motion derived from an action we see that we have found an action that describes the low energy effective field theory in $D=10$. The low energy effective field theory that is derived from the type IIA/B strings is known as type IIA/B super gravity\index{super gravity!type IIA/B}\index{type IIA/B super gravity|see{super gravity, type IIA/B}}.

The Type IIA/B string has a $D=10$ target space as we have argued. This constraint is only realised at the perturbative limit. All the arguments were derived from the perturbative string. The perturbative regime is only valid when the string coupling is small. The string coupling is directly related to the dilaton field $g_s = e^{<\phi>}$. Considering super symmetry in itself however, leads to the conclusion that we can have at most a $D=11$ spacetime. Suppose now that the string coupling is directly related to the radius of a compactified eleventh dimension. We can then derive the low energy effective field theory from the string in the higher dimensional space. This would lead to a $11D$ super gravity\index{super gravity!11@$11D$} theory with one compactified dimension. This indeed turns out to be correct for the Type IIA string. It can be shown that in the limit where the compactified eleventh dimension of the super gravity field theory is reduced to zero by dimensional reduction, we retrieve the $10D$ type IIA super gravity low energy effective action. 
For more details concerning this derivation we refer to \cite{Townsend:1996xj}. 


\subsection{The heterotic string}\index{string!heterotic|(textbf}

The heterotic string was first constructed by \cite{Gross:1985dd} and employs both the bosonic string and the super string. Since the left and right moving sectors are completely independent, we set in the formulation of this thesis, the left moving sector to be super symmetric and the the right moving sector to be purely bosonic. Note that this is mirrored in some standard works\cite{Gross:1985dd, Green:1987sp, Green:1987mn, Polchinski:1998rq, Polchinski:1998rr}. This gives us a ten dimensional theory with remaining degrees of freedom coming from the right moving sector. These remaining degrees of freedom either form a $SO(32)$ or a $\mathbbm{E}_8 \times \mathbbm{E}_8$ gauge group.

In the \mssm\ we find a $N=1$ spacetime super symmetry and a $D=4$ target space or space time. Again there has been considerable effort to reduce the number of dimensions and super symmetries. Although many different ways have been employed, we note here two that have been used extensively. Toroidal compactifications\index{compactifications!toroidal}\index{toroidal compactifications|see{compactifications, toroidal}} are the most widely used compactification scheme, in particular orbifold compactifications\index{compactifications!orbifold}\index{orbifold compactifications|see{compactifications, orbifold}}. They are identified as $\mathbbm{Z}_M$ and $\mathbbm{Z}_M \times \mathbbm{Z}_N$ orbifolds. It has been shown that only a limited number of these types of orbifolds reduce the number of the super symmetries to the number of the \mssm.\cite{Quevedo:1996sv} Of particular interest for phenomenology are the $\mathbbm{Z}_2 \times \mathbbm{Z}_2$ orbifold compactifications\index{compactifications!Z2@$\mathbbm{Z}_2 \times \mathbbm{Z}_2$}. As this thesis is concerned with this last scheme of compactification we refer to chapter \ref{chap:realistic} for more details.

It has been shown that the $SO(32)$ and $\mathbbm{E}_8 \times \mathbbm{E}_8$ heterotic string theories are related by T-duality\index{duality!T-duality} at all orders in perturbation theory. The reason for this lies in the fact that a compactified dimension allows for the presence of non-vanishing Wilson lines. These Wilson lines can break $SO(32) \to SO(16)\times SO(16)$. A similar configuration can be realised by the $\mathbbm{E}_8 \times \mathbbm{E}_8$ model with a compactified dimension of radius $1/R$. For more detail see \cite{Townsend:1996xj}.

In order to have a unification of the five string theories, heterotic string theory has to have a description in the $11D$ stringy theory, from now on called M-theory\index{M-theory}, as well. This description is known as Ho\^rava-Witten theory\cite{Horava:1996ma, Horava:1996qa}. To fully understand the rational for the exact description of this theory, an explanation of the different dualities of the five string theories is necessary. In section \ref{sec:dualities} we describe these issues in more detail. To obtain the type IIA string from M-theory we have compactified this theory on a circle. If we instead compactify on the orbifold $S^1/Z_2$ or a line segment we find a chiral $N=1$ super gravity theory with $\mathbbm{E}_8$ vector multiplets on each orbifold fixed plane or the boundaries of the line segment. It has been shown that the low energy effective field theory of heterotic $\mathbbm{E}_8 \times \mathbbm{E}_8$ string theory describes the same particle content and it is therefore believed that by compactifying M-theory on the $S^1/\mathbbm{Z}_2$ orbifold, we obtain $\mathbbm{E}_8 \times \mathbbm{E}_8$ heterotic string theory. Similar to the type IIA case again the heterotic string coupling constant is proportional to the length of the line segment or radius.\index{string!heterotic|)}

\section{Dualities}\label{sec:dualities}\index{dualities|(textbf}

We have seen that the dimensional reduction of $11D$ super gravity is type IIA super gravity. In this construction we have compactified the $11D$ theory on a circle. We can also consider compactifying M-theory\index{M-theory} on a $T^2$ torus. We will follow \cite{Townsend:1996xj} in our approach. One would expect that this would reduce the theory to a $9D$ super gravity theory. However since we are dealing with a stringy theory, membranes are naturally in the spectrum of the M-theory. Membranes appear more naturally in type I strings. This type I theory of strings has both open and closed strings. The open strings have end points and one can show that these end points are located on higher dimensional planes called membranes. Since we only consider closed string theories we do not explain them here in more detail. We refer to the books \cite{Polchinski:1998rq,Polchinski:1998rr,Green:1987sp,Green:1987mn} for more details. These branes can wrap the $T^2$ torus just as strings can wrap the circle as explained before. Reducing the size of the torus but maintaining its shape one can show that the wrapping modes become massless and complete the spectrum of the chiral Type IIB string spectrum. We have thus realised a connection, through M-theory with IIA and IIB string theory. The connection can be realised even at the $D=10$ level. It has been shown that the type IIA is T-dual to the type IIB string. The string coupling of the type IIB string is then $g_s^{(B)} = \frac{R_{11}}{R_{10}}$. Since a reparameterisation of the torus interchange $R_{10}$ and $R_{11}$ we can see that the type IIB string with string coupling $g_s$ is dual to the type IIB string with string coupling $1/g_s$. This type of duality is known as S-duality\index{duality!S-duality}\index{S-duality|see{duality, S-duality}}. We see that the type II strings can be seen as different regions of a $11D$ theory. 

In order to see how the heterotic strings are realised by dualities it is necessary to include type I strings. Open string theory can be constructed using the type IIB strings. The Type IIB string diagrams are orientable manifolds. The type I strings are equivalent except for the fact that their diagrams are unorientable. This renders the type I closed string sector anomalous. This can be countered by including the open string sector of the Type I strings. The $SO(32)$ gauge group arises by including Chan-Paton factors at the end points of the open strings. This was found in \cite{Green:1984sg} and already then there was a hint of the heterotic string theories. It can be shown that the T-dual type I theory (Type IA) is equivalent to the Type IIA theory compactified on the orbifold $S^1/\mathbbm{Z}_2$. The fixed points of this orbifold are 8-planes with a $SO(16)$ gauge group. Now again we can lift the Type IIA to M-theory and we find that type IA is M-theory compactified on a cylinder.


Although all string theories are different limits of this M-theory, it has not been possible yet to write down the M-theory explicitly. Consequently we are left with all sorts of different descriptions for the same thing. The benefit of all these different descriptions is that one description suits one purpose better than the other. In Type I string theory the gauge group arises from Chan-Paton factors attached to the end points of the open string. Consequently Type I string theory cannot describe a chiral spectrum. In heterotic string theory we do not find this obstacle as is shown in chapter \ref{chap:realistic} and subsequent chapters. Since the \mssm\ is a chiral theory, heterotic string theory provides a convenient description in the search for phenomenological string theories.

\chapter{The Orbifold Background}\label{chap:orbifold.setup}

In this chapter we set up the orbifold description of the heterotic string. The total string amplitude is a sum over all possible Riemann surfaces, similar to the sum over all Feynman graphs in QFT, moded out by conformal invariance, which relates Riemann surfaces to each other. Only the tree amplitude and the first loop amplitude are well understood in string theory. The tree amplitude can be parameterised as a sphere by a stereographical projection. Since the tree amplitude does not take quantum corrections into account we will focus on the next simplest closed Riemann surface after the sphere, the torus. We discuss its symmetries. We then consider the bosonic string in one space time dimension as a toy model. We continue to extend this discussion to the higher dimensional case. This leads us to the fermionization and bosonization on the world sheet torus similar to what we have seen in section \ref{sec:bosonization}. We follow \cite{Kiritsis:1997hj} in our approach. We first consider this equivalence in the one dimensional target space case after which we extend this to the two dimensional background case. We continue to introduce the construction of orbifolds and we explain the $S^1/\mathbbm{Z}_2$ orbifold in detail. We define twists on the two dimensional target space. Having dealt with the higher dimensional orbifolds we introduce shifts on the compactified dimensions. Again we start with the one dimensional case which we generalise to the two dimensional case.

\section{The Torus}\index{torus, the|(}

In this section we show the symmetries of the torus that are derived from conformal invariance of the string and reparameterisation invariance of the torus. Since the surface of the one-loop string amplitude is conformally invariant we can set the surface area to $1$. If we pick coordinates $\sigma_1, \sigma_2 \in [0,1]$ the area of the torus is $1$ if the determinant of the torus metric is one. We parameterise the torus metric by a single complex number $\tau = \tau_1 + i \tau _2$ with $\tau_2 \ge 0$. The metric of the torus is then defined as
\begin{equation}\label{eq:torus.metric}
g^{ij} = \frac{1}{\tau_2}\begin{pmatrix} |\tau|^2 & -\tau_1\\ -\tau_1 & 1 \end{pmatrix}.
\end{equation}
In this parameterisation $\tau$ is known as the complex structure or modulus of the torus and is usually called the Teichm\"uller parameter\index{Teichm\"uller parameter}. 

\begin{figure}[t]
\begin{center}
\includegraphics[width=0.6\textwidth]{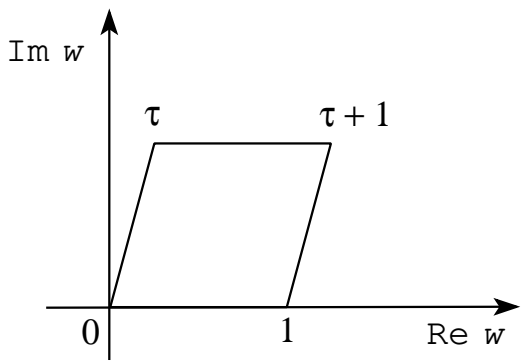}
\caption{The torus as a quotient of the complex plane.}\label{fig:torus}
\end{center}
\end{figure}

We will now consider the symmetries of the torus. We can realise the torus by identifying points $w$ on the complex plane with 
\begin{align}
w &\to w+1, & w &\to w + \tau
\end{align}
as is done in figure \ref{fig:torus}.

It is then easy to see that the translation $\tau \to \tau + 1$ realises the same torus. We have thus found one symmetry of the torus. A similar type of reformation of the torus leads to the second symmetry. In the first case we have moved the defining vector $\tau$ by a value of $1$. The second reformation results in the transformation $\tau \to \frac{\tau}{\tau + 1}$.
We can now realise the two orthogonal modular transformations\index{modular!transformations}
\begin{subequations}\label{eq:mod}
\begin{eqnarray}
\tau &\to& \tau + 1,\\
\tau &\to& -\frac{1}{\tau}.
\end{eqnarray}
\end{subequations}
Another way to write these transformations exhibits the modular group more clearly. We can write equations \eqref{eq:mod} as
\begin{equation}
\tau' = \frac{a\tau + b}{c\tau + d},
\end{equation}
where we can put the numbers $a,b,c,d$ in a matrix formulation
\begin{equation}
A = \begin{pmatrix} a & b\\ c& d \end{pmatrix}.
\end{equation}
The matrix A has integer entries and has determinant $1$ and forms the group $SL(2,\mathbbm{Z})$. Since a change of sign does to affect the modular transformation we find that the modular group\index{modular!group} is $PSL(2,\mathbbm{Z})=SL(2,\mathbbm{Z})/\mathbbm{Z}_2$.\index{torus, the|)}

\section{The String on a Compact Target Space}\index{string!compact|(}

In this section we describe the compactified bosonic string and the free fermionic string. We start by setting the bosonic string in one compactified space dimension. We calculate the partition function starting with the action. We then extend this analysis to the two dimensional case. We employ a similar tactic with the free fermionic string. Again we derive the partition function of a string with two holomorphic and two anti-holomorphic free fermions on the world sheet starting with the action.

\subsection{The bosonic string}\label{sec:boson}\index{string!compact, bosonic|(}

We discuss the compactification of the bosonic string on one compact dimension. We set the string for convenience in a one dimensional target space and do not consider any anomalies. The field $X$ takes its values on the circle with radius $R$. The Wick rotated Polyakov action for the string in a one-dimensional background is similar to equation \eqref{eq:polyakov1}
\begin{equation}\label{eq:bosonaction}
S = \frac{1}{4\pi} \int d^2\sigma \sqrt{g} g^{ij} \partial_iX \partial_j X,
\end{equation}
where $X$ is again a map from the world sheet to the target space and $g$ is the metric on the world sheet.
Using equation \eqref{eq:torus.metric} we evaluate the partition function of this configuration
\begin{equation}
Z(R) = \int \mathcal{D}X e^{-S}.
\end{equation} 
To evaluate this path integral we consider fluctuations of the classical field $X_{class}$
\begin{equation}
X = X_{class} + \chi.
\end{equation}
Since the classical field is periodic when it goes around the two non-contractible loops of the torus we find that the classical equation of motion of the field $X_{class}$ is realised when
\begin{equation}
X_{class} = 2\pi R\ (n\sigma_1 + m\sigma_2).
\end{equation}
The action is separated into a quantum and a classical contribution. We can therefore write the partition function $Z(R)$ as
\begin{equation}
Z(R) = \sum_{m,n \in \mathbbm{Z}} e^{-S_{m,n}} \int D\chi e^{-S(\chi)},
\end{equation}
where
\begin{equation}
S_{m,n} = \frac{\pi R^2}{\tau_2} |m - n\tau|^2,
\end{equation}
which leaves us to solve the quantum contribution. We expand the quantum field $\chi$ in terms of eigenfunctions $\psi_{m_1,m_2}$ of the Laplacian, defined by the equations of motion for the field $X$. The quantum contribution of the action is written in terms of the Fourier components and the eigenvalues $\lambda_{m_1,m_2}$ of the eigenfunctions $\psi_{m_1,m_2}$ of the Laplacian. Putting everything together we get
\begin{equation}
\int D\chi e^{-S(\chi)} = \frac{2\pi R}{\prod_{m_1,m_2}' \lambda_{m_1,m_2}^{1/2}} = \frac{2\pi R}{\sqrt{\det'\Box}},
\end{equation}
where the product runs over the eigenvalues of all but the zero modes. Using $\zeta$-regularisation we find that the determinant is \cite{Ginsparg:1988ui}
\begin{equation}
\text{$\det$}'\Box = 4 \pi^2\tau_2 \eta^2(\tau) \bar{\eta}^2(\bar{\tau}),
\end{equation}
with the Dedekind eta function\index{Dedekind eta function|see{eta function, Dedekind}}\index{eta function, Dedekind} defined as
\begin{align}
\eta &= q^{\frac{1}{24}} \prod_{m > 0} \left( 1 - q^m \right),&  q&=e^{2\pi i\tau},
\end{align}
and we find that the partition function for the bosonic string on a one dimensional compactified target space is
\begin{equation}\label{eq:boson}
Z(R) = \frac{R}{\sqrt{\tau_2} |\eta|^2} \sum_{m,n \in \mathbbm{Z}} e^{-\frac{\pi R^2}{\tau_2} |m - n\tau|^2}.
\end{equation}
We have written the partition function in the winding mode representation. Performing a Poisson resummation on $m$ we write this in the momentum mode representation. We obtain, using the formula for Poisson resummation\cite{Kiritsis:1997hj}
\begin{equation}\label{eq:t20}
\sum_{m_i\in Z}e^{-\pi m_im_jA_{ij}+\pi B_im_i}= \frac{1}{\sqrt{\det A}} \sum_{m_i\in Z}e^{-\pi(m_k+iB_k/2) (A^{-1})_{kl}(m_l+iB_l/2)},
\end{equation}
the partition function 
\begin{equation}
Z(R) \equiv \frac{\Gamma_{1,1}(R)}{|\eta|^2} =  \sum_{m,n \in \mathbbm{Z}} \frac{ q^{\frac{P_L^2}{2}} \bar{q}^{\frac{P_R^2}{2}}}{\eta\bar{\eta}},
\end{equation}
where
\begin{align}
P_L &= \frac{1}{\sqrt{2}} \left( \frac{m}{R} + nR \right), & P_R &= \frac{1}{\sqrt{2}} \left( \frac{m}{R} - nR \right).
\end{align}
The modulus of the compactified dimension $R$ has not been fixed. The quantum contribution of the partition function is encoded in the $\eta$ functions while the classical contribution is described by the left and right moving momenta. T-duality is now realised by interchanging simultaneously $R\to 1/R$ and $m \leftrightarrow n$. 

We have given the partition function of the bosonic string on one compactified dimension. We extend this discussion to the higher dimensional case.

Let us therefore parameterise the background by its metric $G_{ij}$ and its anti-sym\-me\-tric tensor $B_{ij}$, where $i,j$ runs over the number of target space dimensions. For simplicity we require the entire background to be compact. We parameterise the world sheet in terms of its metric $g_{ab}$ and its anti-symmetric tensor $\epsilon_{ab}$. The action for this system is
\begin{equation}
S = \frac{1}{4\pi} \int d^2 \sigma \sqrt{\det g}\ g^{ab} G_{ij} \partial_aX^i\partial_bX^j + \epsilon^{ab}B_{ij} \partial_aX^i\partial_bX^j.
\end{equation}
We find that the partition function of the bosonic string in this configuration is
\begin{equation}\label{eq:boson2}
Z_{d,d}(G,B) = \frac{\sqrt{\det G}}{\left(\sqrt{\tau_2}\eta\bar{\eta}\right)^d} \sum_{\overrightarrow{m},\overrightarrow{n}} e^{-\frac{\pi\left( G_{ij} + B_{ij}\right)}{\tau_2} \left[m_i - n_i\tau\right]\left[m_j + n_j\bar{\tau}\right]}.
\end{equation}
Again we can do a Poisson resummation using equation \eqref{eq:t20}, which leads to the momentum representation of the partition function
\begin{equation}\label{eq:gammadd}
Z_{d,d}(G,B) \equiv \frac{\Gamma_{d,d}(G,B)}{\eta^d\bar{\eta}^d} = \sum_{\overrightarrow{m},\overrightarrow{n} \in \mathbbm{Z}^d} \frac{q^{\frac{P_L^2}{2}} \bar{q}^{\frac{P_R^2}{2}}}{\eta^d\bar{\eta}^d},
\end{equation}
where
\begin{subequations}\label{eq:momentagammadd}
\begin{eqnarray}
P^2_{L,R} &=& P^i_{L,R} G_{ij} P^j_{L,R},\\
P^i_L &=& \frac{G^{ij}}{\sqrt{2}} \left[ m_j + \left(B_{jk} + G_{jk}\right)n_k\right],\\
P^i_R &=& \frac{G^{ij}}{\sqrt{2}} \left[ m_j + \left(B_{jk} - G_{jk}\right)n_k\right],
\end{eqnarray}
\end{subequations}
where the $-$ sign difference in the left and right moving momenta is due to the fact that when we expand the product of equation \eqref{eq:boson2} we find that the anti-symmetric tensor carries a factor $i$ in front of it while the background metric does not. We have therefore realised an expression for the partition function of the compactified bosonic string in terms of both the winding modes as well as the momentum modes.\index{string!compact, bosonic|)}

\subsection{The free fermionic string}\index{string!compact, fermionic|(}

We discuss the partition function of two free Majorana-Weyl fermions on the torus. Although we can pursue a similar method as employed for the bosonic string, we use a different method here and follow \cite{Ginsparg:1988ui} to which we refer for more intricate details.
The action of two free left and right moving fermions is
\begin{equation}
S = \frac{1}{8\pi} \int d^2z\ \  \psi^i \bar{\partial} \psi^i+ \bar{\psi}^i \partial \bar{\psi}^i,
\end{equation}
where $i$ runs over the number of fermions. Again we are interested in the partition function of these free fermions. It can be shown that this becomes 
\begin{eqnarray}
Z = \int e^{-S} &=& tr\ e^{2\pi i \tau_1 P}e^{-2\pi \tau_2 H}\nonumber\\
         &=& q^{-\frac{c}{24}} \bar{q}^{\frac{\bar{c}}{24}}\ tr\ q^{L_0} \bar{q}^{\bar{L}_0}\label{eq:trace},
\end{eqnarray}
where
\begin{align}\label{eq:laurent}
L_n &= \oint \frac{dz}{2\pi i} z^{n+1} T(z), & \bar{L}_n &= \oint \frac{d\bar{z}}{2\pi i} \bar{z}^{n+1} \bar{T}(\bar{z})
\end{align}
are the Laurent series expansion of the holomorphic ($T(z)$) and anti-holomorphic ($\bar{T}(\bar{z})$) components of the energy-momentum tensor of the system. We are left with solving the trace over the operators $L_0$ and $\bar{L_0}$. In order to evaluate the trace we need to define the spin structure on the torus or equivalently define the boundary conditions of the fermions when transported around the non-contractible loops of the torus. When we evaluate the vacuum energy of the different configurations, we find that equation \eqref{eq:trace} represents the sector where the fermions are anti-periodic. It can be shown that the boundary conditions are changed when we do a modular transformation as defined in equations \eqref{eq:mod}. We therefore need to consider the other sectors as well. 

The $c=\frac{1}{2}$ representations of the Virasoro algebra, which is the algebra realised by the Laurent modes of the energy momentum tensor written in equation \eqref{eq:laurent}, for anti-periodic fermions can be identified with the direct sum of the highest weight states with $h=0$ and $h=\frac{1}{2}$, where $h$ is the conformal weight of the states. The vacuum state is the highest weight state with $h=0$. Since we are only interested in counting the representations of the Virasoro algebra once, we need to project out half of the states. A similar argument holds for the periodic fermions.

Let us separate the periodic and the anti-periodic sector in equation \eqref{eq:trace}. We include the vacuum state in the partition function which leads to the expansion
\begin{eqnarray}
( q \bar{q} )^{-\frac{1}{24}}\ tr\ q^{L_0} \bar{q}^{\bar{L}_0} &=& (q\bar{q})^{-\frac{1}{24}}\ tr_{A\bar{A}}\ \frac{1}{2} \left( 1 + (-1)^F \right) q^{L_0} \bar{q}^{\bar{L}_0}\nonumber\\
 &+& (q\bar{q})^{-\frac{1}{24}}\ tr_{P\bar{P}}\ \frac{1}{2} \left( 1 \pm (-1)^F \right) q^{L_0} \bar{q}^{\bar{L}_0}.\label{eq:pptrace}
\end{eqnarray}
In this expression we see that there still remains the choice of the second trace in the periodic sector. We choose the sign to be positive from here on. We discuss the significance of the sign in section \ref{sec:fermionic.heterotic}. We are thus left with the task to evaluate the traces. It can be shown that
\begin{subequations}\label{eqs:theta}
\begin{align}
q^{-\frac{1}{24}}\ 	tr_A \ 		q^{L_0} &=   q^{-\frac{1}{24}} \prod_{n=0}^\infty \left( 1 + q^{n+1/2} \right)^2 	&= \frac{\theta_3}{\eta},\\
q^{-\frac{1}{24}}\ 	tr_A \ (-1)^F 	q^{L_0} &=   q^{-\frac{1}{24}} \prod_{n=0}^\infty \left( 1 - q^{n+1/2} \right)^2 	&= \frac{\theta_4}{\eta},\\
2 q^{-\frac{1}{24}}\  	tr_P \ 		q^{L_0} &= 2 q^{\frac{1}{12}}  \prod_{n=0}^\infty \left( 1 + q^{n}     \right)^2 	&= \frac{\theta_2}{\eta},\\
2 q^{-\frac{1}{24}}\ 	tr_P \ (-1)^F 	q^{L_0} &= 2 q^{\frac{1}{12}}  \prod_{n=0}^\infty \left( 1 - q^{n}     \right)^2 	&= \frac{\theta_1}{\eta},
\end{align}
\end{subequations}
where $\theta_i$ are the well known Jacobi theta functions\index{Jacobi theta functions|see{theta functions, Jacobi}}\index{theta functions, Jacobi} \cite{erdlyi:1953}
\begin{equation}\label{eq:thetadefinition}
\theta[^{a^I}_{b^I}] = \sum_{n\in \mathbbm{Z}} q^{\frac{\left(n-{\frac{a^I}{2}}\right)^2}{2}} e^{2\pi i\left(v-{\frac{b^I}{2}}\right) \left(n-{\frac{a^I}{2}}\right)},
\end{equation}
and we define the Jacobi theta functions as
\begin{align}\label{eq:jacobishort}
\theta[^1_1] &= \theta_1 ,& \theta[^1_0] &= \theta_2, & \theta[^0_0] &= \theta_3 ,& \theta[^0_1] &= \theta_4.
\end{align}
These formulae should be complex conjugated for the right moving sector. When we evaluate $\frac{\theta_1}{\eta}$ we find that this vanishes, but we nevertheless keep its formal description as it will be important for later use. One more difficulty arises now in the evaluation of the path integral in the periodic sector. 

The Fourier expansion of the periodic fermion is
\begin{equation}
i \psi(z) = \sum_{n \in \mathbbm{Z}} \psi_n z^{-n}
\end{equation}
and it can be shown that the commutation relation for the Fourier components of the fermion is
\begin{equation}
\{ \psi_n, \psi_m \} = \delta_{m+n,0}.
\end{equation}
Acting on the vacuum with the zero mode $\psi_0$, we find that this does not change the eigenvalue of $L_0$ leading to a degenerate vacuum consisting of two states. We come back to this issue in section \ref{sec:susybackground}.

The partition function of two holomorphic fermionic fields and two anti-holomorphic fermionic fields on the torus is \cite{Ginsparg:1988ui}
\begin{equation}\label{eq:fermionN}
Z = \frac{1}{2} \sum_{a,b=0,1} \left| \frac{\theta[^a_b]}{\eta} \right|^2.
\end{equation}
using the shorthand notation defined in equation \eqref{eq:jacobishort}.

\section{Bosonization and Fermionization}\index{bosonization|(}\index{fermionization|(}\label{sec:bosonization2}

In section \ref{sec:bosonization} we have shown that bosons can be interchanged with two fermions in a conformal field theory. In this section we show that this equivalence is also realised in the orbifold description of the string. 

In equation \eqref{eq:fermionN} we found for the partition function of two holomorphic fermionic fields and two anti-holomorphic fermionic fields on the torus is \cite{Ginsparg:1988ui}
\begin{equation}
Z = \frac{1}{2} \sum_{a,b=0,1} \left| \frac{\theta[^a_b]}{\eta} \right|^2.
\end{equation}
Applying a Poisson resummation on the partition function we obtain
\begin{equation}
 \left| \frac{\theta[^a_b]}{\eta} \right|^2 = \frac{1}{\eta\bar{\eta}} \frac{1}{\sqrt{2\tau_2}} \sum_{m,n \in \mathbbm{Z}} \exp \left[ -\frac{\pi}{2\tau_2}\left| n-b + \tau(m-a) \right|^2 + i\pi mn \right].
\end{equation}
We can then relabel $n\to n+b$ and $m\to m+a$ since $a,b \in \mathbbm{Z}$. When we consider now the whole partition function we find
\begin{multline}
\frac{1}{2} \sum_{a,b=0,1} \left| \frac{\theta[^a_b]}{\eta} \right|^2 = \\
\frac{1}{\eta\bar{\eta}} \sum_{a,b=0,1} \frac{1}{2 \sqrt{2\tau_2}} \sum_{m,n \in \mathbbm{Z}} \exp \left[ -\frac{\pi}{2\tau_2}\left| n + \tau m \right|^2 + i\pi (m+a)(n+b) \right].
\end{multline}
The summation over $b$ gives a factor of $2$ and requires $m+a \in 2\mathbbm{Z}$. We therefore find that the partition function of one holomorphic fermionic field and one anti-holomorphic fermionic field on the torus is
\begin{equation}\label{eq:nonselfdual}
Z\left(\frac{1}{\sqrt{2}}\right) = \frac{1}{\sqrt{2 \tau_2} |\eta|^2} \sum_{m,n \in \mathbbm{Z}} e^{-\frac{\pi}{2 \tau_2} |m - n\tau|^2}.
\end{equation}
We see that this is identical to equation \eqref{eq:boson} with the radius fixed at the maximal symmetry point $R = 1/\sqrt{2}$. Note that this value is not the self dual point under T-duality. 

We show the boson fermion equivalence in the case where the boson is compactified on a complex torus and follow \cite{Gregori:1999ny}. The complex torus can be parameterised by two complex moduli\index{moduli} in the same way the circle is parameterised by its radius. The moduli $(T=T_1 + iT_2,U=U_1+ iU_2)$ of the complex torus are written in terms of the metric $G_{ij}$ and the anti-symmetric tensor $B_{ij}$
\begin{align}
G_{ij} &= \frac{T_2}{U_2} \begin{pmatrix} 1 & U_1 \\ U_1 & |U|^2 \end{pmatrix}, & B_{ij} &= \begin{pmatrix} 0 & T_1 \\ -T_1 & 0 \end{pmatrix}.
\end{align}
The $\Gamma_{2,2}$ lattice\index{gamma@$\Gamma_{2,2}$ lattice|see{lattice, $\Gamma_{2,2}$}}\index{lattice! $\Gamma_{2,2}$} from equation \eqref{eq:gammadd} can now be written as
\begin{multline}\label{eq:gamma22}
\Gamma_{2,2}(T,U) = \sum_{\overrightarrow{m},\overrightarrow{n} \in \mathbbm{Z}^2} \exp \Bigg[ -\frac{\pi \tau_2}{T_2 U_2} \left| m_1U -m_2 + T(n_1 +Un_2) \right|^2 \\
+ 2\pi i \bar{\tau} (m_1n_1 + m_2n_2) \Bigg].
\end{multline}
Fixing the moduli at the the maximal symmetry point\index{self-dual point} $(T,U)=(i,i)$ leads to 
\begin{multline}
\Gamma_{2,2}(i,i) = \sum_{\overrightarrow{m},\overrightarrow{n} \in \mathbbm{Z}^2} \exp \Bigg[ \pi \tau_2 \left| m_1 i -m_2 + i(n_1 +in_2) \right|^2 \\
+ 2\pi i \bar{\tau} (m_1n_1 + m_2n_2) \Bigg].
\end{multline}
We find that $B_{ij}$ vanishes at this point in the moduli space and the metric $G_{ij}$ takes the form
\begin{equation}
G_{ij} = \begin{pmatrix} 1 & 0 \\ 0 & 1 \end{pmatrix},
\end{equation}
from which it is clear that, since equation \eqref{eq:nonselfdual} holds for the partition function of one holomorphic and one anti-holomorphic fermion field, we find that 
\begin{equation}
\frac{\Gamma_{2,2}(i,i)}{|\eta|^4} = \frac{1}{2} \sum_{a,b=0,1} \left| \frac{\theta[^a_b]}{\eta} \right|^4.
\end{equation}
At the left hand side we have described a bosonic string compactified on a torus of which we have fixed the moduli at the maximal symmetry point $(T,U) = (i,i)$ in the moduli space. On the right hand side we have described a fermionic string of which the fermionic fields can propagate freely on the string. This equivalence therefore shows that at the maximal symmetry point on the torus we can exchange bosons with free fermions. 
\index{bosonization|)}\index{fermionization|)}

\section{Orbifolds and Shifts}\label{sec:orbifoldsandshifts}

In this section we introduce the techniques for constructing orbifolds and shifts in a geometrical formulation. We start by defining the orbifold from a general manifold. We discuss the simplest example and extend this discussion to the bosonic string. We derive the bosonic string compactified to this simplest orbifold. We move on to the higher dimensional case, namely the target space of a critical string with four extended dimensions. We then introduce shifts on the target manifold. We start with the description of shifts in the one dimensional background after which we discuss shifts on the two dimensional torus representing part of the full target space of the critical string.

\subsection{The orbifold construction}\index{orbifolds|(}

Consider a manifold $\mathcal{M}$ with a discrete group action $G: \mathcal{M}\to \mathcal{M}$. The action of the discrete group on points on the manifold is $g \in G: x \to gx$. An orbifold is now defined\index{orbifolds!definition} as the quotient space $\mathcal{M}/G$. In general an orbifold is not a manifold. An orbifold remains a manifold however when the group acts freely, i.e. when there are no fixed points under the group action. One easy example of an orbifold is $S^1/ \mathbbm{Z}_2$. The group elements of $\mathbbm{Z}_2$ act on the points of $S^1$ as $g \in G=\mathbbm{Z}_2: x \to gx = -x$. This group action has two fixed points and the quotient space therefore defines an orbifold. It is easy to see that this quotient space can be represented by a line segment.

Let us now proceed to a group action on the torus. Without the introduction of the discrete group we have the following picture. Going around a non-contractible loop around the torus requires the field $X$ not to change. We now define the discreet group $G: \mathcal{T} \to \mathcal{T}$, where $\mathcal{T}$ represents the target space. In this context it is therefore possible that the field $X$, when it goes around a non-contractible loop, does not go back to its original value but goes to $gX$, which is in the quotient space identical to $X$. Since modular invariance interchanges the two non-contractible loops we see that any modular invariant partition function with an orbifold as target space has four sectors. For each non-contractible loop we have two sectors: on one sector we act on the fields with the group action and on the other we do not act on the fields.

We consider the simplest example of an orbifold, namely $S^1/\mathbbm{Z}_2$. The action for the free boson on the torus compactified on $S^1$ is described in equation \eqref{eq:bosonaction}. We see that this action is invariant under $X \to -X$ which defines an orbifold $S^1/\mathbbm{Z}_2$ as target space. Similar to the derivation of the free fermionic partition function the partition function for the free boson is written as
\begin{multline}
Z_{orb}(R) = (q\bar{q})^{-\frac{1}{24}}\ tr_{(+)}\ \frac{1}{2} (1+g) q^{L_0} \bar{q}^{\bar{L}_0} \\
+ (q\bar{q})^{-\frac{1}{24}}\ tr_{(-)}\ \frac{1}{2} (1+g) q^{L_0} \bar{q}^{\bar{L}_0},
\end{multline}
where $tr_{(\pm)}$ describes the untwisted and the twisted sector\index{twisted sector} respectively. The untwisted sector is defined as the sector where the field $X \to X$ around the non-contractible loop in the space direction on the torus. The twisted sector is the sector where $X\to gX=-X$ around the non-contractible loop in the space direction on the torus. We now evaluate the twisted and untwisted sector separately. 

Note that the untwisted-untwisted sector is identical to the bosonic partition function as derived in section \ref{sec:boson}. We are therefore left with the untwisted-twisted sector. Since the group action $g$ brings $X \to -X$ we find 
\begin{equation}
(q\bar{q})^{-\frac{1}{24}}\ tr_{(+)}\ \frac{1}{2}\ g\ q^{L_0} \bar{q}^{\bar{L}_0} = \frac{1}{2} \frac{(q\bar{q})^{-\frac{1}{24}}}{\prod_{n=1}^{\infty} (1 + q^n)(1+\bar{q}^n)} = \left| \frac{\eta}{\theta_2} \right|.
\end{equation}
Similarly we find for the twisted sector
\begin{multline}
(q\bar{q})^{-\frac{1}{24}}\ tr_{(-)}\ \frac{1}{2} (1+g) q^{L_0} \bar{q}^{\bar{L}_0} = \\
\left[ \frac{(q\bar{q})^{-\frac{1}{24}}}{\prod_{n=1}^{\infty} (1 - q^{n-1/2})(1 - \bar{q}^{n-1/2})} + \frac{(q\bar{q})^{-\frac{1}{24}}}{\prod_{n=1}^{\infty} (1 + q^{n-1/2})(1 + \bar{q}^{n-1/2})} \right]\\
 = \left|\frac{\eta}{\theta_4} \right| + \left|\frac{\eta}{\theta_3} \right|,
\end{multline}
leading to
\begin{equation}\label{eq:orbifoldtwist}
Z_{orb}(R) = \frac{1}{2} Z_{circ}(R) + \left| \frac{\eta}{\theta_2} \right| + \left|\frac{\eta}{\theta_4} \right| + \left|\frac{\eta}{\theta_3} \right|.
\end{equation}
We see that the twisted sector does not depend on the modulus of the circle $S^1$, while the untwisted sector does. By analysing only the twisted sector we can set the radius to \emph{any} value we like. Choosing the maximal symmetry point therefore does not change the structure of the twisted sector. We therefore move from \emph{any} point in the moduli space to the maximal symmetry point. 

We can write equation \eqref{eq:orbifoldtwist} as
\begin{equation}
Z_{orb}(R) = \frac{1}{2} \sum_{h,g=0,1} \frac{\Gamma_{1,1}[^h_g]}{|\eta|^2},
\end{equation}
where $\Gamma_{1,1}[^0_0] = \Gamma_{1,1}(R)$ and
\begin{equation}
\Gamma_{1,1}[^h_g] = 2 \frac{|\eta|^3}{|\theta[^{1-h}_{1-g}]|},\ \ \ (h,g) \ne (0,0).
\end{equation}
We fix the radius at the maximal symmetry point. If we use the identity $\theta_2\theta_3\theta_4=2\eta^3$ we find
\begin{equation}
Z_{orb}\left(\frac{1}{\sqrt{2}}\right) = \frac{1}{4} \left[ \sum_{a,b=0,1} \left| \frac{\theta[^a_b]}{\eta} \right|^2 + 2\frac{|\theta_3\theta_4|}{|\eta|^2}  + 2\frac{|\theta_2\theta_3|}{|\eta|^2} + 2\frac{|\theta_2\theta_4|}{|\eta|^2} \right].
\end{equation}
We see that at the maximal symmetry point the orbifold description from equation \eqref{eq:orbifoldtwist} is exactly equivalent to the fermionic description written down in equation \eqref{eq:fermionN}. By moving from any point in the moduli space to the maximal symmetry point we have not changed the structure of the twisted sector. It is fairly straight forward to rewrite this as
\begin{equation}
Z_{orb}\left(\frac{1}{\sqrt{2}}\right) = \frac{1}{2} \sum_{g,h =0,1} \frac{1}{2} \sum_{a,b=0,1} \left| \frac{\theta[^a_b]\theta[^{a+h}_{b+g}]}{\eta\bar{\eta}} \right|.
\end{equation}

We consider the torus moded out by a $\mathbbm{Z}_2$ symmetry. Along similar lines we find that this results in
\begin{equation}\label{eq:orbpoint}
Z_{orb}(T,U) = \frac{1}{2} \sum_{h,g=0,1} \frac{\Gamma_{2,2}[^h_g](T,U)}{|\eta|^4},
\end{equation}
with the twist parameterised by $h,g$. The partition function at an arbitrary point in the moduli space is similarly written as\cite{Gregori:1999ny}
\begin{equation}
\Gamma_{2,2}[^h_g] = 4  \frac{|\eta|^6}{|\theta[^{1+h}_{1+g}]\theta[^{1-h}_{1-g}]|} ,\ \ \ (h,g) \ne (0,0).
\end{equation}
and $\Gamma_{2,2}[^0_0](T,U) = \Gamma_{2,2}(T,U)$ as written in equation \eqref{eq:gamma22}. We have thus given the partition function for the bosonic string compactified on an orbifold $T^2 / \mathbbm{Z}_2$ \index{orbifolds|)} for all moduli $(T,U)$. We stress the point here that the structure of the twisted sector at \emph{any} point in the moduli space has not changed by fixing the moduli at the maximal symmetry point similar to the one dimensional compactification. The chiral structure of the $\mathbbm{Z}_2 \times \mathbbm{Z}_2$ orbifolds necessarily comes from the twisted sector as we show in chapter \ref{chap:fermform}. By considering the chiral structure at the maximal symmetry point we have therefore considered the chiral structure of all models connected to it by a change of the moduli of the internal space.

\subsection{The shifts}\index{shifts|(}

In the previous section we have constructed the string propagating on a torus. This has lead to winding and momentum modes. The winding modes represent whole loops around the compactified dimension. We can however also consider transporting the string halfway around the compactified dimension. This transportation is known as a shift, or more specific a $\mathbbm{Z}_2$ shift as two subsequent shifts are identical to the identity.

From equation \eqref{eq:fermionization} we see that a shift around half the compactified dimension is realised by
\begin{align}
X(z) &\to X(z)+\pi\ \  \leftrightarrow\ \  \psi(z) \to -\psi(z)
\end{align}
and similarly for the right moving sector. Again in the case of the torus we have two non-contractible loops, which are interchanged by modular invariance. To construct a modular invariant partition function that incorporates shifts, we therefore have to consider the action of the shift on all possible directions on the string world sheet leading to four terms, each describing the shift along either of the two non-contractible loops. We consider only symmetric shifts, i.e. the shift is introduced on both the left and right moving sector simultaneously.

We first consider the bosonic string compactified to one dimension. The partition function for this system is described in \eqref{eq:boson}. Adding the shift along the non-contractible loops is then realised as
\begin{equation}
Z(R) = \frac{1}{2} \sum_{p,q=0,1} Z[^p_q](R),
\end{equation}
where
\begin{equation}
Z[^p_q](R) = \frac{R}{\sqrt{\tau_2} |\eta|^2} \sum_{m,n \in \mathbbm{Z}} \exp \left[-\frac{\pi R^2}{\tau_2} \left|(m+\frac{q}{2}) - (n+\frac{p}{2})\tau\right|^2\right].
\end{equation}
The factor $\frac{p}{2}$ is necessary to avoid a relabelling of the indices. In that sense it is clear that this factor realises the $\mathbbm{Z}_2$ shift. Performing a Poisson resummation we can set the partition function in the momentum representation. We obtain
\begin{equation}\label{eq:gamma11shift}
Z[^p_q](R) = \sum_{m,n \in \mathbbm{Z}} (-1)^{qm} \frac{ q^{\frac{P_L^2}{2}} \bar{q}^{\frac{P_R^2}{2}}}{\eta\bar{\eta}},
\end{equation}
where
\begin{align}
P_L &= \frac{1}{\sqrt{2}} \left( \frac{m}{R} + (n+\frac{p}{2})R \right), & P_R &= \frac{1}{\sqrt{2}} \left( \frac{m}{R} - (n+\frac{p}{2})R \right).
\end{align}
Fixing the radius at the maximal symmetry point and performing a Poisson resummation leads to the partition function
\begin{equation}
Z\left(\frac{1}{\sqrt{2}}\right) = \frac{1}{2} \sum_{p,q=0,1} \frac{1}{2} \sum_{a,b=0,1} \left| \frac{\theta[^{a+p}_{b+q}]}{\eta} \right|^2.
\end{equation}

We extend this discussion to the torus. In the case of the torus we find two independent shifts since there are two non-contractible loops. We find that the full partition function is written as
\begin{equation}
Z(T,U) = \frac{1}{4} \sum_{p_1,q_1=0,1} \sum_{p_2,q_2=0,1} \frac{\Gamma_{2,2}\left[\begin{matrix} p_1 & p_2\\ q_1 & q_2\end{matrix}\right](T,U)}{|\eta|^4}.
\end{equation}

Applying these shifts, we can find the partition function for the bosonic string compactified on a torus, using equation \eqref{eq:gamma22}, at an arbitrary point in the moduli space fairly easy. Using the momentum representation as displayed in equation \eqref{eq:gammadd} and equation \eqref{eq:gamma22}, we follow along the same lines as equation \eqref{eq:gamma11shift} and find
\begin{multline}\label{eq:shiftmomentum}
\Gamma_{2,2}\left[\begin{matrix} p_1 & p_2\\ q_1 & q_2\end{matrix}\right](T,U) = \frac{1}{2} \sum_{\overrightarrow{m},\overrightarrow{n} \in \mathbbm{Z}^2} (-1)^{q_1m_1 + q_2m_2} \\
\exp \Bigg[ -\frac{\pi \tau_2}{T_2 U_2} \bigg| m_1U - m_2  + T\left\{(n_1+\frac{p_1}{2}) +U(n_2+\frac{p_2}{2})\right\} \bigg|^2 \\
+ 2\pi i \bar{\tau} \left\{m_1(n_1+\frac{p_1}{2}) + m_2(n_2+\frac{p_2}{2})\right\} \Bigg].
\end{multline}
Again we fix the moduli at the maximal symmetry point $(T,U) = (i,i)$ and we obtain
\begin{equation}\label{eq:fermpoint}
\Gamma_{2,2}\left[\begin{matrix} p_1 & p_2\\ q_1 & q_2\end{matrix}\right](i,i) = \frac{1}{2}\sum_{a,b=0,1} 
\left|\theta[^{a+p_1}_{b+q_1}]
\theta[^{a+p_2}_{b+q_2}] \right|^2.
\end{equation}
We see that at the maximal symmetry point the bosonic partition function is \emph{exactly} identical to the fermionic partition function. If the fer\-mi\-onic partition function is therefore known, we can easily move away from the maximal symmetry point to \emph{any} point in the moduli space using the identity presented in equation \eqref{eq:fermpoint}. In the following chapters we construct from the free fermionic description the partition function at the maximal symmetry point. Using equation \eqref{eq:fermpoint} we then move to \emph{any} point in the moduli space. As the chiral structure does not depend on the moduli we select the maximal symmetry point to analyse the chiral structure. This structure is \emph{completely} determined by the twisted sector and is therefore equivalent for \emph{all} partition functions related to each other by a change of the moduli of the internal space. 

\chapter{The Fermionic String}\label{chap:fermionic}

In this chapter we set up the free fermionic construction of the heterotic string. The partition function at an arbitrary point in the moduli space is derived. The description of the heterotic string at the self dual point under T-duality follows. We write the partition function at the self-dual point in the most general way which enables us to derive constraints on the form of the partition function. We continue to derive all the necessary and sufficient constraints for the construction of the free fermionic model of the heterotic string. Having derived the tools for the construction we end this chapter by summarising them.

\section{The Heterotic String on a Compact Space}\index{string!heterotic!compact|(}\label{sec:fermionic.heterotic}

In this section we derive the partition function of the heterotic string at the self dual point under T-duality. We follow \cite{Kiritsis:1997hj, Ginsparg:1988ui} initially after which \cite{Antoniadis:1988wp} proves helpful. We derive firstly the partition function at an arbitrary point in the moduli space after which we descend to the self-dual point. We write the partition function at the self-dual point in a general manner which enables us to derive constraints on the form of the partition function induced by modular invariance. We then construct the free fermionic models of the heterotic string.

We construct the partition function for the heterotic string in a $D=4$ super symmetric extended target space and a $D=6$ compactified internal space. We recall that a free compactified boson is from equation \eqref{eq:boson}
\begin{equation}
Z(R) = \frac{R}{\sqrt{\tau_2} |\eta|^2} \sum_{m,n \in \mathbbm{Z}} e^{-\frac{\pi R^2}{\tau_2} |m - n\tau|^2}.
\end{equation}
When we take the limit $R \to \infty$ we find that since the exponent is completely suppressed apart from the $m=n=0$, the partition function is
\begin{equation}
\lim_{R\to \infty}Z(R) = \lim_{R\to \infty} \frac{R}{\sqrt{\tau_2} |\eta|^2}.
\end{equation}
Since it was expected that the partition function would diverge with the volume of the space, we remove this divergence to get the correct partition function for one boson in an extended space
\begin{equation}
Z_B = \frac{1}{\sqrt{\tau_2} |\eta|^2}.
\end{equation}
We can now construct the extended sector of the heterotic string. We write the full partition function as
\begin{equation}
Z = Z_{ext}\ Z_{compact},
\end{equation}
where we focus on $Z_{ext}$. The extended sector consists only of extended bosons, since the super partners of the bosons are free fermions of the world sheet. We write the extended partition function in the light cone gauge fixing two bosonic coordinates
\begin{equation}
Z_{ext} = \left[ \frac{1}{\sqrt{\tau_2} |\eta|^2} \right]^2.
\end{equation}
We are left with describing the compactified partition function. In the case of a super symmetric target space the eight left moving free super partners of the bosonic coordinates are separated\footnote{The word separated will become clear later on.} from the remaining degrees of freedom. The six left and right moving bosons are compactified as are the additional 16 degrees of freedom on the right moving side. This leads to the compact partition function
\begin{equation}\label{eq:Zcompact}
Z_{compact} =  \frac{1}{2} \sum_{a,b=0,1} \cc{a}{b} \frac{\theta[^a_b]^4}{\eta^4} \frac{\Gamma_{6,6+16}}{\eta^6\bar{\eta}^{6+16}}.
\end{equation}
If we would take $\cc{a}{b} = 1, \forall a,b \in \{0,1\}$ we do not find space-time super symmetry. The reason for this is that the bosonic and fermionic part of the partition function do not cancel. We can realise cancellation by using the Jacobi identity\cite{Angelantonj:2002ct}
\begin{equation}
\theta_3^4 - \theta_4^4 -\theta_2^4 \pm \theta_1^4 = 0.
\end{equation}
The partition function of the compactified heterotic string is  realised as
\begin{equation}\label{eq:n4partition}
Z = \frac{1}{2\tau_2 |\eta|^4} \sum_{a,b=0,1} (-1)^{a+b+\mu ab} \frac{\theta[^a_b]^4}{\eta^4} \frac{\Gamma_{6,22}}{\eta^6\bar{\eta}^{22}},
\end{equation}
where $\mu$ determines the sign of $\theta_1$ in the partition function. The physical relevance of $\mu$ in this context may not be clear immediately. In deriving the partition function for the free fermions in equation \eqref{eq:pptrace}, we made a choice by projecting out half of the anti-periodic fermionic states. This choice formally should also be done with the periodic fermions. Since the periodic fermions realise the spacetime fermions, this projects out either the \emph{up} of the \emph{down} states and therefore effectively sets the chirality of the super symmetry\index{chirality!super symmetry}. The value of $\mu$ thus sets the chirality of the space time fermions.

The general $\Gamma_{6,6+16}$ lattice\index{lattice! $\Gamma_{6,6+16}$|(} depends on $6\times 22$ moduli: the metric $G_{ij}$ and the antisymmetric
tensor $B_{ij}$  of the six dimensional internal space, as well as the
Wilson lines $Y^I_i$ that appear in the  2d-world--sheet action
\begin{eqnarray}
S &=& \frac{1}{4\pi} \int d^2\sigma \sqrt{g}g^{ab}G_{ij}\partial_a
X^i\partial_bX^j + \frac{1}{4\pi} \int d^2\sigma \epsilon^{ab}
B_{ij}\partial_aX^i\partial_bX^j\nonumber\\
  & & + \frac{1}{4\pi} \int d^2\sigma \sqrt{g} \sum_I \psi^I \Big[ \bar{\nabla}
 + Y^I_i\bar{\nabla}X^i \Big]\bar{\psi}^I.\label{eq:gamma622action}
\end{eqnarray}
Here $i$ runs over the internal coordinates and $I$ runs over the extra $16$
right--moving degrees of freedom described by $\bar{\psi}^I$.
We can evaluate the partition function for this action along the same lines as done in section \ref{sec:boson}. We find that the partition function at an arbitrary point in the moduli space is
\begin{eqnarray}\label{eq:internal}
\Gamma_{6,6+16} &=& \frac{(\det G)^3}{\tau_2^3}\sum_{m,n} \exp\bigg\{ -\pi
\frac{T_{ij}}{\tau_2}[m^i + n^i\tau][m^j + n^j\bar{\tau}]\bigg\}\nonumber\\
& & \times \frac{1}{2}\sum_{\gamma,\delta} \prod_{I=1}^{16}\exp\Big[-i\pi n^i
\big(m^j + n^j \bar{\tau}\big)Y^I_iY^I_j\Big]\nonumber\\
& & \times \bartheta{\gamma}{\delta}\big(Y^I_i
(m^i+n^i\bar{\tau})|\tau\big),\label{eq:gamma622winding}
\end{eqnarray}
where $T_{ij} = G_{ij} + B_{ij}$.

Equation \eqref{eq:gamma622winding} is the winding mode representation of the
partition function. Using a
Poisson resummation we can put it  in the momentum representation form
\begin{equation}
\Gamma_{6,22} = \sum_{P, \bar{P}, Q} \exp\bigg\{\frac{i\pi\tau}{2}P_iG^{ij}P_j
-\frac{i\pi\bar{\tau}}{2}\bar{P}_iG^{ij}\bar{P}_j - i\pi\bar{\tau}\hat{Q}^I\hat
{Q}^I\bigg\},
\end{equation}
with
\begin{subequations}
\begin{eqnarray}
P_i &=& m_i +B_{ij}n^j + \frac{1}{2}Y^I_iY^I_jn^j + Y^I_iQ^I + G_{ij}n^j,\\
\bar{P}_i &=& m_i +B_{ij}n^j + \frac{1}{2}Y^I_iY^I_jn^j + Y^I_iQ^I - G_{ij}
n^j,\\
\hat{Q}^I &=& Q^I + Y^I_in^i.
\end{eqnarray}
\end{subequations}
The charge momenta $Q^I$ are induced by the right--moving fermions
$\bar{\psi}^I$ which  appear explicitly in the $\theta$--functions from equation \eqref{eq:thetadefinition} where the charge momentum  $Q^I=(n-{\frac{a^I}{2}})$.\index{lattice! $\Gamma_{6,6+16}$|)}\index{string!heterotic!compact|)}

\subsection{The fermionic partition function}

In the fermionic construction we fix the moduli at the self-dual point as is shown in section \ref{sec:bosonization2}. The general first order partition function for an heterotic string compactified to four dimensions where the compact dimensions are at the self-dual point under T-duality is therefore written as \cite{Antoniadis:1987rn}
\begin{equation}\label{eq:partition}
Z = \int \left[ \frac{d\tau d\bar{\tau}}{\tau_2^2} \right] Z_B^2 \sum_{spin str.} \cc{\alpha}{\beta} \prod_{f=1}^{64} Z_F \genfrac{[}{]}{0pt}{}{\alpha_f}{\beta_f},
\end{equation}
where we have as in equations \eqref{eqs:theta}
\begin{subequations}
\begin{align}
Z_F \genfrac{[}{]}{0pt}{}{0}{0} &= \sqrt{\frac{\theta_3}{\eta}}, &  Z_F \genfrac{[}{]}{0pt}{}{0}{1} &= \sqrt{\frac{\theta_4}{\eta}},\\
Z_F \genfrac{[}{]}{0pt}{}{1}{0} &= \sqrt{\frac{\theta_2}{\eta}}, & Z_F \genfrac{[}{]}{0pt}{}{1}{1} &= \sqrt{\frac{\theta_1}{\eta}}.
\end{align}
\end{subequations}
These formulae should be complex conjugated for the right-movers.

The supercurrent\index{supercurrent} is realised on the world sheet non-linearly\cite{Antoniadis:1986az}
\begin{equation}\label{eq:supercurrent}
T_F = \psi^\mu \partial X_\mu + i \chi^I y^I \omega^I, \ \ \ I \in \{1,\ldots,6\},
\end{equation}
where the $\{\chi^I, y^I, \omega^I\}$ are $18$ real free fermions transforming as the adjoint representation of $SU(2)^6$. When we transport a right-moving free fermion around a non-contractible loop it generally transforms as
\begin{equation}
\bar{\phi}^a \to R(\alpha)^a_{\phantom{a}b} \bar{\phi}^b,
\end{equation}
where $R(\alpha)^a_{\phantom{a}b}$ should be orthogonal and should leave the energy\--mo\-men\-tum current unchanged. For a left-moving fermion it is somewhat more complicated as we need to consider the super current \eqref{eq:supercurrent} as well. We can  realise the following transformation
\begin{eqnarray}
\psi^\mu &\to& - \delta_\alpha \psi^\mu,\\
\phi^a   &\to& L(\alpha)^a_{\phantom{a}b} \phi^b,
\end{eqnarray}
where $\phi^b \neq \psi^\mu$. Since the fermions are transported along a non-con\-trac\-ti\-ble loop $\alpha \in \pi_1(M)$, we see that $R(\alpha)$ and $L(\alpha)$ are matrix representations of $\alpha$. When the fundamental group $\pi_1(M)$ is Abelian we can diagonalise $R(\alpha)$ and $L(\alpha)$ simultaneously in some basis consisting in total of $64$ fermions. We can therefore realise any configuration of boundary conditions by a set of $k$ real fermions and $l$ complex fermions, where $k + 2l=64$
\begin{equation}\label{eq:vector}
\alpha = \left\{ \alpha(f_1^r), \ldots, \alpha(f_k^r); \alpha(f_1^c), \ldots, \alpha(f_l^c) \right\}.
\end{equation}

In this thesis we will use a slightly different notation\index{notation}. We will label the fermionized left-moving internal coordinates using equations \eqref{eq:mayoranaweylfermions} and \eqref{eq:fermionization} as
\begin{equation}
\frac{1}{\sqrt{2}} \left( y^I + i \omega^I\right) = e^{iX^I}
\end{equation}
and similarly for the right moving side. The super partners of the left-moving bosons are labelled as $\chi^I$. The super partners of the light-cone gauge fixed bosons are labelled as $\psi^\mu_{1,2}$. The extra $16$ degrees of freedom are labelled as $\bar{\psi}^{1,\ldots,5}, \bar{\eta}^{1,2,3}, \bar{\phi}^{1,\ldots,8}$ and are all complex fermions. We also write the boundary condition vector in such a way that only the periodic fermions are listed in the vector. The other fermions are considered anti-periodic. 

The fermions transform under parallel transport as
\begin{equation}\label{eq:boundarycondition}
f \to - e^{i \pi \alpha(f)}\ f,
\end{equation}
where the minus sign is introduced by convention.

We focus on the modular properties of the partition function of equation \eqref{eq:partition}. The partition function should be invariant under the modular transformations defined in equations \eqref{eq:mod}. Acting with these transformations on the partition function and requiring invariance leads to two constraints on the spin-structure constants $\cc{\alpha}{\beta}$
\begin{subequations}\label{subeq:constraints}
\begin{eqnarray}
\cc{\alpha}{\beta} &=& e^{i \frac{\pi}{4} \left( \alpha \cdot \alpha + \mathbbm{1} \cdot \mathbbm{1} \right) }\ \cc{\alpha}{\beta-\alpha+\mathbbm{1}} \label{eq:constraints1},\\
\cc{\alpha}{\beta} &=& e^{i \frac{\pi}{2} \alpha \cdot \beta}\ \cc{\beta}{\alpha}^*\label{eq:constraints2},
\end{eqnarray}
\end{subequations}
where we have introduced the vector $\mathbbm{1}$ where all fermions are periodic. These constraints are completely due to the transformation properties of the $\theta$-functions in equation \eqref{eq:partition}. Another constraint arises when we incorporate higher order loops, notably the two-loop contribution. This leads to a third constraint.
\begin{equation}\label{eq:constraints3}
\cc{\alpha}{\beta}\cc{\alpha'}{\beta'} = \delta_{\alpha}\ \delta_{\alpha'}\  e^{-i \frac{\pi}{2} \alpha \cdot \alpha'}\ \cc{\alpha}{\beta + \alpha'} \cc{\alpha'}{\beta' + \alpha}.
\end{equation}
We have defined the inner product of two vectors as a Lorentzian inner product\index{inner product}
\begin{equation}\label{eq:innerproduct}
\alpha \cdot \beta = \left\{ \frac{1}{2} \sum_{R,L} + \sum_{C,L} - \frac{1}{2} \sum_{R,R} - \sum_{C,R} \right\} \alpha(f) \beta(f),
\end{equation}
where $R$ and $C$ stand for real and complex respectively. $R$ and $L$ stand for right and left moving fermions respectively. We have defined the space-time spin statistics index $\delta_\alpha$ as
\begin{equation}\label{eq:spacetimespinindex}
\delta_{\alpha} = \begin{cases} 1 & \iff \alpha(\psi^\mu_{1,2}) = 0, \\ -1 & \iff \alpha(\psi^\mu_{1,2}) = 1.  \end{cases}
\end{equation}

\section{The Free Fermionic Formulation}

In this section we derive the sufficient and necessary constraints for constructing a free fermionic model. We follow \cite{Antoniadis:1988wp} closely in our approach. We describe the Hilbert space in terms of basis vectors with the boundary conditions of the free fermions on the world sheet. Continuing with the derivation of the constraints on these basis vectors, we move on to show the relation of the space of boundary conditions and the Hilbert space. We end this section by deriving the charge of a state under a $U(1)$ gauge group together with the frequencies of the states in the spectrum.

\subsection{Model building constraints}\label{sec:modelbuilding}

We construct from the constraints \eqref{subeq:constraints} -- \eqref{eq:constraints3} the necessary constraints on the vectors $\alpha$ and on the spin statistics coefficients $\cc{\alpha}{\beta}$. Combining equations \eqref{eq:constraints2} and \eqref{eq:constraints3} implies by setting $\alpha' = \alpha$ and $\beta = 0$
\begin{equation}\label{eq:c00}
\cc{\alpha}{0}^2 = \delta_\alpha \cc{\alpha}{0} \cc{0}{0},
\end{equation}
which means we can solve $\cc{\alpha}{0}=\delta_\alpha$, where we have normalised $\cc{0}{0}=1$ and have discarded the option where $\cc{\alpha}{0}=0$. We define a set of vectors $\Xi$\index{xi@$\Xi$}
\begin{equation}\label{eq:xi}
\Xi = \left\{ \alpha\ \Bigg|\ \cc{\alpha}{0} = \delta_\alpha \right\}.
\end{equation}
It can be shown that this set is an Abelian additive group if we define the group action to be the standard addition of the boundary conditions for every fermion separately. If we take $\Xi$ to be finite, which means that all boundary conditions in the vector $\alpha$ are rational, we see that $\Xi$ is isomorphic to
\begin{equation}
\Xi = \bigoplus_{i=1}^k \mathbbm{Z}_{N_i} ,
\end{equation}
which means that $\Xi$ is generated by a set $\{ b_1 , \ldots, b_k\}$ called a set of basis vectors\index{basis vectors}, such that 
\begin{equation}\label{eq:basisvectorfinite}
\sum_{i=1}^k m_i b_i = 0 \iff m_i=0 \mod N_i\ \ \forall i ,
\end{equation}
where $N_i$ is the smallest positive integer where $N_i b_i = 0$ or the order of $\mathbbm{Z}_{N_i}$. It can be shown that the vector $\mathbbm{1}$ can be realised by the first basis vector $b_1$. In this thesis we will take $b_1=\mathbbm{1}$ for convenience. We rewrite equation \eqref{eq:constraints3} for the case where $\alpha, \beta, \gamma \in \Xi$
\begin{equation}\label{eq:alphagamma}
\cc{\alpha}{\beta+\gamma} = \delta_\alpha\ \cc{\alpha}{\beta} \cc{\alpha}{\gamma}.
\end{equation}

The basis vectors $b_i$ are necessarily constrained by modular invariance. Using equation \eqref{eq:constraints1} we find that in the case of the heterotic string in four dimensions by setting $\beta=\alpha$
\begin{equation}\label{eq:alphaalpha}
\cc{\alpha}{\alpha} = - e^{\frac{i\pi}{4} \alpha \cdot \alpha} \cc{\alpha}{\mathbbm{1}}.
\end{equation}
Using equation \eqref{eq:alphagamma} we find that since $\beta$ generates a finite group of order $N_\beta$ we can write
\begin{eqnarray}\label{eq:alphabetabetaalpha}
\cc{\alpha}{\beta} &=& \delta_\alpha\  e^{\frac{2\pi i}{N_\beta} n}\nonumber\\
 &=& \delta_\beta\ e^{\frac{i \pi}{2} \alpha \cdot \beta}\ e^{\frac{2\pi i}{N_\alpha} m},
\end{eqnarray}
where we have used equation \eqref{eq:constraints2} for the second equality. Using equation \eqref{eq:constraints2} and raising it to the power $N_{ij}$, which is the least common multiple of $N_i$ and $N_j$ we find using equation \eqref{eq:alphagamma}
\begin{equation}
\exp \left[ i \frac{\pi}{2} N_{ij} b_i \cdot b_j \right] = \left( \delta_{b_i} \delta_{b_j} \right)^{N_{ij}}.
\end{equation}
Since the right hand side is always $1$ we find that 
\begin{equation}\label{eq:basisvectorodd}
N_{ij}\ b_i \cdot b_j = 0 \mod 4.
\end{equation}
If we set $i=j$ the constraint \eqref{eq:basisvectorodd} holds for $N_i$ is odd. In the case where $N_i$ is even we find an even stronger constraint using equation \eqref{eq:constraints1}
\begin{equation}\label{eq:basisvectoreven}
N_i\ b_i^2 = 0 \mod 8,\ \ \ \text{when $N_i$ even}.
\end{equation}

\subsection{The spectrum}

We show how to construct the spectrum of a model defined by its basis vectors $b_i$.

We can write equation \eqref{eq:partition} using \cite{Polchinski:1998rr}
\begin{equation}
Z_F \genfrac{[}{]}{0pt}{}{\alpha}{\beta} = \text{Tr}_\alpha \left[ q^H \exp(\pi i \beta \cdot F_\alpha) \right]
\end{equation}
as
\begin{equation}
Z = \int \left[ \frac{d\tau d\bar{\tau}}{\tau_2^2} \right] Z_B^2 \sum_{spin str.} \cc{\alpha}{\beta} \text{Tr}_{\mathscr{H}_\alpha} \left[ q^{H_\alpha} \exp(\pi i \beta \cdot F_\alpha) \right],
\end{equation}
where $H_\alpha$ is the Hamiltonian in the Hilbert space sector $\mathscr{H}_\alpha$ defined by the vector $\alpha = \sum_i n_i b_i$. The inner product $\beta \cdot F_\alpha$ is similar to equation \eqref{eq:innerproduct}
\begin{equation}
\beta \cdot F_\alpha = \left\{ \frac{1}{2} \sum_{R,L} + \sum_{C,L} - \frac{1}{2} \sum_{R,R} - \sum_{C,R} \right\} \beta(f) F_\alpha(f),
\end{equation}
where the fermion number operator $F$\index{fermion number operator} is
\begin{equation}\label{eq:fermionnumberoperator}
F(f) = \begin{cases} 1, \\ -1,\ \  \text{if fermion is complex conjugate}. \end{cases}
\end{equation}
We write the partition function as a sum over sectors. We note that since $b_i$ are generators of a discrete group $\mathbbm{Z}_{N_i}$ and since equation \eqref{eq:alphagamma} holds, the sum in the partition function is finite and we find
\begin{multline}
Z = \int \left[ \frac{d\tau d\bar{\tau}}{\tau_2^2} \right] Z_B^2 \sum_{\alpha \in \Xi} \delta_\alpha \text{Tr} \Bigg\{ \prod_{b_i} \bigg( \delta_\alpha \cc{\alpha}{b_i} e^{i \pi b_i \cdot F_\alpha} + \ldots \\
\ldots + \Big\{ \delta_\alpha \cc{\alpha}{b_i} e^{i \pi b_i \cdot F_\alpha} \Big\}^{N_i-1} + 1 \bigg) e^{i\pi \tau H_\alpha} \Bigg\}.
\end{multline}
We see that only states are counted that realise a generalised GSO projection\index{generalised GSO projection|see{GSO projection, generalised}}\index{GSO projection, generalised}
\begin{equation}\label{eq:gsoconstraint}
e^{i \pi b_i \cdot F_\alpha} |S>_\alpha = \delta_\alpha \cc{\alpha}{b_i}^* |S>_\alpha,
\end{equation}
which leads to the full Hilbert space\index{Hilbert space} of states
\begin{equation}
\mathscr{H} = \bigoplus_{\alpha \in \Xi}\  \prod_{i=1}^k \left\{ e^{i \pi b_i \cdot F_\alpha} = \delta_\alpha \cc{\alpha}{b_i}^* \right\} \mathscr{H}_\alpha.
\end{equation}

The mass formula\index{mass formula} of any state in the subspace or sector $\mathscr{H}_\alpha$ of the Hilbert space $\mathscr{H}$ is given by the zero-moment Virasoro gauge conditions
\begin{subequations}\label{eq:mass}
\begin{eqnarray}
M_L^2 &=& - c_L + \frac{\alpha_L \cdot \alpha_L}{8} + \text{freq.}\\
M_R^2 &=& - c_R + \frac{\alpha_R \cdot \alpha_R}{8} + \text{freq.}
\end{eqnarray}
\end{subequations}
where $\alpha_{L,R}$ are the boundary conditions defined by the vector $\alpha = \sum n_i b_i$ for the left and right moving fermions respectively. In the case of the heterotic string, we set in this thesis $c_L = \frac{1}{2}$ and $c_R = 1$. Naturally mass level matching is realised $M_L^2=M_R^2$ in the spectrum.

\subsection{U(1) charges}

In this section we derive the $U(1)$ charges $Q(f)$, for the unbroken Cartan generators of the four dimensional gauge group that are in one to one correspondence with the U(1) currents $f^* f$ for each complex fermion $f$. We follow \cite{Kawai:1987ah} in our approach.

We consider a single complex free left moving fermion with boundary condition.
\begin{equation}
\lambda(\sigma_1 + 2\pi, t) = e^{-2\pi i \nu} \lambda(\sigma_1,t),
\end{equation}
where $\nu$ is known as the frequency of the fermion and $t=-i\sigma_2$. $\lambda(\sigma_1,t)$ has the normal mode expansion
\begin{equation}
\lambda(\sigma_1,t) = \sum_{n=1}^\infty \left\{ b_{n+\nu-1}e^{-i(n+\nu-1)(\sigma_1+ t)} + d^\dagger_{n-\nu}e^{i(n-\nu)(\sigma_1+t)} \right\}.
\end{equation}
We find that the Hamiltonian for this system is given by
\begin{multline}
\mathscr{H}_\nu = \sum_{n=1}^\infty \left[ (n+\nu-1) b^\dagger_{n+\nu-1}b_{n+\nu-1} + (n-\nu)d^\dagger_{n-\nu}d_{n-\nu} \right] + \\
\frac{1}{2} \left( \nu^2 -\nu+\frac{1}{6} \right).
\end{multline}
The last term in the Hamiltonian $\mathscr{H}_\nu$ is the vacuum energy arising from the normal ordering of the operators. When we fill the negative energy states we find the vacuum charge of the system or the $U(1)$ charge induced by the left moving complex fermion $\lambda$. Using Riemann $\zeta$-function regularisation we find
\begin{eqnarray}
Q_\nu &=& \sum_{n=-\infty}^0 \left( n+ \nu -1 \right)^0 = \sum_{n=0}^\infty \left( n+1-\nu \right)^0 = \zeta(0,1-\nu) \nonumber\\
 & = &- B_1(1-\nu) = \nu-\frac{1}{2},
\end{eqnarray}
where $B_1$ is the first Bernoulli polynomial. We are now ready to compute the frequency of fermions depending on their boundary conditions as defined by equation \eqref{eq:boundarycondition}. We have
\begin{align}
f &\to - e^{i \pi \alpha(f)}\ f, & f^* &\to - e^{-i \pi \alpha(f)}\ f^*.
\end{align}
It is then easy to see that the frequency\index{frequency} for the fermions is given by
\begin{align}\label{eq:frequency}
\nu_f &= \frac{1+\alpha(f)}{2}, & \nu_{f^*} = \frac{1-\alpha(f)}{2}.
\end{align}
However we are able to write the frequency more concisely due to the periodicity of the phases. Using the fermion number operator $F$ we write the fermion frequency as
\begin{equation}
\nu = \frac{1 + \alpha(f)}{2} + F.
\end{equation}
Using this equality we find that the $U(1)$ charge\index{u1@$U(1)$ charge} $Q_\nu(f)$, for the unbroken Cartan generators of the four dimensional gauge group that are in one to one correspondence with the U(1) currents $f^* f$ for each complex fermion $f$ is
\begin{eqnarray}\label{eq:u1charge}
Q_\nu(f) &=& \frac{1 + \alpha(f)}{2} + F - \frac{1}{2} \nonumber\\
 &=& \frac{1}{2}\alpha(f) + F.
\end{eqnarray}
We have described all the necessary formulae for constructing a free fermionic heterotic string model and for calculating its spectrum and the charges of the unbroken Cartan generators generating a $U(1)$ gauge group.

\section{The Rules Of Construction}\label{sec:construction}

In this section we recall the rules derived in the previous sections for the construction of an heterotic string model that is invariant under modular transformations. We separate the overview into two parts. The first part recalls the necessary conditions for the creation of the Hilbert space by means of the basis vectors as well as the constraints that apply to the GSO coefficients or spin structure constants. The second part describes the necessary tools for calculating the spectrum.

\subsection{The foundation}\label{sec:foundation}

In the construction of any free fermionic model, two ingredients are required. The first is a set of basis vectors that define $\Xi$, the space of all states. We then need to isolate all the physical states. This is done by means of the GSO projections defined in equation \eqref{eq:gsoconstraint}. This requires the configuration of the GSO coefficients $\cc{\alpha}{\beta}$. These coefficients are constrained by modular invariance of the partition function. We recall the constraints for these two ingredients.

\subsubsection{The basis vectors}

The basis vectors $b_i$ consist of a set of numbers that realise the boundary conditions of each left or right moving fermion similar to equation \eqref{eq:vector}
\begin{equation}
b_i = \{ \alpha(\psi^\mu_{1,2}), \ldots, \alpha(\omega^6)\ |\ \alpha(\bar{y}^1), \ldots, \alpha(\bar{\phi}^8) \},
\end{equation}
where 
\begin{equation}
f \to - e^{i\pi\alpha(f)} f.
\end{equation}

To obtain a modular invariant model, the basis vectors are required to obey the following constraints
\begin{quote}
\begin{subequations}\label{subeq:basisconstraints}
From equation \eqref{eq:basisvectorfinite} we require
\begin{equation}
\sum_{i=1}^k m_i b_i = 0 \iff m_i=0 \mod N_i\ \ \forall i, 
\end{equation}
where $N_i$ is the smallest positive integer where $N_i b_i = 0$. From equation \eqref{eq:basisvectorodd} we require
\begin{equation}
N_{ij}\ b_i \cdot b_j = 0 \mod 4,
\end{equation}
where $N_{ij}$ is the least common multiplier of $N_i$ and $N_j$. From equation \eqref{eq:basisvectoreven} we require
\begin{equation}
N_i\ b_i \cdot b_i = 0 \mod 8,
\end{equation}
when $N_i$ even. In this thesis we take the first basis vector $b_1$ for reasons explained in section \ref{sec:modelbuilding} to be
\begin{equation}
b_1 = \mathbbm{1},
\end{equation}
which means that all fermions have periodic boundary conditions.
\end{subequations}
\end{quote}
For these constraints we have used the following definition for the inner product. We have defined the inner product as in equation \eqref{eq:innerproduct}
\begin{equation}
\alpha \cdot \beta = \left\{ \frac{1}{2} \sum_{R,L} + \sum_{C,L} - \frac{1}{2} \sum_{R,R} - \sum_{C,R} \right\} \alpha(f) \beta(f),
\end{equation}
where $R$ and $C$ stand for real and complex respectively. $R$ and $L$ stand for right and left moving fermions respectively.

\subsubsection{The coefficients}

We have defined the total space of states $\Xi$. In this space there is only a limited number of physical states that realise a generalised GSO constraint. From equation \eqref{eq:gsoconstraint} we find
\begin{equation}
e^{i \pi b_i \cdot F_\alpha} |S>_\alpha = \delta_\alpha \cc{\alpha}{b_i}^* |S>_\alpha,
\end{equation}
where $|S>_\alpha$ is a state in the sector $\alpha \subset \Xi$ where $\alpha = \sum n_i b_i$ is a superposition of the basis vectors $b_i$. The fermion number operator $F$ is defined as in equation \eqref{eq:fermionnumberoperator}
\begin{equation}
F(f) = \begin{cases} 1 \\ -1,\ \  \text{if fermion is complex conjugate}. \end{cases}
\end{equation}
and the space-time spin statistics index is defined as in equation \eqref{eq:spacetimespinindex}
\begin{equation}
\delta_{\alpha} = \begin{cases} 1 & \iff \alpha(\psi^\mu_{1,2}) = 0, \\ -1 & \iff \alpha(\psi^\mu_{1,2}) = 1 . \end{cases}
\end{equation}
To fully define the GSO projection we need to set the GSO coefficients $\cc{\alpha}{b_i}$. These coefficients are required to obey a set of constraints
\begin{quote}
\begin{subequations}\label{subeq:gsoconstraints}
From equation \eqref{eq:alphabetabetaalpha} we have
\begin{eqnarray}
\cc{\alpha}{\beta} &=& \delta_\alpha\ e^{\frac{2\pi i}{N_\beta} n}\nonumber\\
 &=& \delta_\beta\ e^{\frac{i \pi}{2} \alpha \cdot \beta}\ e^{\frac{2\pi i}{N_\alpha} m}.
\end{eqnarray}
From equation \eqref{eq:constraints2} we require
\begin{equation}
\cc{\alpha}{\beta} = e^{i \frac{\pi}{2} \alpha \cdot \beta}\ \cc{\beta}{\alpha}^*.
\end{equation}
From equation \eqref{eq:alphaalpha} we require
\begin{equation}
\cc{\alpha}{\alpha} = - e^{\frac{i\pi}{4} \alpha \cdot \alpha} \cc{\alpha}{\mathbbm{1}}.
\end{equation} 
From equation \eqref{eq:alphagamma} we require
\begin{equation}
\cc{\alpha}{\beta+\gamma} = \delta_\alpha\ \cc{\alpha}{\beta} \cc{\alpha}{\gamma}.
\end{equation}
\end{subequations}
\end{quote}	

\subsection{The spectrum}

Having defined all the basis vectors together with their GSO coefficients we are ready to construct the spectrum explicitly. Here we list the necessary formulae for analysing the complete spectrum.

\begin{quote}
\begin{subequations}\label{subeq:spectrum}
From equation \eqref{eq:mass} we find that the mass of any state in the Hilbert space is for the heterotic string with a super symmetric left moving sector
\begin{subequations}
\begin{eqnarray}
M_L^2 &=& - \frac{1}{2} + \frac{\alpha_L \cdot \alpha_L}{8} + N_L,\\
M_R^2 &=& - 1 + \frac{\alpha_R \cdot \alpha_R}{8} + N_R,
\end{eqnarray}
\end{subequations}
where $M_L^2 = M_R^2$. The value of the oscillators are calculated using the frequencies of the states $N_{L,R} = \sum \nu_{L,R}$. From equation \eqref{eq:frequency} we find for the frequencies
\begin{align}
\nu_f &= \frac{1+\alpha(f)}{2}, & \nu_{f^*} = \frac{1-\alpha(f)}{2}.
\end{align}
When a $U(1)$ gauge group is realised by a free fermion we find that the charge under this $U(1)$ is given by equation \eqref{eq:u1charge}
\begin{equation}
Q_\nu(f) = \frac{1}{2}\alpha(f) + F.
\end{equation}
where as usual $F$ is the fermion number operator and $\alpha$ is the boundary condition for the fermion generating the $U(1)$ gauge group.
\end{subequations}
\end{quote}

We have discussed all the necessary tools for the construction of any heterotic free fermionic model.


\chapter{Realistic Models}\label{chap:realistic}

In this chapter we discuss the construction of realistic models. We start with the simplest possible model that does not exhibit a super symmetric target space. We show that adding one additional basis vector introduces super symmetry on the background. The starting set of basis vectors for any realistic model is then discussed in detail. We isolate the spinorial representations of a $SO(10)$ gauge group. This leads us to two different advanced semi-realistic models. We derive the spectrum of a model where the gauge group is enhanced and of a model where the gauge group is not enhanced.
This chapter is based in part on \cite{Cleaver:2002ps}.

\section{The $N=4$ Models}

In this section we derive the spectrum of the most simple models. We start with the description of a non super symmetric free fermionic model after which we introduce a super symmetry generator basis vector. We show that adding this vector gives rise to a super symmetric spectrum. The gauge group of the two examples is discussed. We show the mechanism for the breaking of the gauge group by adding more basis vectors.

\subsection{The non super symmetric background}

We describe the simplest possible free fermionic model. We show the massless particle content and the gauge group of the model.

From equation \eqref{subeq:basisconstraints} we see that the simplest possible model is constructed using only the vector $\mathbbm{1}$ as mentioned in section \ref{sec:modelbuilding}
\begin{equation}\label{eq:1a}
\mathbbm{1} = \{ \psi^{1,2}, \chi^{1,\ldots,6}, y^{1,\ldots,6}, \omega^{1,\ldots,6} | \bar{y}^{1,\ldots,6}, \bar{\omega}^{1,\ldots,6}, \bar{\psi}^{1,\ldots,5},  \bar{\eta}^{1,2,3}, \bar{\phi}^{1,\ldots,8} \}.
\end{equation}
In this vector we have used the following notation. We have written the boundary condition vector in such a way that only the periodic fermions are listed in the vector. We have labeled the fermionized left-moving internal coordinates using equations \eqref{eq:mayoranaweylfermions} and \eqref{eq:fermionization} as
\begin{equation}
\frac{1}{\sqrt{2}}\left( y^I + i \omega^I \right) = e^{iX^I}
\end{equation}
and similarly for the right moving side. The super partners of the left-moving bosons are labeled as $\chi^I$. The super partners of the light-cone gauge fixed bosons are labeled as $\psi^\mu_{1,2}$. The extra $16$ degrees of freedom are labeled as $\bar{\psi}^{1,\ldots,5}, \bar{\eta}^{1,2,3}, \bar{\phi}^{1,\ldots,8}$ and are all complex fermions. The other fermions are considered anti-periodic. It is easy to check that this vector obeys all rules described in equations \eqref{subeq:basisconstraints}.

The space of states $\Xi$ as defined in equation \eqref{eq:xi} contains two sectors. We describe the two sectors as the $\mathbbm{1}$ sector and the $0$ or $NS$ sector. To fully specify the model under consideration we need to define the spin statistics or GSO coefficients. We use the latter formulation from now on. The GSO coefficients for this model are $\cc{NS}{NS}\equiv\cc{0}{0}$, $\cc{NS}{\mathbbm{1}}$, $\cc{\mathbbm{1}}{NS}$, $\cc{\mathbbm{1}}{\mathbbm{1}}$. Using equations \eqref{subeq:gsoconstraints} we find that
\begin{subequations}
\begin{eqnarray}
\cc{NS}{NS} 			&=& - e^{\frac{i\pi}{4} \alpha_0 \cdot \alpha_0} \cc{NS}{\mathbbm{1}}  =  - \cc{NS}{\mathbbm{1}} = 1,\\
\cc{NS}{\mathbbm{1}} 		&=& e^{i\frac{\pi}{2} \alpha_0 \cdot \alpha_\mathbbm{1}} \cc{\mathbbm{1}}{NS}^* =  \delta_\mathbbm{1} = -1,\\
\cc{\mathbbm{1}}{NS} 		&=& \delta_\mathbbm{1} = -1,\\
\cc{\mathbbm{1}}{\mathbbm{1}} 	&\equiv & -1.
\end{eqnarray}
\end{subequations}
We see that the value for $\cc{NS}{NS}\equiv\cc{0}{0}=1$ is consistent with equation \eqref{eq:c00} as is to be expected. It is also clear that the only sign we need to set by hand is the $\cc{\mathbbm{1}}{\mathbbm{1}}$ coefficient. The rest is either fixed by modular invariance or fixed by definition \eqref{eq:xi}. We will therefore list only the relevant GSO coefficients from now on.

The states in the spectrum have a mass described by equation \eqref{eq:mass}. We find that the $\mathbbm{1}$ sector does not contain massless states. The only sector which produces massless states is the $NS$ sector. To realise mass level matching we need to introduce oscillators. From equation \eqref{eq:frequency} we find that any complex creation operator acting on the vacuum realises a frequency of 
\begin{align}
\nu_f &= \frac{1+\alpha(f)}{2} = \frac{1}{2}, & \nu_{f^*} = \frac{1-\alpha(f)}{2} = \frac{1}{2}.
\end{align}
The massless states are separated into four subsectors 
\begin{subequations}\label{eq:nonsusyferm}
\begin{align}
S_1 &: \psi^\mu_{1,2}\ \partial \bar{X}^\nu \ket{0}, 		& S_2 &: \phi^a\ \partial \bar{X}^\nu \ket{0},\\
S_3 &: \psi^\mu_{1,2}\ \bar{\phi}^b \bar{\phi}^{b'} \ket{0}, 	& S_4 &: \phi^a\ \bar{\phi}^b \bar{\phi}^{b'} \ket{0},
\end{align}
\end{subequations}
where $a$ runs over all the left moving free fermions apart from $\psi^\mu_{1,2}$ and $b$ and $b'$ run over all the right moving free fermions. We can also realise states with $M^2 = -\frac{1}{2} < 0$ which are described by
\begin{equation}\label{eq:tachyon}
T_1 : \bar{\phi}^b \ket{0},
\end{equation}
which are known as tachyonic states. This is a well known result of non super sym\-me\-tric string theories. To find the physical spectrum we need to impose the GSO projection described in equation \eqref{eq:gsoconstraint}. We find that all states remain in the spectrum. We have thus found in the $S_1$ subsector a symmetric tensor, the graviton, an anti-symmetric tensor  and its trace the dilaton. The states in the $S_2$ subsector generate gauge bosons of a $SU(2)^6$ gauge group. Similarly the states in the $S_3$ subsector realise the gauge bosons of a $SO(44)$ gauge group. The states in the $S_4$ subsector are scalars in the adjoint representation of $SU(2)^6 \times SO(44)$. This completes the whole analysis of the heterotic string in a non super symmetric background.

\subsection{The super symmetric background}\label{sec:susybackground}

In this section we describe a free fermionic model with a $N=4$ super symmetric background in detail. We show the massless particle content of the model and its gauge group.

In equation \eqref{eq:Zcompact} we have seen that separating the eight left moving free super partners of the bosonic coordinates from the rest of the compact partition function results in a super symmetric target space. We do the same in the fermionic formulation by defining the vector $S$ in addition to the vector $\mathbbm{1}$, where
\begin{equation}\label{eq:sa}
S = \{ \psi^{1,2}, \chi^{1,\ldots,6} \}.
\end{equation}
Again it is easy to check that this vector obeys all the rules listed in equation \eqref{subeq:basisconstraints}. With the addition of the vector $S$ we have introduced a bigger space of states $\Xi$. We find the additional sectors $S$, $\mathbbm{1}+S$.

We finalise the specification of the whole model by setting the GSO coefficients 
\begin{equation}\label{eq:gsocoeff1}
\cc{\alpha}{\beta} = \bordermatrix{
       &1  &S  \cr
   1   &-1  &-1  \cr
   S   &-1  &-1
  }.
\end{equation}
We could have chosen the coefficients differently but we will argue later on in this section that this choice does not have any physical effect.

Using equation \eqref{eq:mass} we find that in the sector $\mathbbm{1}+S$ there are no massless states. We therefore focus on the new states in the $S$ sector. In the $S$ sector we find states of a new kind. Since the $\psi^{1,2}, \chi^{1,\ldots,6}$ fermions are periodic they form a degenerate ground state. Consider the mode expansion for the free fermion\cite{Ginsparg:1988ui, Polchinski:1998rr}
\begin{equation}
i \psi(z) = \sum \psi_n z^{-n-\frac{1}{2}},
\end{equation}
where the sum is over $\mathbbm{Z}$ for periodic fermions and over $\mathbbm{Z}+\frac{1}{2}$ for anti-periodic fermions. We find that the anti-commutator for the fermionic modes $\psi_n$ is given by
\begin{equation}
\{ \psi_n, \psi_m\} = \delta_{n+m,0}
\end{equation}
We see that the zero mode $\psi_0$ takes a ground state into another ground state due to $\{ \psi_n, \psi_0\} = 0$ for $n>0$. We label the two states as $\ket{\pm}$, where $\ket{+} = \ket{0}$ and $\ket{-} = \psi_0 \ket{0}$ leading to $F\ket{+}=0$ and $F\ket{-}=-1$, where $F$ is the fermion number operator as in equation \eqref{eq:fermionnumberoperator}. Since the periodic fermions contribute to the massless spectrum we find using above notation the following massless states
\begin{subequations}\label{eq:ssector}
\begin{eqnarray}
S_5 &:& \ket{\pm}_1\ket{\pm}_2\ket{\pm}_3\ket{\pm}_4\  \partial \bar{X}^\nu \ket{0},\\
S_6 &:& \ket{\pm}_1\ket{\pm}_2\ket{\pm}_3\ket{\pm}_4\  \bar{\phi}^b \bar{\phi}^{b'} \ket{0},
\end{eqnarray}
\end{subequations}
where we have complexified the real fermions $\psi^{1,2}$, $\chi^{1,2}$, $\chi^{3,4}$ and $\chi^{5,6}$.

Using the GSO projection from equation \eqref{eq:gsoconstraint} we find that the following states remain in the spectrum
\begin{subequations}
\begin{align}
S_1 &: \psi_\mu^{1,2}\ \partial \bar{X}^\nu \ket{0}, 		& S_2 &: \chi^a\ \partial \bar{X}^\nu \ket{0},\\
S_3 &: \psi_\mu^{1,2}\ \bar{\phi}^b \bar{\phi}^{b'} \ket{0}, 	& S_4 &: \chi^a\ \bar{\phi}^b \bar{\phi}^{b'} \ket{0},
\end{align}
\end{subequations}
where $a$ runs over $1,\ldots,6$ and $b$ and $b'$ as in equations \eqref{eq:nonsusyferm}. Additionally the states
\begin{subequations}\label{eq:combinatorics}
\begin{eqnarray}
S_5 &:& \left[ \binom{4}{1} + \binom{4}{3} \right]\  \partial \bar{X}^\nu \ket{0},\\
S_6 &:& \left[ \binom{4}{1} + \binom{4}{3} \right]\  \bar{\phi}^b \bar{\phi}^{b'} \ket{0},
\end{eqnarray}
\end{subequations}
remain in the spectrum, where $\binom{4}{3}$ means that $3$ of the $4$ degenerate states are in the down state $\ket{-}$. In the second line we can separate the degeneracy induced by the spacetime spinor $\psi^\mu_{1,2}$. We separate
\begin{equation}
\left[ \binom{4}{1} + \binom{4}{3} \right]\ = \binom{1}{0} \left[ \binom{3}{1} + \binom{3}{3} \right]\ + \binom{1}{1} \left[ \binom{3}{0} + \binom{3}{2} \right],
\end{equation}
where an additional degeneracy of $4$ is apparent next to the spacetime \emph{up} and \emph{down} spinor state.

One important feature of these super symmetric models is that the tachyonic states found in equation \eqref{eq:tachyon} all drop out.

If we were to set the GSO coefficients as
\begin{equation}
\cc{\alpha}{\beta} = \bordermatrix{
       &1  &S  \cr
   1   &-1  &1  \cr
   S   &1  &1
  }.
\end{equation}
we find that the states in the $NS$ sector remain the same but the states in the $S$ sector become
\begin{subequations}
\begin{eqnarray}
S_5 &:& \left[ \binom{4}{1} + \binom{4}{2} + \binom{4}{4} \right]\  \partial \bar{X}^\nu \ket{0},\\
S_6 &:& \left[ \binom{4}{0} + \binom{4}{2} + \binom{4}{4} \right]\  \bar{\phi}^b \bar{\phi}^{b'} \ket{0}.
\end{eqnarray}
\end{subequations}
Again we can isolate the spacetime spinor $\psi^\mu_{1,2}$, leading again to a degeneracy of $4$. The configuration of \emph{down} states is different from the configuration induced by the GSO coefficients presented in equation \eqref{eq:gsocoeff1}. We therefore find that $\cc{\mathbbm{1}}{S}$ sets the number of down states $\ket{-}$ in the $S$ sector. This sign choice is directly related to the value of $\mu$ from equation \eqref{eq:n4partition}\cite{Gregori:1999ny, Faraggi:2004rq}
\begin{equation}\label{eq:susychirality}
\mu=\frac{1}{2}{\left(1-\cc{\mathbbm{1}}{S}\right)}.
\end{equation}
Since the choice of chirality is a convention, we have therefore not found a physically different model with the different configuration of the GSO coefficients. A notable result of this model is that, again, the tachyonic states from equation \eqref{eq:tachyon} are projected out. 

The degeneracy of the space time spinors is identified as the result of an $N=4$ super symmetric background. We therefore identify the states in the $S_5$ subsector as the super partners of the gravitons called gravitinos. Similarly we identify the states in the $S_6$ subsector as the super partners of the $SO(44)$ gauge bosons, the $SO(44)$ gauginos. 

The gauge group of this model is again realised by the states in the $S_2$ and $S_3$ subsector. Since the states in the $S_2$ subsector do not have super partners we identify the induced gauge group not as a space time gauge group. The states in the $S_3 \subset NS$ subsector generate a $SO(44)$ space time gauge group with an $N=4$ super symmetry where the super partners are generated by the sector $NS+S=S$. This is a general feature of free fermionic models. In a super symmetric model where the super symmetry is generated by the vector $S$ all the super partners of a particular sector $\alpha$ are generated by the sector $\alpha + S$.

\section{A $N=1$ Model: the \nahe\ Set}\label{sec:nahe}

We discuss a model that describes a $N=1$ space time with spinorial representations of a SO(10) gauge group. This model is known in the literature to be constructed using the \nahe\ set. In \cite{Antoniadis:1988tv} the flipped $SU(5)$ free fermionic models were analysed. In \cite{Antoniadis:1987rn} an initial start was made in the construction of the set used for the flipped $SU(5)$ model. In \cite{Faraggi:1993ac} a subset of the flipped $SU(5)$ basis vector set was identified as relevant for the construction of phenomenologically interesting models. This subset was labeled the \nahe\ set. 
We start by defining the model. We then set the GSO coefficients after which we discuss some interesting properties of the model in detail. 

The \nahe\ set consists of a set of five basis vectors. The vectors are $\mathbbm{1}$, $S$, defined in equation \eqref{eq:1a} and equation \eqref{eq:sa}, and the vectors $b_1$, $b_2$ and $b_3$
\begin{subequations}\label{eq:nahetwist}
\begin{eqnarray}
b_1 &=& \{ \psi^{1,2}, \chi^{1,2}, y^{3,\ldots,6}\ |\ \bar{y}^{3,\ldots,6}, \bar{\psi}^{1,\ldots,5}, \bar{\eta}^1 \},\\
b_2 &=& \{ \psi^{1,2}, \chi^{3,4}, y^{1,2}, \omega^{5,6}\ |\ \bar{y}^{1,2}, \bar{\omega}^{5,6}, \bar{\psi}^{1,\ldots,5}, \bar{\eta}^2 \},\\
b_3 &=& \{ \psi^{1,2}, \chi^{3,4}, \omega^{1,\ldots,4}\ |\ \bar{\omega}^{1,\ldots,4}, \bar{\psi}^{1,\ldots,5}, \bar{\eta}^3 \}.
\end{eqnarray}
\end{subequations}
Note that these vectors $b_i$ are labeled as such to reflect the notation of the literature best. It will be clear from the context whether $b_i$ are general basis vectors or the ones listed above. Since the basis vectors listed above induce a $\mathbbm{Z}_2 \times \mathbbm{Z}_2$ twist as will become clear later on, this notation will also be used later on for other vectors that induce the same type of twisting. The addition of these three vectors have enlarged the space of states $\Xi$ considerably. We therefore do not list all the new sectors. We however isolate some new sectors which prove to be enlightening for further analysis of more complex models.

We set the GSO coefficients 
\begin{equation}\label{eq:phasesmodel1}
\cc{\alpha}{\beta} = \bordermatrix{
 	& 1 & S & & b_1 & b_2 & b_3 \cr
 1	& 1 & 1 & & -1  &  -1 & -1  \cr
 S	& 1 & 1 & &  1  &   1 &  1  \cr
	&   &   & &     &     &     \cr
 b_1	& -1& -1& & -1  &  -1 & -1  \cr
 b_2	& -1& -1& & -1  &  -1 & -1  \cr
 b_3	& -1& -1& & -1  &  -1 & -1  \cr}.
\end{equation}

The space time vector bosons come from the $NS$ sector. In this model additional vector bosons can come from the $\xi = \mathbbm{1} + b_1 + b_2 + b_3$ sector. The $NS$ sector gives rise to a $SO(6)^3 \times SO(16) \times SO(10)$. With the GSO coefficients given in equation \eqref{eq:phasesmodel1} we find that the states in the $\xi$ sector remain in the spectrum and give similar to the construction of the $S$ sector in equation \eqref{eq:ssector}
\begin{equation}
\xi : \left[ \binom{8}{0} + \binom{8}{2} + \binom{8}{4} + \binom{8}{6} + \binom{8}{8} \right] \psi_\mu^{1,2} \ket{0}
\end{equation}
vector bosons. This enhances the $SO(16) \to \mathbbm{E}_8$ as the $\xi$ sector contains $128$ states, the spinorial representation of $SO(16)$ and the $NS$ sector contains $120$ states, the vectorial representation of $SO(16)$. We therefore find the gauge group $SO(6)^3 \times SO(10) \times \mathbbm{E}_8$.

The additional basis vectors $b_i$ induce additional GSO projections on the states in the $S$ sector. We find that $b_1$ reduces the number of states in the $S_5 \subset S$ sector to
\begin{equation}
S_5 : \left[ \binom{2}{1} \right] \left[ \binom{2}{0} + \binom{2}{2} \right]\  \partial \bar{X}^\nu \ket{0}.
\end{equation}
The $b_2$ vector reduces the number of states in the $S_5 \subset S$ sector to
\begin{equation}
S_5 : \left[ \binom{1}{1} \binom{1}{0} \binom{1}{0} \binom{1}{0} + \binom{1}{0} \binom{1}{1} \binom{1}{1} \binom{1}{1} \right]\  \partial \bar{X}^\nu \ket{0}.
\end{equation}
The $b_3$ vector does not reduce the number of states any further. We find that there is a degeneracy of $1$ of the space time spinorial \emph{up} and \emph{down} state. We have therefore reduced the number of space time super symmetries to $N=1$.

The \nahe\ set divides the 44 right-moving and 20 left-moving real internal
fermions in the following way: ${\bar\psi}^{1,\cdots,5}$ are complex and
produce the observable $SO(10)$ symmetry; ${\bar\phi}^{1,\cdots,8}$ are
complex and produce the hidden $\mathbbm{E}_8$ gauge group;
$\{{\bar\eta}^1,{\bar y}^{3,\cdots,6}\}$, $\{{\bar\eta}^2,{\bar y}^{1,2}
,{\bar\omega}^{5,6}\}$, $\{{\bar\eta}^3,{\bar\omega}^{1,\cdots,4}\}$
give rise to the three horizontal $SO(6)$ symmetries. The left-moving
$\{y,\omega\}$ states are also divided into the sets $\{{y}^{3,\cdots,6}\}$,
$\{{y}^{1,2}
,{\omega}^{5,6}\}$, $\{{\omega}^{1,\cdots,4}\}$. The left-moving
$\chi^{12},\chi^{34},\chi^{56}$ states carry the super symmetry charges.
Each sector $b_1$, $b_2$ and $b_3$ carries periodic boundary conditions
under $(\psi^\mu\vert{\bar\psi}^{1,\cdots,5})$ and one of the three groups:
$(\chi_{12},\{y^{3,\cdots,6}\vert\-{\bar y}^{3,\cdots6}\},{\bar\eta}^1)$,
 $(\chi_{34},\{y^{1,2},\omega^{5,6}\vert\-{\bar y}^{1,2}\-{\bar\omega}^{5,6}\},\-{\bar\eta}^2)$, $(\chi_{56},\{\omega^{1,\cdots,4}\vert\-{\bar\omega}^{1,
\cdots4}\},{\bar\eta}^3)$.

There are three sectors of special interest in this model. They are the sectors $b_1$, $b_2$ and $b_3$. We will focus on only one as the other two can be dealt with similarly. In the sector $b_1$ we find massless states as is easily verified using equation \eqref{eq:mass}. Similar to the sector $S$ we get a degenerate vacuum. One complication arises here. In the case of the $S$ vector we could pair two left, or right moving fermions to create a complex fermion. Although at the level of the \nahe\ set this is possible as well for the internal fermions $y^i, \omega^i, \bar{y}^i, \bar{\omega}^i$, adding more vectors generally disallows this type of pairing. 

Periodic free fermions have a conformal weight $h=\frac{1}{16}$\cite{Ginsparg:1988ui}. Fields with conformal weight $(h,\bar{h}) = (\frac{1}{16},\frac{1}{16})$ realise an order operator $\sigma$ in a two-dimensional Ising model\cite{mccoy1973aa}. Since in this case the real fermions have periodic boundary conditions we can pair a real left moving fermion with a real right-moving fermion to construct an order operator $\sigma$
\begin{equation}
\sigma^a = \phi^a(z) \bar{\phi}^a(\bar{z}),
\end{equation}
leading to a degenerate vacuum. We find that the unprojected lowest mass states induced by the sector $b_1$ can be written as
\begin{equation}\label{eq:twistedmatterb1}
\ket{S} = \ket{\psi_\mu^{1,2}}_{\pm}\  \ket{\chi^{1,2}}_\pm \prod_{l=3}^6 \ket{\sigma^l}_\pm\  \ket{\bar{\eta}^1}_\pm \prod_{m=1}^5 \ket{\bar{\psi}^m}_\pm \ket{0},
\end{equation}
where we have used the following pairing
\begin{subequations}
\begin{eqnarray}
\psi_\mu^{1,2} &=& \frac{1}{\sqrt{2}} \left( \psi_\mu^1 + i\psi_\mu^2 \right),\\
\chi^{2k-1,2k} &=& \frac{1}{\sqrt{2}} \left( \chi^{2k-1} + i\chi^{2k} \right),\\
\sigma^l &=& y^l \bar{y}^l, 
\end{eqnarray}
\end{subequations}
where the other fermions are already complex by construction. We impose the GSO projection as defined in equation \eqref{eq:gsoconstraint}. We use the combinatorics notation as employed in equation \eqref{eq:combinatorics}. Projecting using the basis vector $\mathbbm{1}$ gives
\begin{equation}\label{eq:1gsob1}
\left[ \binom{12}{0} + \binom{12}{2} + \binom{12}{4} + \binom{12}{6} + \binom{12}{8} + \binom{12}{10} + \binom{12}{12} \right] = 2^{11}
\end{equation}
states in the $b_1$ sector. Projecting with the basis vector $S$ gives
\begin{equation}
\left[ \binom{2}{0} + \binom{2}{2} \right] \left[ \binom{10}{0} + \binom{10}{2} + \binom{10}{4} + \binom{10}{6} + \binom{10}{8} + \binom{10}{10} \right] = 2^{10}
\end{equation}
states in the $b_1$ sector. Projecting with the basis vector $b_1$ does not reduce the number of states. Projecting with the vector $b_2$ reduces the number of states to
\begin{multline}
\binom{1}{0} \binom{1}{0} \left[ \binom{5}{0} + \binom{5}{2} + \binom{5}{4} \right] \left[ \binom{5}{0} + \binom{5}{2} + \binom{5}{4} \right] +\\
\binom{1}{1} \binom{1}{1} \left[ \binom{5}{1} + \binom{5}{3} + \binom{5}{5} \right] \left[ \binom{5}{1} + \binom{5}{3} + \binom{5}{5} \right] = 2^{9}.
\end{multline}
Note that the sign of $\cc{b_1}{b_2}$ determines the number of \emph{up} and \emph{down} states in the vacuum. Projecting with the last basis vector $b_3$ does not reduce the number of states any further. We find that the $b_1$ sector produces spinorial ${\bf 16}$ representations of $SO(10)$. There is a degeneracy of $16$ for the \emph{up} and \emph{down} state of the space time spinor $\psi_\mu^{1,2}$. This degeneracy is interpreted as $16$ families of the spinorial ${\bf 16}$ of $SO(10)$. Since the number of \emph{up} and \emph{down} states in the vacuum is determined by the sign of $\cc{b_1}{b_2}$, we find that this sign determines the chirality of the ${\bf 16}$, i.e. determines whether we find a $\overline{\bf 16}$ or ${\bf 16}$ of $SO(10)$. A similar method can be followed for the sectors $b_2$ and $b_3$. We therefore find $16$ spinorial ${\bf 16}$ of $SO(10)$ coming from each $b_i$ giving a total of $48$ generations.

We have thus found a model with a $SO(6)^3 \times SO(10) \times \mathbbm{E}_8$ gauge group and $N=1$ space time super symmetry. The next step in the construction of a realistic model is to break the $SO(10)$ gauge group and simultaneously reduce the number of families to three. This is usually done using three more basis vectors called $\alpha$, $\beta$ and $\gamma$. In section \ref{sec:advanced} we discuss such models in detail and derive their complete spectra.

\section{Two Advanced Models}\label{sec:advanced}\label{su421ffm}

In this section we discuss two advanced semi realistic models. We first recall the setup of the \nahe\ set after which we break the unification $SO(10)$ to standard model like gauge groups. We then discuss one model in which the gauge group is not enhanced and one in which the gauge group is enhanced.

A model in the free fermionic formulation \cite{Antoniadis:1987rn,Kawai:1987ah} is constructed by
choosing a consistent set of boundary condition basis vectors as explained in section \ref{sec:construction}.
The basis vectors, $b_k$, span a finite
additive group $\Xi=\sum_k{{n_k}{b_k}}$
where $n_k=0,\cdots,{{N_{z_k}}-1}$.
The physical massless states in the Hilbert space of a given sector
$\alpha\subset{\Xi}$, are obtained by acting on the vacuum with
bosonic and fermionic operators and by
applying the generalised GSO projections. The $U(1)$
charges, $Q(f)$, for the unbroken Cartan generators of the four
dimensional gauge group are in one
to one correspondence with the $U(1)$
currents ${f^*}f$ for each complex fermion f, and are from equation \eqref{eq:u1charge}
\begin{equation}
Q(f) = \frac{1}{2}\alpha(f) + F_\alpha(f),
\label{u1charges}
\end{equation}
where $\alpha(f)$ is the boundary condition of the world--sheet fermion $f$
in the sector $\alpha$, and
$F_\alpha(f)$ is the fermion number operator from equation \eqref{eq:fermionnumberoperator}.
For each periodic complex fermion $f$
there are two degenerate vacua ${\vert +\rangle},{\vert -\rangle}$ ,
annihilated by the zero modes $f_0$ and
${{f_0}^*}$ and with fermion numbers  $F(f)=0,-1$, respectively as explained in section \ref{sec:nahe}.

The realistic models in the free fermionic formulation are generated by
a basis of boundary condition vectors for all world--sheet fermions
\cite{Antoniadis:1989zy, Lopez:1993kg, Faraggi:1990ka, Faraggi:1992af, Antoniadis:1990hb, Leontaris:1999ce, Faraggi:1992jr, Faraggi:1994eu}. The basis is constructed in
two stages. The first stage consists of the \nahe\ set
\cite{Antoniadis:1989zy, Lopez:1993kg, Faraggi:1993ac, Faraggi:1997dc, Faraggi:1992jr},
which is a set of five boundary condition basis
vectors, $\{ \mathbbm{1},S,b_1,b_2,b_3 \}$ as explained in section \ref{sec:nahe}.
The gauge group at the level of the \nahe\ set
is $SO(10)\times SO(6)^3\times \mathbbm{E}_8$ with $N=1$ space--time super symmetry.
The vector $S$ is the super symmetry generator and the super partners of
the states from a given sector $\alpha$ are obtained from the sector
$S+\alpha$ as explained in section \ref{sec:susybackground}. The space--time vector bosons that generate the gauge group
arise from the Neveu--Schwarz (NS) sector and from the sector $\zeta
\equiv \mathbbm{1}+b_1+b_2+b_3$.
The NS sector produces the generators of
$SO(10)\times SO(6)^3\times SO(16)$. The sector
$\zeta$
produces the spinorial $\bf 128$ of $SO(16)$ and completes the hidden
gauge group to $\mathbbm{E}_8$. The vectors $b_1$, $b_2$ and $b_3$
produce 48 spinorial $\bf 16$'s of $SO(10)$, sixteen from each sector $b_1$,
$b_2$ and $b_3$. The vacuum of these sectors contains eight periodic
world sheet fermions, five of which produce the charges under the
$SO(10)$ group, while the remaining three periodic fermions
generate charges with respect to the flavour symmetries. Each of the
sectors $b_1$, $b_2$ and $b_3$ is charged with respect to a
different set of flavour quantum numbers, $SO(6)_{1,2,3}$.

The second stage of the basis construction consist of adding three
additional basis vectors to the \nahe\ set.
Three additional vectors are needed to reduce the number of generations
to three, one from each sector $b_1$, $b_2$ and $b_3$.
One specific example is given in table \ref{tab:su421without}.
The choice of boundary
conditions to the set of real internal fermions
${\{y,\omega\vert{\bar y},{\bar\omega}\}^{1,\cdots,6}}$
determines the low energy properties, such as the number of generations.

The $SO(10)$ gauge group is broken to one of its subgroups
$SU(5)\times U(1)$, $SO(6)\times SO(4)$ or
$SU(3)\times SU(2)\times U(1)^2$ by the assignment of
boundary conditions to the set ${\bar\psi}^{1\cdots5}_\frac{1}{2}$:
\begin{subequations}\label{su51so64breakingbc}
\begin{eqnarray}
1.\ b\{{{\bar\psi}_\frac{1}{2}^{1\cdots5}}\}&=& \{\frac{1}{2}\frac{1}{2}\frac{1}{2}\frac{1}{2} \frac{1}{2}\}\Rightarrow SU(5)\times U(1),\label{su51breakingbc}\\
2.\ b\{{{\bar\psi}_\frac{1}{2}^{1\cdots5}}\}&=&\{1 1 1 0 0\}  \Rightarrow SO(6)\times SO(4).
\end{eqnarray}
\end{subequations}

To break the $SO(10)$ symmetry to
$SU(3)_C\times SU(2)_L\times
U(1)_C\times U(1)_L$
both steps, 1 and 2, are used, in two separate basis vectors.
The breaking pattern
$SO(10)\rightarrow SU(3)_C\times SU(2)_L\times SU(2)_R \times U(1)_{B-L}$
is achieved by the following assignment in two separate basis
vectors
\begin{subequations}\label{su3122breakingbc}
\begin{eqnarray}
1.\ b\{{{\bar\psi}_\frac{1}{2}^{1\cdots5}}\}&=&\{1 1 1 0 0\}  \Rightarrow SO(6)\times SO(4),\\
2.\ b\{{{\bar\psi}_\frac{1}{2}^{1\cdots5}}\}&=&\{\frac{1}{2}\frac{1}{2}\frac{1}{2}00\}\nonumber\\
 &\Rightarrow& SU(3)_C\times U(1)_C \times SU(2)_L\times SU(2)_R.
\end{eqnarray}
\end{subequations}

Similarly, the breaking pattern
$SO(10)\rightarrow SU(4)_C\times SU(2)_L\times U(1)_R$
is achieved by the following assignment in two separate basis
vectors
\begin{subequations}\label{su421breakingbc}
\begin{eqnarray}
1.\ b\{{{\bar\psi}_\frac{1}{2}^{1\cdots5}}\}&=&\{1 1 1 0 0\}  \Rightarrow SO(6)\times SO(4),\\
2.\ b\{{{\bar\psi}_\frac{1}{2}^{1\cdots5}}\}&=& \{000\frac{1}{2}\frac{1}{2}\}\Rightarrow SU(4)_C\times SU(2)_L\times U(1)_R.
\end{eqnarray}
\end{subequations}

We comment here that
a recurring feature of some of the three generation free fermionic heterotic
string models is the emergence of a combination of the basis vectors
which extend the \nahe\ set,
\begin{equation}\label{xcomb}
X=n_\alpha\alpha+n_\beta\beta+n_\gamma\gamma,
\end{equation}
for which $X_L\cdot X_L=0$ and $X_R\cdot X_R\ne0$. Such a
combination may produce additional space--time vector
bosons, depending on the choice of GSO phases.
These additional space--time vector bosons
enhance the four dimensional gauge group.
This situation is similar to the presence of the combination
of the \nahe\ set basis vectors $\mathbbm{1}+b_1+b_2+b_3$, which
enhances the hidden gauge group, at the level of the \nahe\ set,
 from $SO(16)$ to $\mathbbm{E}_8$.
In the free fermionic models this type of
gauge symmetry enhancement in the observable sector is,
in general, family universal and is intimately related to the
$\mathbbm{Z}_2\times \mathbbm{Z}_2$ orbifold structure which underlies the
realistic free fermionic models. This will become clear later on in chapters \ref{chap:orbifold} and \ref{chap:fermform}. Such
enhanced symmetries were shown to forbid proton decay
mediating operators to all orders of nonrenormalizable
terms \cite{Faraggi:1994eu}. Below we discuss examples of
models with and without gauge enhancement.

The SU421 symmetry breaking pattern induced by the
boundary condition assignment given in equation \eqref{su421breakingbc}
has an important distinction from the previous symmetry breaking patterns.
As in the previous cases, since it involves a breaking of an
$SO(2n)$ group to $SU(n)\times U(1)$ it contains 1/2 boundary
conditions. As discussed above the observable and hidden
non--Abelian gauge groups arise from the sets of complex world--sheet
fermions $\{{\bar\psi}^{1,\cdots,5}{\bar\eta}^{1,\cdots,3}\}$ and
$\{{\bar\phi}^{1,\cdots,8}\}$, respectively. The breaking pattern
\eqref{su421breakingbc} entails an assignment of $1/2$ boundary
condition to two complex fermions in the observable set,
whereas the symmetry breaking patterns in equations
\eqref{su51so64breakingbc} and \eqref{su3122breakingbc}
involve three
such assignments. On the other hand, the modular invariance
rules \cite{Antoniadis:1987rn, Kawai:1987ah} for the product $b_j\cdot\gamma$, where $b_j$ are
the \nahe\ set basis vectors and $\gamma$ is the basis vector that
contains the $1/2$ boundary conditions, enforces that no other
complex fermion from the observable set has $1/2$ boundary conditions as explained in section \ref{sec:foundation}.
Additionally, the constraint on the product $\gamma\cdot\gamma$ imposes
that either 8 or 12 complex fermions have $1/2$ boundary conditions.
Since, as we saw, only two can have such boundary conditions
from the observable set, it implies that six and only six from
the hidden set must have $1/2$ boundary conditions. This is
in contrast to the other cases that allow assignment of
twelve $1/2$ boundary conditions in the basis vector $\gamma$.
The consequence of having only eight $1/2$ boundary conditions in the
basis vector $\gamma$ is the appearance of additional sectors
that may lead to enhancement of the four dimensional gauge group.

\subsection{A model without enhanced symmetry}\label{noes}

As our first example of a SU421
free fermionic heterotic string model
we consider the model defined in table \ref{tab:su421without} and equation \eqref{phasesmodel1}, specified below in addition to the \nahe\ set explained in section \ref{sec:nahe}.
Also given in table \ref{tab:su421without} are the pairings of
left-- and right--moving real fermions from the set
$\{y,\omega|{\bar y},{\bar\omega}\}$. These fermions are
paired to form either complex, left-- or right--moving fermions,
or Ising model operators, which combine a real left--moving fermion with
a real right--moving fermion. 
\begin{table}
\begin{center}
\begin{tabular}{c|c|ccc|c|ccc|c}
 ~ & $\psi^\mu$ & $\chi^{12}$ & $\chi^{34}$ & $\chi^{56}$ &
        $\bar{\psi}^{1,...,5} $ &
        $\bar{\eta}^1 $&
        $\bar{\eta}^2 $&
        $\bar{\eta}^3 $&
        $\bar{\phi}^{1,...,8} $ \\\hline\hline
  ${\alpha}$  &  0 & 0&0&0 & 1~1~1~0~0 & 0 & 0 & 0 &1~1~1~1~0~0~0~0 \\
  ${\beta}$   &  0 & 0&0&0 & 1~1~1~0~0 & 0 & 0 & 0 &1~1~1~1~0~0~0~0 \\
  ${\gamma}$  &  0 & 0&0&0 & 0~0~0~$\frac{1}{2}$~$\frac{1}{2}$& 0 & 0 & 0 &
  $\frac{1}{2}$~$\frac{1}{2}$~$\frac{1}{2}$~$\frac{1}{2}$~
  $\frac{1}{2}$~$\frac{1}{2}$~0~0
\end{tabular}

\vspace{\baselineskip}
\begin{tabular}{c|c|c|c}
  &   $y^3{y}^6$
      $y^4{\bar y}^4$
      $y^5{\bar y}^5$
      ${\bar y}^3{\bar y}^6$
  &   $y^1{\omega}^5$
      $y^2{\bar y}^2$
      $\omega^6{\bar\omega}^6$
      ${\bar y}^1{\bar\omega}^5$
  &   $\omega^2{\omega}^4$
      $\omega^1{\bar\omega}^1$
      $\omega^3{\bar\omega}^3$
      ${\bar\omega}^2{\bar\omega}^4$ \\
\hline
\hline
$\alpha$& 1 ~~~ 1 ~~~ 1 ~~~ 0  & 1 ~~~ 1 ~~~ 1 ~~~ 0  & 1 ~~~ 1 ~~~ 1 ~~~ 0 \\
$\beta$ & 0 ~~~ 1 ~~~ 0 ~~~ 1  & 0 ~~~ 1 ~~~ 0 ~~~ 1  & 1 ~~~ 0 ~~~ 0 ~~~ 0 \\
$\gamma$& 0 ~~~ 0 ~~~ 1 ~~~ 1  & 1 ~~~ 0 ~~~ 0 ~~~ 0  & 0 ~~~ 1 ~~~ 0 ~~~ 0 \\
\end{tabular}
\end{center}
\caption{Additional basis vectors giving rise to a $SU(4) \times SU(2) \times U(1)$ gauge group without enhancement.}\label{tab:su421without}
\end{table}

The generalised GSO coefficients for the model with basis vectors given in table \ref{tab:su421without} are
\begin{equation}
{\bordermatrix{
 &\mathbbm{1}&S & & {b_1}&{b_2}&{b_3}& & {\alpha}&{\beta}&{\gamma}\cr
\mathbbm{1}&~~1&~~1 & & -1   &  -1 & -1   & &  ~~1     & ~~1   & ~~1   \cr
           S&~~1&~~1 & &~~1   & ~~1 &~~1   & &   -1     &  -1   &  -1   \cr
	      &   &    & &      &     &      & &         &       &       \cr
       {b_1}& -1& -1 & & -1   &  -1 & -1   & &   -1     &  -1   & ~~1   \cr
       {b_2}& -1& -1 & & -1   &  -1 & -1   & &  ~~1     & ~~1   & ~~1   \cr
       {b_3}& -1& -1 & & -1   &  -1 & -1   & &   -1     & ~~1   & ~~i   \cr
	      &   &    & &      &     &      & &          &       &       \cr
     {\alpha}&~~1& -1 & &~~1   &  -1 &~~1   & &  ~~1     & ~~1   & ~~i   \cr
     { \beta}&~~1& -1 & & -1   & ~~1 & -1   & &   -1     &  -1   & ~~i   \cr
     {\gamma}&~~1& -1 & & -1   & ~~1 & -1   & &   -1     & ~~1   & ~~i   \cr}}
\label{phasesmodel1}.
\end{equation}

In matrix \eqref{phasesmodel1} only the entries above the diagonal are
independent and those below and on the diagonal are fixed by
the modular invariance constraints.
Blank lines are inserted to emphasise the division of the free
phases between the different sectors of the realistic
free fermionic models. Thus, the first two lines involve
only the GSO phases of $\cc{\{\mathbbm{1},S \}}{a_i}$. The set
$\{\mathbbm{1},S\}$ generates the $N=4$ model with $S$ being the
space--time super symmetry generator and therefore the phases
$\cc{S}{a_i}$ are those that control the space--time super symmetry
in the super string models. Similarly, in the free fermionic
models, sectors with periodic and anti--periodic boundary conditions,
of the form of $b_i$, produce the chiral generations.
The phases $\cc{b_i}{b_j}$ determine the chirality
of the states from these sectors similar to equation \eqref{eq:susychirality}.

In the free fermionic models
the basis vectors $b_i$ are those that respect the $SO(10)$ symmetry
while the vectors denoted by Greek letters are those that break the
$SO(10)$ symmetry.
As the Standard Model matter states arise from sectors which
preserve the $SO(10)$ symmetry, the phases that fix
the Standard Model charges are, in general,
the phases $\cc{b_i}{a_i}$. 
The phases associated with the basis vectors
$\{\alpha,\beta,\gamma\}$ are associated with exotic physics, beyond the Standard Model.
These phases, therefore, also affect the final four dimensional
gauge symmetry.

The final gauge group of the model defined in table \ref{tab:su421without} and matrix \eqref{phasesmodel1} arises
as follows. In the observable sector the NS boundary conditions
produce gauge group generators for
\begin{equation}
SU(4)_C\times SU(2)_L\times U(1)_R\times U(1)_{1,2}\times SU(2)_3\times
U(1)_{4,5}\times SU(2)_6.
\end{equation}
Here the flavour $SU(2)_{3} \times SU(2)_{6}$ symmetries are generated by
$\{{\bar\eta}^3{\bar\zeta}^3\}$ where $\bar{\zeta}^3=\frac{1}{\sqrt{2}}
(\bar{w}^2+i\bar{w}^4)$. In previous free fermionic models
this group factor breaks to $U(1)^2$, but this is an artifact
of the specific model considered in table \ref{tab:su421without}, and
is not a generic feature of SU421 models.
Thus, the $SO(10)$ symmetry is broken to
$SU(4)\times SU(2)_L\times U(1)_R$, as discussed above,
where,
\begin{equation}
U(1)_R={\rm Tr}\, U(2)_L~\Rightarrow~Q_R=
			 \sum_{i=4}^5Q({\bar\psi}^i).
\label{u1r}
\end{equation}
The flavour $SO(6)^3$ symmetries are broken to $U(1)_{1,2}\times SU(2)_3
\times U(1)_{4,5}\times SU(2)_6$, where $Q_{1,2} = Q(\bar{\eta}_{1,2})$, $Q_{4} = Q(\bar{y}^3\bar{y}^5)$ and $Q_5 = Q(\bar{y}^1\bar{\omega}^5)$.
In the hidden sector the NS boundary conditions produce the generators of
\begin{equation}\label{nshiden}
SU(4)_{H}\times SU(2)_{H_1}\times SU(2)_{H_2}\times SU(2)_{H_3}\times
U(1)_{7,8}
\end{equation}
where $SU(2)_{H_1}\times SU(2)_{H_2}$ and $SU(2)_{H_3}$ arise from the complex
world--sheet fermions
$\{{\bar\phi}^7{\bar\phi}^8\}$ and $\{{\bar\phi}^5{\bar\phi}^6\}$,
respectively; and $U(1)_{7}$ and $U(1)_{8}$
correspond to the combinations of world--sheet charges
\begin{eqnarray}
Q_{7}&=&\sum_{i=5}^6Q({\bar\phi^i}),\label{qh2model1}\\
Q_{8}&=&\sum_{i=1}^4Q({\bar\phi^i}).\label{qh1model1}
\end{eqnarray}

As we discussed in section \ref{su421ffm} the SU421 models
contain additional sectors that may produce space--time
vector bosons and enhance the four dimensional gauge group.
In the model of table \ref{tab:su421without} these are the sectors
$2\gamma$, $\zeta_1\equiv\mathbbm{1}+b_1+b_2+b_3$ and
 $\zeta_2\equiv\mathbbm{1}+b_1+b_2+b_3+2\gamma$. However, due to the
choice of one--loop phases in equation \eqref{phasesmodel1}
all the additional vector bosons from these sectors are
projected out by the GSO projections and there is therefore
no gauge enhancement from these sectors in this model.

In addition to the graviton, dilaton, antisymmetric sector and
spin--1 gauge bosons, the NS sector gives two pairs
of colour triplets, transforming as (6,1,0) under
$SU(4)_C\times SU(2)_L\times U(1)_R$;
three quadruplets of $SO(10)$ singlets with
$U(1)_{1,2,3}$ charges; and three singlets of the entire four
dimensional gauge group.
The states from the sectors $b_j$ $(j=1,2,3)$ produce
the three light twisted generations. These states and their
decomposition under the entire gauge group are shown in
appendix \ref{sec:anomalyfree}. The remaining massless states and their quantum numbers
also appear in appendix \ref{sec:anomalyfree}.

\subsection{A model with enhanced symmetry}\label{es}

We next turn to our second example defined by table \ref{tab:su421with} and matrix \eqref{phasesmodel2}. 
\begin{table}
\begin{center}
\begin{tabular}{c|c|ccc|c|ccc|c}
 ~ & $\psi^\mu$ & $\chi^{12}$ & $\chi^{34}$ & $\chi^{56}$ &
        $\bar{\psi}^{1,...,5} $ &
        $\bar{\eta}^1 $&
        $\bar{\eta}^2 $&
        $\bar{\eta}^3 $&
        $\bar{\phi}^{1,...,8} $ \\
\hline
\hline
 ${\alpha}$  &  0 & 0&0&0 & 1~1~1~0~0 & 0 & 0 & 0 &1~1~1~1~0~0~0~0 \\
 ${\beta}$   &  0 & 0&0&0 & 1~1~1~0~0 & 0 & 0 & 0 &1~1~1~1~0~0~0~0 \\
 ${\gamma}$  &  0 & 0&0&0 & 0~0~0~$\frac{1}{2}$~$\frac{1}{2}$ & 0 & 0 & 0 &
	$\frac{1}{2}$  $\frac{1}{2}$  $\frac{1}{2}$  $\frac{1}{2}$
	$\frac{1}{2}$  $\frac{1}{2}$  0~0\\
\end{tabular}\\
\vspace{\baselineskip}

\begin{tabular}{c|c|c|c}
  &   $y^3{y}^6$
      $y^4{\bar y}^4$
      $y^5{\bar y}^5$
      ${\bar y}^3{\bar y}^6$
  &   $y^1{\omega}^5$
      $y^2{\bar y}^2$
      $\omega^6{\bar\omega}^6$
      ${\bar y}^1{\bar\omega}^5$
  &   $\omega^2{\omega}^4$
      $\omega^1{\bar\omega}^1$
      $\omega^3{\bar\omega}^3$
      ${\bar\omega}^2{\bar\omega}^4$ \\
\hline
\hline
$\alpha$& 0 ~~~ 0 ~~~ 1 ~~~ 1 & 1 ~~~ 0 ~~~ 0 ~~~ 0  & 0 ~~~ 1 ~~~ 0 ~~~ 1 \\
$\beta$ & 1 ~~~ 0 ~~~ 0 ~~~ 0 & 0 ~~~ 0 ~~~ 1 ~~~ 1  & 0 ~~~ 0 ~~~ 1 ~~~ 1 \\
$\gamma$& 0 ~~~ 1 ~~~ 0 ~~~ 1 & 0 ~~~ 1 ~~~ 0 ~~~ 1  & 1 ~~~ 0 ~~~ 0 ~~~ 1 \\
\end{tabular}
\end{center}
\caption{Additional basis vectors giving rise to a $SU(4) \times SU(2) \times U(1)$ gauge group with enhancement.}\label{tab:su421with}
\end{table}

The generalised GSO coefficients for the model described by table \ref{tab:su421with} are
\begin{equation}
{\bordermatrix{
              &\mathbbm{1}&S & & {b_1}&{b_2}&{b_3}&
&{\alpha}&{\beta}&{\gamma}\cr
       \mathbbm{1}&~~1&~~1 & & -1   &  -1 & -1  & & ~~1     & ~~1   & ~~1   \cr
           S&~~1&~~1 & &~~1   & ~~1 &~~1  & &  -1     &  -1   &  -1   \cr
	      &   &    & &      &     &     & &         &       &       \cr
       {b_1}& -1& -1 & & -1   &  -1 & -1  & & ~~1     &  -1   &  -1   \cr
       {b_2}& -1& -1 & & -1   &  -1 & -1  & &  -1     &  -1   &  -1   \cr
       {b_3}& -1& -1 & & -1   &  -1 & -1  & & ~~1     & ~~1   & ~~i   \cr
	      &   &    & &      &     &     & &         &       &       \cr
     {\alpha}&~~1& -1 & &~~1   & ~~1 &~~1  & &  -1     & ~~1   &  -1   \cr
     {\beta} &~~1& -1 & &~~1   &  -1 &~~1  & & ~~1     &  -1   & ~~1   \cr
     {\gamma}& -1& -1 & &~~1   & ~~1 & -1  & &  -1     & ~~1   &  -1   \cr}}
\label{phasesmodel2}.
\end{equation}
The total gauge group of the model defined by table \ref{tab:su421with} and matrix \eqref{phasesmodel2} arises as follows.
In the observable sector the NS boundary conditions produce the
generators of
$SU(4)_C \times SU(2)_L\times U(1)_R\in SO(10)
\times U(1)_{1,2,3}\times U(1)_{4,5,6}$,
while in the hidden sector
the NS boundary conditions produce the generators of
\begin{equation}
SU(4)_{H}\times SU(2)_{H_1}\times SU(2)_{H_2}\times
SU(2)_{H_3}\times U(1)_7\times U(1)_8\, .
\end{equation}
$U(1)_{7}$ and $U(1)_{8}$
correspond to the combinations of the world--sheet charges
given in equations \eqref{qh2model1} and \eqref{qh1model1}, respectively and where $Q_{1,2,3} = Q(\bar{\eta}_{1,2,3})$, $Q_{4} = Q(\bar{y}^3\bar{y}^5)$, $Q_5 = Q(\bar{y}^1\bar{\omega}^5)$ and $Q_6 = Q(\bar{\omega}^2\bar{\omega}^4)$.

The model with enhancement contains two combinations
of non--\nahe\ basis vectors with $X_L\cdot X_L=0$, which
therefore may give rise to additional space--time vector bosons.
The first is the sector ${2\gamma}$.
The second arises from the vector combination given by $\zeta+2\gamma$,
where $\zeta\equiv\mathbbm{1}+b_1+b_2+b_3$.
Both sectors arise from the
\nahe\ set basis vectors plus $2\gamma$ and are therefore
independent of the assignment of periodic boundary
conditions in the basis vectors $\alpha$, $\beta$.
Both are therefore generic for the
pattern of symmetry breaking $SO(10)\rightarrow
SU(4)_C\times  SU(2)_L\times U(1)_R$,
in \nahe\ based models.

In the model without enhancement the additional space--time vector bosons from both
sectors are projected out and therefore there is no gauge
enhancement. In the model with enhancement all the
space--time vector bosons from the sector $2\gamma$
are projected out by the GSO projections and therefore
give no gauge enhancement from this sector.
The sector $\zeta+2\gamma$ may, or may not, give
rise to additional space--time vector bosons, depending on
the choice of GSO phase
\begin{equation}\label{b3gammaphase}
\cc{\gamma}{b_3}=\pm1,
\end{equation}
where with the $+1$ choice all the additional vector bosons are
projected out,
whereas the $-1$ choice gives rise to additional space--time gauge bosons
which are charged with respect to the
$SU(2)_L\times SU(2)_{H_1}$ groups. This enhances the
$SU(2)_L\times SU(2)_{H_1}$ group to $SO(5)$.
Thus, in this case,
the full massless spectrum transforms under the final gauge group,
$SU(4)_C\times SO(5)\times U(1)_R\times
U(1)_{1,2,3}\times U(1)_{4,5,6}\times
SU(4)_{H}\times SU(2)_{H_2}\times SU(2)_{H_3}\times
U(1)_{7,8}$.

In addition to the graviton,
dilaton, antisymmetric sector and spin--1 gauge bosons,
the NS sector gives rise to
three quadruplets of $SO(10)$ singlets with
$U(1)_{1,2,3}$ charges; and three singlets of the entire four
dimensional gauge group.
The states from the sectors $b_j\oplus \zeta+2\gamma~(j=1,2,3)$ produce
the three light generations. The states from these sectors and their
decomposition under the entire gauge group are shown in appendix \ref{sec:anomalous}.

\section{Overview of the Realistic Models}\label{sec:general}

In the previous section we have shown how to construct realistic and semi-realistic models. We have given details of two specific models. In this section we give a recapitulation of the general properties of realistic and semi-realistic models. We describe properties which have been dealt with in detail in the previous sections. We show that the realistic models are $\mathbbm{Z}_2 \times \mathbbm{Z}_2$ orbifolds with symmetric shifts. This section therefore provides a bridge between specific models and the classification of the chiral $\mathbbm{Z}_2 \times \mathbbm{Z}_2$ heterotic string models. 

The notation
and details of the construction of these
models are given in chapter \ref{chap:fermionic} and \cite{Antoniadis:1989zy, Lopez:1993kg, Faraggi:1990ka, Faraggi:1992af, Antoniadis:1990hb, Leontaris:1999ce, Faraggi:1992jr, Faraggi:1992be, Faraggi:1992fa, Faraggi:1993rd, Cleaver:1998sa, Cleaver:1999mw, Faraggi:1993ac, Faraggi:1997dc, Cleaver:2000ds, Cleaver:2001fd}.
In the free fermionic formulation of the heterotic string
in four dimensions all the world-sheet
degrees of freedom  required to cancel
the conformal anomaly are represented in terms of free world--sheet
fermions \cite{Antoniadis:1987rn, Kawai:1987ah}.
In the light-cone gauge the world-sheet field content consists
of two transverse left- and right-moving space-time coordinate bosons,
$X^\mu_{1,2}$ and ${\bar X}^\mu_{1,2}$,
and their left-moving fermionic super partners $\psi^\mu_{1,2}$,
and additional 62 purely internal
Majorana-Weyl fermions, of which 18 are left-moving,
and 44 are right-moving.
In the super symmetric sector the world-sheet super symmetry is realised
non-linearly and the world-sheet supercurrent \cite{Antoniadis:1986az}
is from equation \eqref{eq:supercurrent} given by
\begin{equation}
T_F=\psi^\mu\partial X_\mu+i\chi^Iy^I\omega^I,~(I=1,\cdots,6).
\end{equation}
The $\{\chi^{I},y^I,\omega^I\}~(I=1,\cdots,6)$ are 18 real free
fermions transforming as the adjoint representation of ${\rm SU}(2)^6$.
Under parallel transport around a non-contractible loop on the toroidal
world-sheet the fermionic fields pick up a phase as is shown in equation \eqref{eq:boundarycondition}.
Each set of specified
phases for all world-sheet fermions, around all the non-contractible
loops is called the spin structure of the model. Such spin structures
are usually given in the form of 64 dimensional boundary condition vectors,
with each element of the vector specifying the phase of the corresponding
world-sheet fermion. The basis vectors are constrained by string consistency
requirements and completely determine the vacuum structure of the model.
The physical spectrum is obtained by applying the generalised GSO
projections
\cite{Antoniadis:1987rn, Kawai:1987ah}.

The boundary condition basis vectors defining a typical realistic free
fermionic heterotic string model is constructed in two stages. The
first stage consists of the \nahe\ set, which is a set of five
boundary condition basis vectors, $\{ 1 ,S,b_1,b_2,b_3\}$
\cite{Ferrara:1987tp, Ferrara:1987jr, Ferrara:1987jq, Ferrara:1989jx, Kiritsis:1997dn, Kiritsis:1998en, Faraggi:1993ac, Faraggi:1997dc} and is explained in detail in section \ref{sec:nahe}. The gauge group induced by the \nahe\ set is
${SO}(10) \times {SO}(6)^3 \times \mathbbm{E}_8$ with ${ N}=1$
super symmetry. The space-time vector bosons that generate the
gauge group arise from the Neveu--Schwarz sector and from the
sector $\xi_2\equiv 1+b_1+b_2+b_3$. The Neveu-Schwarz sector
produces the generators of ${SO}(10)\times {SO}(6)^3\times
{SO}(16)$. The $\xi_2$-sector produces the spinorial ${\bf 128}$ of
SO(16) and completes the hidden gauge group to $\mathbbm{E}_8$. 

The second stage of the
construction consists of adding to the
\nahe\ set three (or four) additional basis vectors.
These additional vectors reduce the number of generations
to three, one from each of the sectors $b_1$,
$b_2$ and $b_3$, and simultaneously break the four dimensional
gauge group. The assignment of boundary conditions to
$\{{\bar\psi}^{1,\cdots,5}\}$ breaks SO(10) to one of its subgroups
${\rm SU}(5)\times {\rm U}(1)$ \cite{Antoniadis:1989zy, Lopez:1993kg}, ${\rm SO}(6)\times {\rm SO}(4)$
\cite{Antoniadis:1990hb, Leontaris:1999ce},
${\rm SU}(3)\times {\rm SU}(2)\times {\rm U}(1)^2$ \cite{Faraggi:1990ka, Faraggi:1992af, Faraggi:1992jr, Faraggi:1992be, Faraggi:1992fa, Faraggi:1993rd, Cleaver:1998sa, Cleaver:1999mw},
${\rm SU}(3)\times {\rm SU}(2)^2\times {\rm U}(1)$ \cite{Cleaver:2000ds, Cleaver:2001fd} or
${\rm SU}(4)\times {\rm SU}(2)\times {\rm U}(1)$ \cite{Cleaver:2002ps}.
Similarly, the hidden $\mathbbm{E}_8$ symmetry is broken to one of its
subgroups, and the flavour ${\rm SO}(6)^3$ symmetries are broken
to $U(1)^n$, with $3\le n\le9$.
For details and phenomenological studies of
these three generation string models we refer interested
readers to the original literature and review articles
\cite{Lykken:1995kk, Lopez:1996gy, Faraggi:1997yq, Faraggi:2002nq, Faraggi:2003xw}.

The correspondence of the free fermionic models
with the orbifold construction is illustrated
by extending the \nahe\ set, $\{ 1,S,b_1,b_2,b_3\}$, by at least
one additional boundary condition basis vector \cite{Faraggi:1994pr, Ellis:1998ec}
\begin{equation}\label{vectorx}
\xi_1=(0,\cdots,0\vert{\underbrace{1,\cdots,1}_{{\bar\psi^{1,\cdots,5}},
{\bar\eta^{1,2,3}}}},0,\cdots,0)~.
\end{equation}
With a suitable choice of the GSO projection coefficients the
model possesses an ${\rm SO}(4)^3\times \mathbbm{E}_6\times {\rm U}(1)^2
\times \mathbbm{E}_8$ gauge group
and ${ N}=1$ space-time super symmetry. The matter fields
include 24 generations in the 27 representation of
$\mathbbm{E}_6$, eight from each of the sectors $b_1\oplus b_1+\xi_1$,
$b_2\oplus b_2+\xi_1$ and $b_3\oplus b_3+\xi_1$.
Three additional 27 and $\overline{27}$ pairs are obtained
from the Neveu-Schwarz $\oplus~\xi_1$ sector.

To construct the model in the orbifold formulation one starts
with the compactification on a torus with nontrivial background
fields \cite{Narain:1986jj, Narain:1987am}.
The subset of basis vectors
\begin{equation}\label{neq4set}
\{ 1,S,\xi_1,\xi_2\}
\end{equation}
generates a toroidally-compactified model with ${ N}=4$ space-time
super symmetry and ${\rm SO}(12)\times \mathbbm{E}_8\times \mathbbm{E}_8$ gauge group.
The same model is obtained in the geometric (bosonic) language
by tuning the background fields to the values corresponding to
the SO(12) lattice. 

Adding the two basis vectors $b_1$ and $b_2$ to the set
(\ref{neq4set}) corresponds to the $\mathbbm{Z}_2\times \mathbbm{Z}_2$
orbifold model with standard embedding.
Starting from the $N=4$ model with ${\rm SO}(12)\times
\mathbbm{E}_8\times \mathbbm{E}_8$
symmetry~\cite{Narain:1986jj, Narain:1987am}, and applying the $\mathbbm{Z}_2\times \mathbbm{Z}_2$
twist on the
internal coordinates, reproduces
the spectrum of the free-fermion model
with the six-dimensional basis set
$\{ 1,S,\xi_1,\xi_2,b_1,b_2\}$.
The Euler characteristic of this model is 48 with $h_{11}=27$ and
$h_{21}=3$.

It is noted that the effect of the additional basis vector $\xi_1$ of equation
\eqref{vectorx}, is to separate the gauge degrees of freedom, spanned by
the world-sheet fermions $\{{\bar\psi}^{1,\cdots,5},
{\bar\eta}^{1},{\bar\eta}^{2},{\bar\eta}^{3},{\bar\phi}^{1,\cdots,8}\}$,
from the internal compactified degrees of freedom $\{y,\omega\vert
{\bar y},{\bar\omega}\}^{1,\cdots,6}$.
In the realistic free fermionic
models this is achieved by the vector $2\gamma$ \cite{Ellis:1998ec}, with
\begin{equation}\label{vector2gamma}
2\gamma=(0,\cdots,0\vert{\underbrace{1,\cdots,1}_{{\bar\psi^{1,\cdots,5}},
{\bar\eta^{1,2,3}} {\bar\phi}^{1,\cdots,4}} },0,\cdots,0)~,
\end{equation}
which breaks the $\mathbbm{E}_8\times \mathbbm{E}_8$ symmetry to $SO(16)\times
SO(16)$.
The $\mathbbm{Z}_2\times \mathbbm{Z}_2$ twist induced by $b_1$ and $b_2$
breaks the gauge symmetry to
$SO(4)^3\times SO(10)\times U(1)^3\times SO(16)$.
The orbifold still yields a model with 24 generations,
eight from each twisted sector,
but now the generations are in the chiral ${\bf 16}$ representation
of SO(10), rather than in the ${\bf 27}$ of $\mathbbm{E}_6$. The same model can
be realized with the set
$\{ 1,S,\xi_1,\xi_2,b_1,b_2\}$,
by projecting out the $16\oplus{\overline{16}}$
from the $\xi_1$-sector taking
\begin{equation}\label{changec}
\cc{\xi_1}{\xi_2} \rightarrow -\cc{\xi_1}{\xi_2}.
\end{equation}
This choice also projects out the massless vector bosons in the
128 of SO(16) in the hidden-sector $\mathbbm{E}_8$ gauge group, thereby
breaking the $\mathbbm{E}_6\times \mathbbm{E}_8$ symmetry to
$SO(10)\times U(1)\times SO(16)$.
We can define two ${ N}=4$ models generated by the set
(\ref{neq4set}), ${ Z}_+$ and ${ Z}_-$, depending on the sign
in equation (\ref{changec}). The first, say ${ Z}_+$,
produces the $\mathbbm{E}_8\times \mathbbm{E}_8$ model, whereas the second, say
${ Z}_-$, produces the ${\rm SO}(16)\times {\rm SO}(16)$ model.
However, the $\mathbbm{Z}_2\times
\mathbbm{Z}_2$
twist acts identically in the two models, and their physical characteristics
differ only due to the discrete torsion equation (\ref{changec}).

This analysis confirms that the $\mathbbm{Z}_2\times \mathbbm{Z}_2$
orbifold on the $SO(12)$ lattice is at the core of the realistic
free fermionic models.
To illustrate how the chiral generations are generated in the free
fermionic models we consider the $\mathbbm{E}_6$ model which is produced
by the extended \nahe --set $\{ 1,S,\xi_1,\xi_2,b_1,b_2\}$.

The chirality of the states from a twisted sector
$b_j$ is determined by the free phase $\cc{b_j}{b_i}$.
Since we have a freedom in the choice of the sign of this
free phase, we can get from the sector $(b_i)$ either the
${\bf 27}$ or the $\overline{\bf 27}$. Which of those we obtain in the physical
spectrum depends on the sign of the free phase.
The free phases $\cc{b_j}{b_i}$ also fix the total number
of chiral generations. Since there are two $b_i$ projections
for each sector $b_j$, $i\ne j$ we can use one projections to project
out the states with one chirality and the other projection to
project out the states with the other chirality. Thus, the total
number of generations with this set of basis vectors
is given by
$$8\left(\frac{\cc{b_1}{b_2} + \cc{b_1}{b_3}}{2}\right) +
8\left(\frac{\cc{b_2}{b_1} + \cc{b_2}{b_3}}{2}\right) +
8\left(\frac{\cc{b_3}{b_1} + \cc{b_3}{b_1}}{2}\right).$$
Since the modular invariance rules fix
$\cc{b_j}{b_i}=\cc{b_i}{b_j}$ we get that
the total number of generations is either
24 or 8. Thus, to reduce the number of generation further
it is necessary to introduce additional basis vectors.

To illustrate the reduction to three generations in
the realistic free fer\-mi\-on\-ic models we consider the model in table
\ref{m278}.
\begin{table}
\begin{center}
\begin{tabular}{c|c|ccc|c|ccc|c}
 ~ & $\psi^\mu$ & $\chi^{12}$ & $\chi^{34}$ & $\chi^{56}$ &
        $\bar{\psi}^{1,...,5} $ &
        $\bar{\eta}^1 $&
        $\bar{\eta}^2 $&
        $\bar{\eta}^3 $&
        $\bar{\phi}^{1,...,8} $\\
\hline
\hline
  ${\alpha}$  &  0 & 0&0&0 & 1~1~1~0~0 & 0 & 0 & 0 & 1~1~1~1~0~0~0~0 \\
  ${\beta}$   &  0 & 0&0&0 & 1~1~1~0~0 & 0 & 0 & 0 & 1~1~1~1~0~0~0~0 \\
  ${\gamma}$  &  0 & 0&0&0 &
        $\frac{1}{2}$~$\frac{1}{2}$~$\frac{1}{2}$~$\frac{1}{2}$~$\frac{1}{2}$
          & $\frac{1}{2}$ & $\frac{1}{2}$ & $\frac{1}{2}$ &
                $\frac{1}{2}$~0~1~1~$\frac{1}{2}$~$\frac{1}{2}$~$\frac{1}{2}$~0
\\
\end{tabular}\\
\vspace{\baselineskip}

\begin{tabular}{c|c|c|c}
  &   $y^3{y}^6$
      $y^4{\bar y}^4$
      $y^5{\bar y}^5$
      ${\bar y}^3{\bar y}^6$
  &   $y^1{\omega}^5$
      $y^2{\bar y}^2$
      $\omega^6{\bar\omega}^6$
      ${\bar y}^1{\bar\omega}^5$
  &   $\omega^2{\omega}^4$
      $\omega^1{\bar\omega}^1$
      $\omega^3{\bar\omega}^3$
      ${\bar\omega}^2{\bar\omega}^4$ \\
\hline
\hline
$\alpha$ & 1 ~~~ 0 ~~~ 0 ~~~ 0  & 0 ~~~ 0 ~~~ 1 ~~~ 1  & 0 ~~~ 0 ~~~ 1 ~~~ 1
\\
$\beta$  & 0 ~~~ 0 ~~~ 1 ~~~ 1  & 1 ~~~ 0 ~~~ 0 ~~~ 0  & 0 ~~~ 1 ~~~ 0 ~~~ 1
\\
$\gamma$ & 0 ~~~ 1 ~~~ 0 ~~~ 1  & 0 ~~~ 1 ~~~ 0 ~~~ 1  & 1 ~~~ 0 ~~~ 0 ~~~ 0
\\
\end{tabular}
\end{center}
\caption{Model for the illustration of the reduction to three generations.}\label{m278}
\end{table}
Here the vector $\xi_1$ (\ref{vectorx})
is replaced by the vector $2\gamma$ (\ref{vector2gamma}).
At the level of the \nahe\ set we have 48 generations.
One half of the generations is projected by the vector $2\gamma$.
Each of the three vectors in table \ref{m278}
acts non trivially on the degenerate
vacuum
of the sectors $b_1$, $b_2$ and $b_3$ and reduces the
number of generations in each step
by a half. Thus, we obtain one generation from each sector
$b_1$, $b_2$ and $b_3$.

The geometrical interpretation of the basis vectors
beyond the \nahe\ set is facilitated by taking combinations of the
basis vectors in \ref{m278}, which entails choosing another set
to generate the same vacuum. The combinations
$\alpha+\beta$, $\alpha+\gamma$, $\alpha+\beta+\gamma$ produce
the boundary conditions under the set of internal
real fermions as is displayed in table \ref{m2782}.
\begin{table}
\begin{center}
\begin{tabular}{c|c|c|c}
 ~&   $y^3{y}^6$
      $y^4{\bar y}^4$
      $y^5{\bar y}^5$
      ${\bar y}^3{\bar y}^6$
  &   $y^1{\omega}^5$
      $y^2{\bar y}^2$
      $\omega^6{\bar\omega}^6$
      ${\bar y}^1{\bar\omega}^5$\\\hline
\hline
$\alpha+\beta$ & 1 ~~~ 0 ~~~ 1 ~~~ 1  & 1 ~~~ 0 ~~~ 1 ~~~ 1  \\
$\beta+\gamma$ & 0 ~~~ 1 ~~~ 1 ~~~ 0  & 1 ~~~ 1 ~~~ 0 ~~~ 1  \\
$\alpha+\beta+\gamma$ & 1 ~~~ 1 ~~~ 1 ~~~ 0  & 1 ~~~ 1 ~~~ 1 ~~~ 0  
\end{tabular}\\
\vspace{\baselineskip}

\begin{tabular}{c|c}
  &   $\omega^2{\omega}^4$
      $\omega^1{\bar\omega}^1$
      $\omega^3{\bar\omega}^3$
      ${\bar\omega}^2{\bar\omega}^4$ \\
\hline
\hline
$\alpha+\beta$ 		& 0 ~~~ 1 ~~~ 1 ~~~ 0\\
$\beta+\gamma$ 		& 1 ~~~ 1 ~~~ 0 ~~~ 1\\
$\alpha+\beta+\gamma$ 	& 1 ~~~ 1 ~~~ 1 ~~~ 0
\end{tabular}
\end{center}
\caption{A redefinition of the model listed in table \ref{m278}}\label{m2782}
\end{table}
It is noted that the two combinations $\alpha+\beta$ and $\beta+\gamma$
are fully symmetric between the left and right movers, whereas the
third, $\alpha+\beta+\gamma$, is asymmetric.
The action of the first two combinations
on the compactified bosonic coordinates translates therefore to symmetric
shifts. Thus, we see that reduction of the number of generations
is obtained by further action of symmetric shifts as described in chapter \ref{chap:orbifold.setup}.

Due to the presence of the third combination, the situation
is more complicated. The third combination in table \ref{m2782} is
asymmetric between the left and right movers and therefore
does not have an obvious geometrical interpretation.
In subsequent chapters we perform a complete
classification of all the possible \nahe --based
$\mathbbm{Z}_2\times \mathbbm{Z}_2$ orbifold models with symmetric shifts on the complex
tori, which reveals that three generations are not obtained in this manner.
Three generations are obtained in the free fermionic models
by the inclusion of the asymmetric shift in (\ref{m2782}).
This outcome has profound implications on the type of geometries that
may be related to the realistic string vacua, as well as on the
issue of moduli stabilisation.


	\part{Classification of the Chiral $\mathbbm{Z}_2 \times \mathbbm{Z}_2$ Heterotic String Models}
\chapter{The Classification in the Orbifold Formulation}\label{chap:orbifold}

In this chapter we set up the partition function of the heterotic string using the geometrical description. We start by describing the $N=4$ heterotic string. We break the number of super symmetries using a $\mathbbm{Z}_2\times \mathbbm{Z}_2$ twist. This leads to four subclasses of $\mathbbm{Z}_2\times \mathbbm{Z}_2$ heterotic string models. We isolate the different subclasses and show their partition functions.
This chapter is based on parts of \cite{Faraggi:2003yd, Faraggi:2004rq, Faraggi:2004xnew}.

\section{$N=1$ Heterotic Orbifold Constructions}

In this section we revise the $\mathbbm{Z}_2\times \mathbbm{Z}_2$
heterotic orbifold construction and relate this to
 the free fermionic construction. We isolate the individual conformal blocks
that will facilitate the classification of the models and set up a procedure
to analyse all possible $N=1$ heterotic $\mathbbm{Z}_2 \times \mathbbm{Z}_2 $ models. We start by
describing the procedure to descend from $N=4$ to $N=1$ super symmetric
heterotic vacua.

\subsection{The $N=4$ orbifold models}

The partition function for any  heterotic model via the fermionic construction is from equation \eqref{eq:partition}
\begin{equation}\label{eq:fermionpt}
Z = \frac{1}{\tau_2} \frac{1}{\eta^{12}\bar{\eta}^{24}} \sum_{a,b \in \Xi} c[^a
_b] \frac{1}{2^{M}} \prod_{i=1}^{20} \theta[^{a_i}_{b_i}]^{\frac{1}{2}} \prod
_{j=1}^{44} \bar{\theta}[^{a_j}_{b_j}]^{\frac{1}{2}}.
\end{equation}
In the above  equation $M$ is the number of basis vectors and the
parameters in the $\theta$--functions represent the action of
the vectors. In order to obtain a super symmetric model we need at least two
basis vectors $\{1,S\}$ as explained in section \ref{sec:susybackground}
\begin{eqnarray}
1 &=& \{ \psi^{1,2}, \chi^{1,\ldots,6}, y^{1,\ldots,6}, \omega^{1,\ldots,6} |
\nonumber\\
  & & \bar{y}^{1,\ldots,6}, \bar{\omega}^{1,\ldots,6}, \bar{\psi}^{1,\ldots,5},
 \bar{\eta}^{1,2,3}, \bar{\phi}^{1,\ldots,8} \} ,\label{eq:1}\\
S &=& \{ \psi^{1,2}, \chi^{1,\ldots,6} \} .\label{eq:S}
\end{eqnarray}
The super symmetric GSO projection is induced by the set $S$ for any choice
of the  GSO coefficient
\begin{equation}
\cc{S}{1}=\pm1.
\end{equation}
The corresponding partition function has a factorised left--moving
contribution coming from the sector $S$. We find from equation \eqref{eq:n4partition}
\begin{equation}\label{eq:fermionpt44}
Z_{1,S} = \frac{1}{\tau_2 |\eta |^4 } \frac{1}{2} \sum_{a,b=0}^1
(-1)^{a+b+\mu ab} \frac{\theta[^a_b]^4}{\eta^4}
\frac{\Gamma_{6,6+16}[SO(44)]}{\eta^6\bar{\eta}^{22}},
\end{equation}
where
\begin{equation}
\frac{\Gamma_{6,6+16}[SO(44)]}{\eta^6\bar{\eta}^{22}}=\frac{1}{2} \sum_{c,d}
\frac{\theta[^c_d]^6 \bar{\theta}[^c_d]^{22}}{\eta^6\bar{\eta}^{22}},
\end{equation}
and as in equation \eqref{eq:susychirality}
\begin{equation}
\mu=\frac{1}{2}{\left(1-\cc{\mathbbm{1}}{S}\right)}
\end{equation}
defines the chirality of $N=4$
super symmetry. Therefore, the role of the boundary condition vector $S$ is to
factorise the left--moving contribution \cite{Kiritsis:1997gu},
\begin{equation}\label{eq:Jacobi}
Z^L_{N=4}= \frac{1}{2 } \sum_{a,b=0}^1
(-1)^{a+b+\mu ab}\theta[^a_b](v)\theta[^a_b]^3(0) \sim v^4,
\end{equation}
which is zero with the multiplicity of $N=4$ super symmetry.

The above partition function gives rise to a $SO(44)$ right--moving
gauge group and is the maximally symmetric point in the moduli space of
the Narain $\Gamma_{6,6+16}$ lattice. The  general $\Gamma_{6,6+16}$ lattice
depends on $6\times 22$ moduli, the metric $G_{ij}$ and the antisymmetric
tensor $B_{ij}$  of the six dimensional internal space, as well as the
Wilson lines $Y^I_i$ that appear in the  2d-world--sheet as in equation \eqref{eq:gamma622action}
\begin{eqnarray}
S &=& \frac{1}{4\pi} \int d^2\sigma \sqrt{g}g^{ab}G_{ij}\partial_a
X^i\partial_bX^j + \frac{1}{4\pi} \int d^2\sigma \epsilon^{ab}
B_{ij}\partial_aX^i\partial_bX^j\nonumber\\
  & & + \frac{1}{4\pi} \int d^2\sigma \sqrt{g} \sum_I \psi^I \Big[ \bar{\nabla}
 + Y^I_i\bar{\nabla}X^i \Big]\bar{\psi}^I.
\end{eqnarray}
Here $i$ runs over the internal coordinates and $I$ runs over the extra $16$
right--moving degrees of freedom described by $\bar{\psi}^I$.

The compactified sector of the partition function is given by
$\Gamma_{6,6+16}$
\begin{eqnarray}\label{eq:internal5}
\Gamma_{6,6+16} &=& \frac{(\det G)^3}{\tau_2^3}\sum_{m,n} \exp\bigg\{ -\pi
\frac{T_{ij}}{\tau_2}[m^i + n^i\tau][m^j + n^j\bar{\tau}]\bigg\}\nonumber\\
& & \times \frac{1}{2}\sum_{\gamma,\delta} \prod_{I=1}^{16}\exp\Big[-i\pi n^i
\big(m^j + n^j \bar{\tau}\big)Y^I_iY^I_j\Big]\nonumber\\
& & \times \bartheta{\gamma}{\delta}\big(Y^I_i(m^i+n^i\bar{\tau})|\tau\big),
\end{eqnarray}
where $T_{ij} = G_{ij} + B_{ij}$.

Equation \eqref{eq:internal5} is the winding mode representation of the
partition function. Using a
Poisson resummation we can put it  in the momentum representation form:
\begin{equation}
\Gamma_{6,22} = \sum_{P, \bar{P}, Q} \exp\bigg\{\frac{i\pi\tau}{2}P_iG^{ij}P_j
-\frac{i\pi\bar{\tau}}{2}\bar{P}_iG^{ij}\bar{P}_j - i\pi\bar{\tau}\hat{Q}^I\hat
{Q}^I\bigg\},
\end{equation}
with
\begin{subequations}
\begin{eqnarray}
P_i &=& m_i +B_{ij}n^j + \frac{1}{2}Y^I_iY^I_jn^j + Y^I_iQ^I + G_{ij}n^j,\\
\bar{P}_i &=& m_i +B_{ij}n^j + \frac{1}{2}Y^I_iY^I_jn^j + Y^I_iQ^I - G_{ij}
n^j,\\
\hat{Q}^I &=& Q^I + Y^I_in^i.
\end{eqnarray}
\end{subequations}
The charge momenta $Q^I = n - \frac{a^I}{2}$ are induced by the right--moving fermions
$\bar{\psi}^I$ which  appear explicitly in the $\theta$--functions from equation \eqref{eq:thetadefinition}.

For generic $G_{ij}, B_{ij}$ and for vanishing values for  Wilson lines,
$Y^I_i=0$
one  obtains an $N=4$ model with a gauge group $U(1)^6 \times SO(32)$.
The $U(1)^6$ can be extended to $SO(12)$ by fixing the moduli of the
internal manifold \cite{Faraggi:1994pr, Ellis:1998ec}.

The $N=4$ fermionic construction based on $\{1,S\}$ \eqref{eq:fermionpt44}
has an  extended  gauge group, $SO(44)$. From the lattice construction
point of view, a $N=4$ model with a gauge group $G \subset SO(44)$
can be generated by switching on Wilson lines and fine tune the moduli
of the internal manifold.
Moving from the $SO(44)$ to  $U(1)^6 \times SO(32)$ heterotic point as well
as to the  $U(1)^6 \times \mathbbm{E}_8 \times \mathbbm{E}_8$ point can be realized continuously
\cite{Kiritsis:1997ca}.
The partition function at the $U(1)^6 \times \mathbbm{E}_8 \times \mathbbm{E}_8$ point takes
a simple factorised form
\begin{eqnarray}
\Gamma_{6,6+16} &=& \frac{(\det G)^3}{\tau_2^3}\sum_{m,n} \exp\bigg\{ -\pi
\frac{T_{ij}}{\tau_2}[m^i + n^i\tau][m^j + n^j\bar{\tau}]\bigg\}
\label{eq:internal2}\nonumber\\
& & \times \frac{1}{2}\sum_{\gamma,\delta} \bartheta{\gamma}{\delta}^8
\times  \frac{1}{2}\sum_{h,g} \bartheta{h}{g}^8.
\end{eqnarray}

\subsection{The $N=1$ orbifold models}\label{sec:n1}

To break the number of super symmetries down from $N=4$ to $N=1$ in the
fermionic formulation using a $\mathbbm{Z}_2 \times \mathbbm{Z}_2$ twist, we introduce the vectors $b_1$ and $b_2$.
\begin{eqnarray}
b_1 &=& \{ \chi^{3,4}, \chi^{5,6},\ y^{3,4},y^{5,6}\ |\ \ldots\  \} ,
\label{eq:b1flats3}\\
b_2 &=& \{ \chi^{1,2}, \chi^{5,6}, y^{1,2},\ y^{5,6}\ |\ \ldots\ \} .
\label{eq:b2flats3}
\end{eqnarray}
The $ b_1$ twists the second  and third complex planes (3,4) and (5,6) while
$ b_2$ twists the first  and third (1,2) and (5,6) ones. Thus, $b_1,b_2$
separate the internal lattice into the three complex  planes:
(1,2), (3,4) and (5,6). Note that these basis vectors are linear combinations of the vectors defined in equation \eqref{eq:nahetwist}.

The action of the $ b_i$--twists  fully determines the fermionic
content for the left--moving sector. The dots $ \ \ldots\ $
in $ b_1$, $ b_2$ stand for the $n_1$, $ n_2$ right--moving fermions.
To generate a modular invariant model we can distinguish four options
$n_i$ are either $8$, $16$, $24$ or $32$ real right-moving fermions in
the basis vector $b_i$.

Defining the basis vectors with $8$ real right--moving fermions leads to
massless states in the spectrum in  vectorial representations of the gauge
groups; $16$ real right--moving fermions give rise to
spinorial representations on each plane. Adding either $24$ or $32$
right--moving  fermions would produce massive states in the spectrum.
We discard the last two options. We thus need to introduce
$16$ real fermions ($8$ complex) in the vectors $b_1,b_2$ for the
existence of spinorial representations on the first and second plane.

A suitable choice having section \ref{sec:construction} in mind, is for instance,
\begin{eqnarray}
b_1 &=& \{ \chi^{3,4}, \chi^{5,6}, y^{3,4}, y^{5,6}\
| \bar{y}^{3,4}, \bar{y}^{5,6},\
\bar{\eta}^{1}, \bar{\psi}^{1,\ldots,5} \} ,\label{eq:twistvectors1s3}\\
b_2 &=& \{ \chi^{1,2}, \chi^{5,6}, y^{1,2}, y^{5,6}\
| \bar{y}^{1,2}, \bar{y}^{5,6},\
\bar{\eta}^{2}, \bar{\psi}^{1,\ldots,5} \} .
\label{eq:twistvectors2s3}
\end{eqnarray}
The  $N=1$ partition function based on $\{1,S,b_1,b_2\}$ is
\begin{subequations}
\begin{eqnarray}
Z_{N=1} &=&
\frac{1}{\tau_2 |\eta |^4 } \frac{1}{2}
\sum_{\alpha,\beta} e^{i\pi(a+b+\mu ab)}\nonumber\\
& & \frac{1}{4}\sum_{h_1,h_2,g_1,g_2}\  {\frac{\theta[^a_b]}{\eta}}\
{ \frac{\theta[^{a+h_2}_{b+g_2}]}{\eta}}\
 {\frac{\theta[^{a+h_1}_{b+g_1}]}{\eta}}\
{\frac{\theta[^{a-h_1-h_2}_{b-g_1-g_2}]}{\eta}} \nonumber\\
&  & \times  \frac{1}{2}\sum_{\gamma,\delta}
 { \frac{\Gamma_{6,6}\left[^{\gamma,h_1,h_2}_{\delta,g_1,g_2}\right]}{
{\eta^6 {\bar\eta}^6}}} \times {\frac{Z_{\eta}\left[^{\gamma,h_1,h_2}_{\delta,g_1,g_2}\right]}{
{\bar \eta}^7}} \nonumber\\
& & \times { \frac{Z_{18}\left[^{\gamma}_{\delta} \right]}{\bar \eta^{9}}}\ e^{i\pi\varphi_L},\\
\Gamma_{6,6} \left[^{\gamma,h_1,h_2}_{\delta,g_1,g_2}\right]&=&
\Big|\theta[^{\gamma}_{\delta}]\theta[^{\gamma+h_2}_{\delta+g_2}] \Big|^2
\Big|\theta[^{\gamma}_{\delta}]\theta[^{\gamma+h_1}_{\delta+g_1}]\Big|^2\
\Big|\theta[^{\gamma}_{\delta}]\theta[^{\gamma-h_1-h_2}_{\delta-g_1-g_2}]
\Big|^2
\label{eq:selfdual},\\
Z_{\eta}\left[^{\gamma,h_1,h_2}_{\delta,g_1,g_2}\right] &=& \bar{\theta}[^{\gamma+h_2}_{\delta+g_2}]
\bar{\theta}[^{\gamma+h_1}_{\delta+h_2}]\
\bar{\theta}[^{\gamma-h_1-h_2}_{\delta-g_1-g_2}]^5,\\
Z_{18}\left[^{\gamma}_{\delta} \right] &=& \bar{\theta}[^{\gamma}_{\delta}]^{9}\label{eq:gaugesectors3},
\end{eqnarray}
\end{subequations}
In equation \eqref{eq:selfdual} the internal manifold is twisted and
thereby separated explicitly into three planes.
The above model is the minimal  $\mathbbm{Z}_2 \times \mathbbm{Z}_2$ with  $N=1$ super symmetry
and massless spinorial representations in the same $SO(10)$ group coming
from the first and/or from the second plane. The number of families
depends on the choice of the phase $\varphi_L$.
The freedom of this phase arises from the different possible choices
of the modular invariant GSO coefficients  $c[^{v_i}_{v_j}]$.
The maximal number of the families for this model is $32$.
Introducing internal  shifts, associated in part to  $\varphi_L$,
can reduce this number as we will discuss below.

We could have chosen the boundary conditions for different right-moving
fermions. This would lead to spinorial representations on each plane, but the
group to which they would belong would differ in each plane. As we require
spinors in the same group we have discarded this option. Choosing an overlap
with more than $6$ complex fermions in the right-moving sector between the
vectors $b_1$ and $b_2$ leads to a $SO(14)$ gauge group and is discussed in detail in section \ref{sec:sv2}.

In order to have spinors in the spectrum on \emph{all} three planes we need to
separate at least an $SO(16)$ (or $\mathbbm{E}_8$) from the $\Gamma_{6,22}$ lattice . We
therefore introduce the additional vector
\begin{equation}\label{eq:z}
z = \{ \bar{\phi}^{1,\ldots,8} \}
\end{equation}
to the set. It is easy to see that this vector is consistent with the rules written down in section \ref{sec:foundation}. With this vector the partition function for the gauge sector
\eqref{eq:gaugesectors3} modifies to
\begin{equation}\label{eq:gamma013}
Z_{18}\left[^{\gamma}_{\delta} \right] =
 \frac{1}{2} \sum_{h_z,g_z} \bar{\theta}[^{\gamma}_{\delta}]^{}
\bar{\theta}[^{\gamma+h_z}_{\delta+g_z}]^{8}.
\end{equation}
We can further  separate out the internal  $\Gamma_{6,6}$ lattice by
introducing the additional vector
\begin{equation}\label{eq:internalseparation}
e = \{ y_{1,\ldots,6}, \omega_{1,\ldots,6}
| \bar{y}_{1,\ldots,6}, \bar{\omega}_{1,\ldots,6} \} ,
\end{equation}
which modifies the  $\Gamma_{6,6}$ lattice in \eqref{eq:selfdual} to
\begin{multline}\label{eq:gamma6,6}
\Gamma_{6,6} \left[^{\gamma,h_1,h_2}_{\delta,g_1,g_2}\right]=
 \frac{1}{2} \sum_{h_e,g_e}
\Big|\theta[^{\gamma+h_e}_{\delta+g_e}]
\theta[^{\gamma+h_e+h_2}_{\delta+g_e+g_2}] \Big|^2
\Big|\theta[^{\gamma+h_e}_{\delta+g_e}]
\theta[^{\gamma+h_e+h_1}_{\delta+g_e+g_1}]\Big|^2\\
\Big|\theta[^{\gamma+h_e}_{\delta+g_e}]
\theta[^{\gamma+h_e-h_1-h_2}_{\delta+g_e-g_1-g_2}]\Big|^2.
\end{multline}
In the above $\{1,S,b_1,b_2,e, z\}$ construction the gauge group of the
observable sector becomes either $SO(10) \times U(1)^3$
or $\mathbbm{E}_6 \times U(1)^2$ and the hidden sector necessarily is $SO(16)$ or $\mathbbm{E}_8$
depending on the generalised GSO coefficients, (the choice of the phase
$\varphi_L$), while the gauge group from the $\Gamma_{6,6}$ lattice becomes
$G_L=SO(6)\times U(1)^3$.

So far the construction of the $N=1$ models is generic. The only
requirement we are imposing is the presence of
\emph{spinors on all three planes}. We denote this subclass of models 
the $S^3$ models.
In a $N=1$ model the spinors could be replaced by vectorial
representations of the observable gauge group. This replacement gives rise
to three additional classes of models which we denote by  $S^2V$, $SV^2$
and $V^3$. 
The condition of spinorial representations arising from
each one of the $\mathbbm{Z}_2 \times \mathbbm{Z}_2$ orbifold planes together with the complete
separation of the internal manifold is synonymous to having a well defined
hidden gauge group. In subsequent sections we discuss these four classes of models. We define two subclasses of the $S^2V$
models. 
When we separate one $SO(8)$ we
find that spinorial representations of a second $SO(10)$ are present in the
spectrum. We also find in this subclass of models massless states
on the third plane. Part of the massless states form vectorial representations
of the observable $SO(10)$.
Note that when we separate a second $SO(8)$ from the hidden sector we end up in
the $S^3$ subclass of $\mathbbm{Z}_2 \times \mathbbm{Z}_2$ heterotic orbifold models as is
discussed in \cite{Faraggi:2004rq, Faraggi:2003yd}. We refer to the class where no $SO(8)$ gauge group is separated from the hidden sector as the simple $S^2V$ models. The models that have a $SO(8)$ gauge group separated are referred to as the extended $S^2V$ models.

\section{The $S^3$ Orbifold Models}\label{sec:n1general}

In the class of $\mathbbm{Z}_2 \times \mathbbm{Z}_2$ orbifold models, the internal manifold is
broken into three planes. The hidden gauge group is necessarily $\mathbbm{E}_8$ or
$SO(16)$  broken to any subgroup by Wilson lines (at the $N=4$ level).
In order to  classify all possible $S^3$ models, it is necessary to
consider all possible basis vectors consistent with the rules from section \ref{sec:foundation}.
Namely:
\begin{eqnarray}
z_1 &=& \{ \bar{\phi}^{1,\ldots,4} \}  \label{eq:z1},\\
z_2 &=& \{ \bar{\phi}^{5,\ldots,8} \}  \label{eq:z2},\\
e_i &=& \{ y_i, \omega_i | \bar{y}_i, \bar{\omega}_i \} ,\
\ i \in \{1,2,3,4,5,6\}.
\end{eqnarray}
The $z_1,z_2$ vectors allow for a breaking of hidden $\mathbbm{E}_8$ or $SO(16)$ to
$SO(8) \times SO(8)$ depending on the modular  coefficients.
As we discuss below this splitting of the hidden
gauge group has important consequences in the
classification of the $S^3$ class of models by the number of generations.
The introduction of $e_i$ vectors is  necessary in order to obtain all possible
internal shifts which also induces all possible  modification to the number
of generations.


The partition function for the $N=1$, $S^3$ model based on
$\{1,\-S,\-e_i,\-z_1,\-z_2,\-b_1,\-b_2\}$ is
\begin{subequations}
\begin{eqnarray}
Z_{N=1} &=&
\frac{1}{\tau_2 |\eta |^4 } \frac{1}{2}
\sum_{\alpha,\beta} e^{i\pi(a+b+\mu ab)}\nonumber\\
&  & \frac{1}{4}\sum_{h_1,h_2,g_1,g_2}\  {\frac{\theta[^a_b]}{\eta}}\
{\frac{\theta[^{a+h_1}_{b+g_1}]}{\eta}}\
{\frac{\theta[^{a+h_2}_{b+g_2}]}{\eta}}\
{\frac{\theta[^{a-h_1-h_2}_{b-g_1-g_2}]}{\eta}} \nonumber\\
&  & \times 
\sum_{p_i,q_i}
\frac{\Gamma_{2,2} \left[^{h_1|p_1,p_2}_{g_1|q_1,q_2} \right]}{\eta^2 {\bar\eta}^2}
\frac{\Gamma_{2,2} \left[^{h_2|p_3,p_4}_{g_2|q_3,q_4} \right]}{\eta^2 {\bar\eta}^2}
\frac{\Gamma_{2,2} \left[^{-h_1-h_2|p_5,p_6}_{-g_1-g_2|q_5,q_6} \right]}{\eta^2 {\bar\eta}^2}
\nonumber\\
&  & \times \frac{1}{8}\sum_{\gamma,\gamma',\xi,\delta,\delta',\zeta}
\frac{Z_{\eta}\left[^{\gamma,h_1,h_2}_{\delta,g_1,g_2} \right]}{{\bar \eta}^7}
\times \frac{Z_{2}\left[^{\gamma}_{\delta} \right]}{{\bar\eta}^{}}\nonumber\\
&  & \times \frac{Z_{16}\left[^{\gamma',\xi}_{\delta', \zeta} \right]}{{\bar\eta}^{8}} \ e^{i\pi\varphi_L}\label{eq:N1partition},\\
Z_{\eta}\left[^{\gamma,h_1,h_2}_{\delta,g_1,g_2}\right] &=&
\bar{\theta}[^{\gamma+h_2}_{\delta+g_2}]
\bar{\theta}[^{\gamma+h_1}_{\delta+h_2}]\
\bar{\theta}[^{\gamma-h_1-h_2}_{\delta-g_1-g_2}]^5,\\
Z_{2}\left[^{\gamma}_{\delta} \right] &=&
\bar{\theta}[^{\gamma}_{\delta}]^{},\\
Z_{16}\left[^{\gamma',\xi}_{\delta',\zeta} \right] &=&
\bar{\theta}[^{\gamma'}_{\delta'}]^{4}
\bar{\theta}[^{\gamma'+\xi}_{\delta'+\zeta}]^{4}.
\end{eqnarray}
\end{subequations}
The $\Gamma_{6,6}$ lattice of $N=4$ is twisted by $h_i,g_i$, thus in the
$N=1$ case separated into three (2,2) planes. The contribution of each
of these planes in $N=1$ partition function is written in terms of twists
$h_i,g_i$ and shifts $p_i, q_i$ on the $\Gamma_{2,2}$ lattice. The
expressions of those lattices at the maximal symmetry point is \cite{Faraggi:2004rq} like equations \eqref{eq:orbpoint} and \eqref{eq:fermpoint}
\begin{equation}\label{eq:fermpoints3}
\Gamma_{2,2} \left[^{h|p_i,p_j}_{g|q_i,q_j} \right]|_{f.p}=
\frac{1}{4}\sum_{a_i,b_i,a_j,b_j} e^{i\pi\phi_1 +i\pi\phi_2}
\left|\theta[^{a_i}_{b_i}]\theta[^{a_i+h}_{b_i+g}]
\theta[^{a_j}_{b_j}]\theta[^{a_j+h}_{b_j+g}] \right|,
\end{equation}
where the phases
\begin{align}
\phi_i&=a_iq_i+b_ip_i+q_ip_i,& \phi_j&=a_jq_j+b_jp_j+q_jp_j,
\end{align}
define the two shifts of the $\Gamma_{2,2}$ lattice. At the generic point
of the moduli space the  shifted $\Gamma_{2,2}$ lattice depends on the moduli
$(T,U)$, keeping however identical modular transformation properties as
those of the fermionic point.

For non-zero twist, $(h,g)\ne (0,0)$,
$\Gamma_{2,2}$ is independent of the moduli $T,U$ and thus it is
identical to that of \eqref{eq:fermpoints3} constructed at the fermionic
point\cite{Gregori:1999ny,Gregori:1999ns}.
Thus for non-- zero twist, $(h,g)\ne (0,0)$,
\begin{equation}\label{eq:twisted}
\Gamma_{2,2} \left[^{h|p_i,p_j}_{g|q_i,q_j} \right]_{(T,U)}|_{(h,g)\ne (0,0)}=
\Gamma_{2,2} \left[^{h|p_i,p_j}_{g|q_i,q_j} \right]|_{f.p}.
\end{equation}

For zero twist, $(h,g)=(0,0)$, the momentum and winding modes are moduli
dependent and are shifted  by $q_i,q_j$ and $p_i, p_j$, we find from equation \eqref{eq:shiftmomentum}
\begin{multline}
\Gamma_{2,2} \left[^{0|p_i,p_j}_{0|q_i,q_j} \right]_{(T,U)} =
\sum_{\overrightarrow{m},\overrightarrow{n} \in Z} e^{i\pi \{ m_1q_i + m_2
 q_j \} } \\
 \exp \Bigg\{ 2\pi i \bar{\tau} \left[ m_1 \left( n_1+\frac{p_i}{2}
\right) + m_2 \left( n_2+\frac{p_j}{2}\right) \right] \\
 -\frac{\pi \tau_2}{T_2 U_2} \left| m_1 U - m_2 + T(n_1+\frac{p_i}{2})
+ TU(n_2+\frac{p_j}{2}) \right|^2 \Bigg\}.
\end{multline}
The phase  $\varphi_L$ is determined by the chirality of the super symmetry
as well as by the other modular coefficients similar to equation \eqref{eq:susychirality}
\begin{equation}
\varphi_L(a,b) =  \frac{1}{2}\sum_{i,j}\left( 1-c[^{v_i}_{v_j}] \right) \alpha_i\beta_j,
\end{equation}
where $\alpha_i$ and $\beta_j$ are the upper-- and lower--  arguments in
$\theta$--functions corresponding to the  boundary conditions in the two
directions of the world sheet torus and which are associated to the basis
vectors $v_i$ and $v_j$ of the fermionic construction. The only freedom
which remains in the general $S^3$ $N=1$ model is therefore
the choice of the  generalised GSO projection coefficients
$c[^{v_i}_{v_j}]=\pm1$. The space of models is  classified according to that
choice  which determines at the end the phase $\varphi_L$.
We have in total $55$  independent choices for $c[^{v_i}_{v_j}]$ that can
take the values $\pm1$.
Thus, the total number of  models in this restricted class of $N=1$ models
is $2^{55}$. We classify all these  models according to the
values of the GSO coefficients.

The \nahe\ model is an example of a model of the general $S^3$, $N=1$
deformed fermionic $N=1$ model. More precisely we can write the \nahe\ set
basis vectors as a linear combination of
basis vectors $\{1,\-S,\-e_i,\-z_1,\-z_2,\-b_1,\-b_2\}$ which define the general $S^3$
$N=1$ model as mentioned in section \ref{sec:n1}
\begin{eqnarray}
b_1^\text{\nahe } &=& S + b_1,\\
b_2^\text{\nahe } &=& S + b_2 + e_5 + e_6, \\
b_3^\text{\nahe } &=& 1 + b_1 + b_2 + e_5 + e_6 + z_1 + z_2.
\end{eqnarray}
We see that the \nahe\ set is included in these models as mentioned in section
\ref{sec:n1}.


\section{The Simple $S^2V$ Orbifold Models}\label{sec:orbis2v1}

In this section we present the $N=1$ $\mathbbm{Z}_2 \times \mathbbm{Z}_2$ heterotic
string orbifold models that exhibit spinorial representations on
the first two planes. We write down the partition functions of the simple $S^2V$ models.

Without separating a $SO(8)$ from the $N=1$ partition function we
introduce the free shifts on the internal manifold. In the fermionic language
this is accomplished by means of the vectors $e_i$
\begin{equation}\label{eq:shifts}
e_i = \{ y^i, \omega^i | \bar{y}^i, \bar{\omega}^i \} ,\ \ i \in \{1,
2,3,4,5,6\}.
\end{equation}
The partition function for this subclass of models generated using the basis
set $\{1, S, b_1, b_2, e_i\}$ is
\begin{subequations}\label{eq:partitions2v}
\begin{eqnarray}
Z_{N=1} &=&
\frac{1}{\tau_2 |\eta |^4 } \frac{1}{2}
\sum_{a,b} e^{i\pi(a+b+\mu ab)}\nonumber\\
&  & \frac{1}{4}\sum_{h_1,h_2,g_1,g_2}\  \frac{\theta[^a_b]}{\eta}\
\frac{\theta[^{a+h_2}_{b+g_2}]}{\eta}\
\frac{\theta[^{a+h_1}_{b+g_1}]}{\eta}\
\frac{\theta[^{a-h_1-h_2}_{b-g_1-g_2}]}{\eta} \nonumber\\
&  & \times  
\sum_{p_i,q_i}
\frac{\Gamma_{2,2} \left[^{h_1|p_1,p_2}_{g_1|q_1,q_2} \right]}{
\eta^2 {\bar\eta}^2}
\frac{\Gamma_{2,2} \left[^{h_2|p_3,p_4}_{g_2|q_3,q_4} \right]}{
\eta^2 {\bar\eta}^2}
\frac{\Gamma_{2,2} \left[^{-h_1-h_2|p_5,p_6}_{-g_1-g_2|q_5,q_6} \right]}{
\eta^2 {\bar\eta}^2}
\nonumber\\
&  & \times  \frac{1}{2}\sum_{\gamma,\delta}
\frac{Z_{\eta}\left[^{\gamma,h_1,h_2}_{\delta,g_1,g_2}\right]}{
{\bar \eta}^7} \times
\frac{Z_{18}\left[^{\gamma}_{\delta} \right]}{{\bar\eta}^{9}}\
e^{i\pi\varphi_L}, \\
Z_{\eta}\left[^{\gamma,h_1,h_2}_{\delta,g_1,g_2}\right] &=&
\bar{\theta}[^{\gamma+h_2}_{\delta+g_2}]\bar{\theta}[^{\gamma+h_1}_{\delta+h_2}]\ \bar{\theta}[^{\gamma-h_1-h_2}_{\delta-g_1-g_2}]^5,\\
Z_{18}\left[^{\gamma}_{\delta} \right] &=& \bar{\theta}[^{\gamma}_{\delta}]^{9}.\label{eq:gaugesectors2v}
\end{eqnarray}
\end{subequations}
The $\Gamma_{6,6}$ lattice of $N=4$ is twisted by $h_i,g_i$, thus in the
$N=1$ case separated into three (2,2) planes. The contribution of each
of these planes in the $N=1$ partition function is written in terms of twists
$h_i,g_i$ and shifts $p_i, q_i$ on the $\Gamma_{2,2}$ lattice. The
expressions of those lattices at the self dual point (free fermionic
point) is similar to equation \eqref{eq:fermpoints3}
\begin{equation}\label{eq:fermpoints2v}
\Gamma_{2,2} \left[^{h|p_i,p_j}_{g|q_i,q_j} \right]|_{f.p}=
\frac{1}{4}\sum_{a_i,b_i,a_j,b_j} e^{i\pi\phi_1 +i\pi\phi_2}
\left|\theta[^{a_i}_{b_i}]\theta[^{a_i+h}_{b_i+g}]
\theta[^{a_j}_{b_j}]\theta[^{a_j+h}_{b_j+g}] \right|,
\end{equation}
where the phases
\begin{align}
\phi_i& =a_iq_i+b_ip_i+q_ip_i,& \phi_j& =a_jq_j+b_jp_j+q_jp_j,
\end{align}
define the two shifts of the $\Gamma_{2,2}$ lattice. At the generic point
of the moduli space the  shifted $\Gamma_{2,2}$ lattice depends on the moduli
$(T,U)$, keeping however identical modular transformation properties as
those at the fermionic point.

\section{The Extended $S^2V$ Orbifold Models}

In this section we present the $N=1$ $\mathbbm{Z}_2 \times \mathbbm{Z}_2$ heterotic
string orbifold models that exhibit spinorial representations on
the first two planes with a broken hidden gauge group. We write down the partition functions of the extended $S^2V$ models.

Separating a $SO(8)$ from the hidden gauge group and introducing the free
shifts on the internal manifold is realised in the fermionic language by the
vectors
\begin{eqnarray}
e_i &=& \{ y^i, \omega^i | \bar{y}^i, \bar{\omega}^i \} ,\ \ i \in \{1,
2,3,4,5,6\},\label{eq:shifts2}\\
z_2 &=& \{ \bar{\phi}^{5,\ldots,8} \}. \label{eq:so82}
\end{eqnarray}
The partition function for this subclass of models generated by the basis set
$\{1,S, b_1, b_2, e_i, z_2\}$ is similar to equations \eqref{eq:partitions2v}. The difference lies in equation \eqref{eq:gaugesectors2v}
and reads for this subclass of models
\begin{equation}
Z_{18}\left[^{\gamma, \xi}_{\delta, \zeta}\right] = \frac{1}{2} \sum_{\xi,
\zeta} \bar{\theta}[^{\gamma}_{\delta}]^{5} \bar{\theta}[^{\xi}_{\zeta}]^{4}.
\end{equation}
We note that in this subclass of the $\mathbbm{Z}_2 \times \mathbbm{Z}_2$ orbifolds the hidden
$SO(18)$ identified as $Z_{18} = Z_{10} Z_{8}$ is broken to a $SO(10) \times
SO(8)$ by means of the $z_2$ vector. In this subclass of models we find that a
priori spinorial representations of this second $SO(10)$ can appear in the
massless spectrum. We will expand on this issue further in section \ref{sec:s2v2matter}.


\chapter{The Classification in the Fermionic Formulation}\label{chap:fermform}

In this chapter we use the direct translation of the geometrical description of the heterotic super string in a free fermionic description to classify the four subclasses of the $\mathbbm{Z}_2 \times \mathbbm{Z}_2$ heterotic string models. We use the methods and techniques developed in chapter \ref{chap:realistic}. We discuss each of the subclasses $S^3$, $S^2V$, $SV^2$ and $V^3$ of models in detail. We give their basis set for the free fermionic formulation. The gauge group and the observable matter spectrum is derived. This formulations allows for the classification of the chiral $\mathbbm{Z}_2 \times \mathbbm{Z}_2$ heterotic string models.
This chapter is based on parts of \cite{Faraggi:2003yd, Faraggi:2004rq, Faraggi:2004xnew}.

\section{The $S^3$ Free Fermionic Models}\label{sec:ferms3}

A model in the free fermionic formulation of the heterotic super string
is determined by a set of
basis vectors, associated with the phases picked up by the fermions
when parallelly transported along
non-trivial loops and a set of coefficients associated with GSO
projections as described in section \ref{sec:construction}.
The free fermions in the light-cone gauge in the traditional notation
are:
$\psi^\mu, \chi^i,y^i, \omega^i, i=1,\dots,6$ (left movers)
and  $\bar{y}^i,\bar{\omega}^i, i=1,\dots,6$,
$\psi^A, A=1,\dots,5$, $\bar{\phi}^\alpha, \alpha=1,8$ (right movers).
The chiral $\mathbbm{Z}_2 \times \mathbbm{Z}_2$ fermionic models under consideration with spinorial representations on each plane are generated by a set of
12 basis vectors
\begin{equation}
B=\{v_1,v_2,\dots,v_{12}\},
\end{equation}
where
\begin{eqnarray}
v_1=1&=&\{\psi^\mu,\
\chi^{1,\dots,6},y^{1,\dots,6},\omega^{1,\dots,6}|\bar{y}^{1,\dots,6},
\bar{\omega}^{1,\dots,6},\bar{\eta}^{1,2,3},
\bar{\psi}^{1,\dots,5},\bar{\phi}^{1,\dots,8}\},\nn\\
v_2=S&=&\{\psi^\mu,\chi^{1,\dots,6}\},\nn\\
v_{2+i}=e_i&=&\{y^{i},\omega^{i}|\bar{y}^i,\bar{\omega}^i\}, \ i=1,\dots,6,\nn\\
v_{9}=b_1&=&\{\chi^{34},\chi^{56},y^{34},y^{56}|\bar{y}^{34},\bar{y}^{56},
\bar{\eta}^1,\bar{\psi}^{1,\dots,5}\},\label{basis}\\
v_{10}=b_2&=&\{\chi^{12},\chi^{56},y^{12},y^{56}|\bar{y}^{12},\bar{y}^{56},
\bar{\eta}^2,\bar{\psi}^{1,\dots,5}\},\nn\\
v_{11}=z_1&=&\{\bar{\phi}^{1,\dots,4}\},\nn\\
v_{12}=z_2&=&\{\bar{\phi}^{5,\dots,8}\}.\nn
\end{eqnarray}
The vectors $1,S$ generate a $N=4$ super symmetric model as explained in section \ref{sec:susybackground}. The vectors
$e_i,i=1,\dots,6$ give rise to
all possible symmetric shifts of internal fermions
($y^i,\omega^i,\bar{y}^i,\bar{\omega}^i$) while $b_1$ and $b_2$
stand for  the $\mathbbm{Z}_2\times \mathbbm{Z}_2$ orbifold twists. The remaining fermions
not affected by the action
of the previous vectors are $\bar{\phi}^i,i=1,\dots,8$ which normally give
rise to the hidden sector gauge group.
 The vectors $z_1,z_2$ divide these eight fermions in two sets of
four which in the $\mathbbm{Z}_2\times \mathbbm{Z}_2$ case is the maximum consistent
partition function \cite{Antoniadis:1987rn, Kawai:1987ah}.
This is the most general basis, with symmetric shifts for the internal
fermions, that is compatible with
Kac--Moody level one $SO(10)$ embedding.

The associated projection coefficients are denoted by
$\cc{v_i}{v_j}, i,j=1,\dots,12$ and can take the values $\pm1$.
They
are related by modular invariance
$\cc{v_i}{v_j}=\exp\{i\frac{\pi}{2}v_i\cdot v_j\} \cc{v_j}{v_i}$
and $\cc{v_i}{v_i}=\exp\{i\frac{\pi}{4}v_i\cdot v_i\} \cc{v_j}{1}$
leaving $2^{66}$ independent coefficients. Out of them, the requirement
of $N=1$ super symmetric spectrum fixes
(up to  a phase convention) all $\cc{S}{v_i},i=1,\dots,12$. Moreover,
without loss of generality we can set $\cc{1}{1}=-1$,
and leave the rest 55 coefficients free. For more details we refer to chapter \ref{chap:fermionic}. Therefore, a simple counting
gives  $2^{55}$ (that is approximately $10^{16.6}$) distinct models in
the class under consideration.
In the following we study this class of models by deriving  analytic
formulas for the gauge group and the
spectrum and then using these formulas for the classification.

\subsection{The gauge group}\label{sec:n4gauges3}

We describe the gauge configuration of the models defined by the basis
vectors $\{1,\-S,\-e_i,\-z_1,\-z_2,\-b_1,\-b_2\}$. For
this purpose we start with a simplification and separate out the internal
manifold using equation \eqref{eq:internalseparation}. As the twisting vectors
$b_1$ and $b_2$ are used to break the $SO(16) \to SO(10) \times U(1)^3$ we will
firstly describe the configuration without these vectors. The gauge group
induced by the vectors $\{1,S, e, z_1, z_2\}$ without enhancements is.
\begin{equation}
G = SO(16) \times SO(8)_1 \times SO(8)_2 \times SO(12),
\end{equation}
where the internal manifold is described by $SO(12)$ and the hidden sector by
$SO(8) \times SO(8)$ and the observable by $SO(16)$. By choosing the GSO
coefficients the SO(16) can enhance either to $\mathbbm{E}_8$ or mix with the other
sectors
producing either $SO(24)$ or $SO(32)$. Similarly the $SO(8) \times SO(8)$ can
enhance either to $SO(16)$ or $\mathbbm{E}_8$ or mix with the observable or internal
manifold gauge group. This leads to enhancements of the form $SO(20)$ or
$SO(24)$.
The exact form depends only on the three GSO coefficients $\cc{e}{z_1}$,
$\cc{e}{z_2}$, $\cc{z_1}{z_2}$. We have shown the results in table
\ref{tab:gauge}.

\begin{table}
\begin{center}
\begin{tabular}{c c c | r}
$\cc{z_1}{z_2}$ & $\cc{e}{z_1}$ & $\cc{e}{z_2}$ & Gauge group G\\\hline
+ & + & + & $\mathbbm{E}_8    \times SO(28) $\\
+ & - & + & $SO(24) \times SO(20) $\\
+ & + & - & $SO(24) \times SO(20) $\\
+ & - & - & $SO(32) \times SO(12) $\\
- & + & + & $SO(16) \times SO(16) \times SO(12)$\\
- & - & + & $SO(16) \times SO(16) \times SO(12)$\\
- & + & - & $SO(16) \times SO(16) \times SO(12)$\\
- & - & - & $\mathbbm{E}_8    \times \mathbbm{E}_8    \times SO(12)$\\
\end{tabular}
\caption{The configuration of the gauge group of the $N=4$ theory.
We have separated a priori the internal and the hidden and observable
gauge group using the vectors $e$ and $z_i$. Introducing the other
vectors $e_i$ and $b_i$ only induce breaking of these groups.}
\label{tab:gauge}
\end{center}
\end{table}

Proceeding to the complete model $\{1,S,e_i,z_1,z_2,b_1,b_2\}$
we break these gauge groups to their subgroups.
Imposing the shifts $e_i$ we can break the internal gauge group down to its
Cartan generators by a suitable choice of the coefficients. By a suitable
choice we can break $SO(20) \to SO(8) \times U(1)^6$.

When we also include the twists we break $SO(16) \to SO(10) \times U(1)^3$ and
$\mathbbm{E}_8 \to \mathbbm{E}_6 \times U(1)^2$. Similarly we can break $SO(24) \to SO(10) \times
U(1)^3 \times SO(8)$ and $SO(32) \to SO(10) \times U(1)^3 \times SO(8) \times
SO(8)$. Enhancements can subsequently occur of the form $SO(8) \times U(1)
\subset SO(32) \to SO(10)$ or $SO(8) \times SO(8) \times U(1) \subset SO(32)
\to SO(18)$. We find possible enhancements of the form $SO(10) \times SO(8)
\subset SO(32) \to SO(18)$.

In table \ref{tab:gauge} we notice that the coefficient $\cc{z_1}{z_2}$
distinguishes between the $SO(32)$ models and the $\mathbbm{E}_8 \times \mathbbm{E}_8$ models.
Since we require complete separation of the gauge group into a well defined
observable and
 hidden gauge group, we set the coefficient $\cc{z_1}{z_2}=-1$ in the
classification.

Gauge bosons arise from the
following four sectors at the $N=1$ level.
\begin{equation}
G=\{0,z_1,z_2,z_1+z_2,x\},
\end{equation}
where
\begin{equation}
x=1+S+\sum_{i=1}^6e_i+\sum_{k=1}^2 z_k=\{{\bar{\eta}^{123},\bar{\psi}^{12345}}\}\ .
\end{equation}
The $0$ sector gauge bosons give rise to the gauge group at the $N=1$ level
\begin{equation}
SO(10)\times {U(1)}^3 \times SO(8) \times SO(8).
\end{equation}
The $x$ gauge bosons when present lead to enhancements of the
traditionally called observable sector (the sector that includes
$SO(10)$) while
the $z_1+z_2$ sector can enhance the hidden sector ($SO(8)\times SO(8)$).
However, the $z_1, z_2$ sectors accept oscillators that can
 also give rise to mixed type gauge bosons and completely reorganise
the gauge group.
The appearance of mixed states is in general controlled by the
phase $\cc{z_1}{z_2}$. The choice $\cc{z_1}{z_2}=+1$ allows for
mixed gauge bosons
and leads to the gauge groups presented in Table \ref{gga}.
\begin{table}
\begin{center}
\begin{sideways}
\begin{tabular}{|c|c|c|c|c|c|c|c|c|}
\hline
$\cc{z_1}{z_2}$&$\cc{b_1}{z_1}$&$\cc{b_2}{z_1}$&$\cc{b_1}{z_2}$&$\cc{b_2}{z_2}$&$\cc{e_1}{z_1}$&$\cc{e_2}{z_2}$
&$\cc{e_1}{e_2}$&Gauge group\\
\hline
$+$&$+$&$+$&$+$&$+$&$+$&$+$&$+$&$SO(10)\times{SO(18)}\times{U(1)^2}$\\
\hline
$+$&$+$&$+$&$+$&$+$&$-$&$-$&$+$&$SO(10)\times{SO(9)}^2\times{U(1)^3}$\\
\hline
$+$&$+$&$+$&$+$&$+$&$-$&$+$&$+$&$SO(10)^2\times{SO(9)}\times{U(1)^2}$\\
\hline
$+$&$+$&$+$&$+$&$-$&$+$&$+$&$+$&$SO(10)^3\times{U(1)}$\\
\hline
$+$&$-$&$-$&$-$&$-$&$+$&$+$&$+$&$SO(26)\times{U(1)}^3$\\
\hline
$-$&$+$&$+$&$+$&$+$&$+$&$+$&$+$&$E_6\times{U(1)^2}\times{E_8}$\\
\hline
$-$&$-$&$+$&$-$&$+$&$+$&$+$&$+$&$E_6\times{U(1)^2}\times{SO(16)}$\\
\hline
$-$&$-$&$+$&$+$&$-$&$+$&$+$&$+$&$E_6\times{U(1)^2}\times{SO(8)\times{SO(8)}}$\\
\hline
$-$&$+$&$+$&$+$&$+$&$+$&$+$&$-$&$SO(10)\times{U(1)}^3\times{E_8}$\\
\hline
$-$&$+$&$+$&$+$&$+$&$-$&$-$&$-$&$SO(10)\times{U(1)}^3\times{SO(16)}$\\
\hline
\end{tabular}
\end{sideways}
\caption{Typical enhanced gauge groups  and
associated projection coefficients for a generic model generated
by the basis \eqref{basis} (coefficients not included
equal to +1 except those fixed by space-time super symmetry
and conventions).}\label{gga}
\end{center}
\end{table}

The choice  $\cc{z_1}{z_2}=-1$ eliminates all mixed gauge
bosons and there are a few possible enhancements:
${SO(10)}\times{U(1)}\to{\mathbbm{E}_6}$ and/or ${SO(8)}^2\to\{SO(16), \mathbbm{E}_8\}$.
The $x$ sector gauge bosons survive only when the GSO coefficients are such that
\begin{equation}
\cc{x}{z_k} = \cc{x}{e_i} = 1,
\end{equation}
and the states in the sector $x$ remain. We can rewrite this as
\ba
&&\sum_{j=1,i\ne j}^6 \oo{e_i}{e_j}+\sum_{k=1}^2\oo{e_i}{z_k}=0\ \mod 2  \ , \ i=1,\dots,6,\label{gb1}\\
&&
\sum_{j=1}^6\oo{e_j}{z_k}=0\ \mod 2\ , \ k=1,2,\label{gb2}
\ea
where we have introduced the notation
\ba
\cc{v_i}{v_j}=e^{i \pi \oo{v_i}{v_j}}\,\ ,\  \oo{v_i}{v_j}=0,1,
\ea
and one of the constraints in \eqref{gb1} and \eqref{gb2}  can be dropped because of linear independence with the rest.

As far as the ${SO(8) \times SO(8)}$ is concerned, we have using a similar line of thought as before, the following
possibilities
\begin{subequations}\label{gs3}
\begin{eqnarray}
\text{(i)} & & \oo{e_i}{z_1}=\oo{b_a}{z_1}=0\  \forall\  i=1,\dots,6,\ a=1,2,\\
\text{(ii)} & & \oo{e_i}{z_2}=\oo{b_a}{z_2}=0\  \forall\  i=1,\dots,6,\  a=1,2,\\
\text{(iii)} & & \oo{e_i}{z_1+z_2}=\oo{b_a}{z_1+z_2}=0\  \forall\
i=1,\dots,6,\  a=1,2.\label{ge}
\end{eqnarray}
\end{subequations}
Depending on which of the above equations are true the enhancement
is
\begin{subequations}
\begin{eqnarray}
&&\mbox{both } \text{(i)} \mbox{ and } \text{(ii)}  \Longrightarrow {\mathbbm{E}_8},\\
&&\mbox{one of } \text{(i)} \mbox{ or } \text{(ii)} \mbox{ or } \text{(iii)} \Longrightarrow {SO(16)},\\
&&\mbox{none of } \text{(i)} \mbox{ or } \text{(ii)} \mbox{ or }
\text{(iii)} \Longrightarrow {SO(8)}\times{SO(8)}.
\end{eqnarray}
\end{subequations}
In the following we will restrict to the case $\cc{z_1}{z_2}=-1$ as
this is the more promising phenomenologically.

\subsection{Observable matter spectrum}\label{sec:s3matter}

The untwisted sector matter is common to all models
and consists of six vectorials of $SO(10)$ and  12 non-Abelian gauge
group singlets.
In models where the gauge group  enhances to $\mathbbm{E}_6$ extra matter
comes from the $x$ sector
giving rise to six  $\mathbbm{E}_6$ fundamental reps ($\bf 27$).

Chiral twisted matter arises from several sectors. In section \ref{sec:nahe} we saw that twisted matter comes from the sector $S+b_1$. From equation \eqref{eq:twistedmatterb1} we find that the states in the sector $S+b_1$ are
\begin{equation}\label{eq:b1}
\ket{S}_{S + b_1} = \ket{\psi_\mu^{1,2}}_{\pm}\  \ket{\chi^{1,2}}_\pm \prod_{l=3}^6 \ket{\sigma^l_y}_\pm\  \ket{\bar{\eta}^1}_\pm \prod_{m=1}^5 \ket{\bar{\psi}^m}_\pm\ket{0},
\end{equation}
where we have used the following pairing
\begin{subequations}
\begin{eqnarray}
\psi_\mu^{1,2} &=& \frac{1}{\sqrt{2}} \left( \psi_\mu^1 + i\psi_\mu^2 \right),\\
\chi^{2k-1,2k} &=& \frac{1}{\sqrt{2}} \left( \chi^{2k-1} + i\chi^{2k} \right),\\
\sigma^l_y &=& y^l \bar{y}^l, 
\end{eqnarray}
\end{subequations}
where the other fermions are already complex by construction. 

The vector $S+b_1 + e_3$ is
\begin{equation}
S + b_1 + e_3 = \{\psi_\mu^{1,2},\chi^{12},\omega^{3}, y^{4},y^{56}\ |\ \bar{\omega}^{3},\bar{y}^{4},\bar{y}^{56}, \bar{\eta}^1,\bar{\psi}^{1,\dots,5}\}.
\end{equation}
The states in this sector are
\begin{equation}
\ket{S}_{S + b_1 + e_3} = \ket{\psi_\mu^{1,2}}_{\pm}\  \ket{\chi^{1,2}}_\pm \ket{\sigma^3_\omega}_\pm\ \prod_{l=4}^6 \ket{\sigma^l_y}_\pm\  \ket{\bar{\eta}^1}_\pm \prod_{m=1}^5 \ket{\bar{\psi}^m}_\pm\ket{0},
\end{equation}
where we have used the following pairing
\begin{subequations}
\begin{eqnarray}
\psi_\mu^{1,2} &=& \frac{1}{\sqrt{2}} \left( \psi_\mu^1 + i\psi_\mu^2 \right),\\
\chi^{2k-1,2k} &=& \frac{1}{\sqrt{2}} \left( \chi^{2k-1} + i\chi^{2k} \right),\\
\sigma^l_y &=& y^l \bar{y}^l,\\
\sigma^m_\omega &=& \omega^m \bar{\omega}^m, 
\end{eqnarray}
\end{subequations}
where the other fermions are already complex by construction. It is then clear that by adding the basis vector $e_i$, where $i \in \{3,\ldots,6\}$, to the vector $b_1$ gives rise to new massless chiral twisted matter.

We therefore find that the chiral twisted matter arises from the following sectors
\begin{subequations}\label{ss}
\begin{eqnarray}
B_{pqrs}^{(1)}&=&S+b_1+p\,e_3+q\,e_4 +
r\,e_5 + s\,e_6+(x),\label{eq:ss1}\\
B_{pqrs}^{(2)}&=&S+b_2+p\,e_1+q\,e_2 +
r\,e_5 + s\,e_6+(x) ,\\
B_{pqrs}^{(3)}&=& S + b_3+ p\,e_1+q\, e_2 +r\,e_3+ s\,e_4+(x),
\end{eqnarray}
\end{subequations}
where $b_3=b_1+b_2+x$. These are 48 sectors (16 sectors per orbifold
plane) and we choose to label them
using  the plane number $i$ (upper index) and  the integers
$p, q, r, s=\{0,1\}$ (lower index) corresponding to the
coefficients of the appropriate shift vectors. Note that for a particular
orbifold plane $i$ only
four shift vectors can be added to the twist vector $b_i$ (the ones that
have non empty intersection)
the other two give rise to massive states. Each of the above sectors
(\ref{ss}) can produce
a single spinorial of $SO(10)$ (or fundamental of $\mathbbm{E}_6$ in the case of
enhancement). Since the $\mathbbm{E}_6$ model
spectrum is in one to one correspondence with the $SO(10)$ spectrum in
the following we use the
name spinorial meaning the $\bf 16$ of $SO(10)$ and in the case of
enhancement the $\bf27$ of $\mathbbm{E}_6$.

One of the  advantages of our formulation is that it allows to extract
generic  formulas regarding the
number and the chirality of each spinorial. This is important because
it allows an algebraic
treatment of the entire class of models without deriving each model
explicitly. 

We now concentrate on one sector. We take for convenience the 
sector $S+b_1$ thereby setting $p,q,r,s=0$ in equation \eqref{eq:ss1}. The $S+b_1$ sector is described in 
equation \eqref{eq:b1}. The projection induced by $\mathbbm{1}$ sets the overall sign to be equal to $+1$ as 
explained in equation \eqref{eq:1gsob1}. The GSO constraints of the $e_i$ 
vectors are
\begin{equation}\label{eq:GSOb1ei}
GSO_{e_i} \equiv \delta_{S+b_1}\cc{S+b_1}{e_i} = -\cc{S+b_1}{e_i} = -\cc{e_i}{S+b_1} .
\end{equation}
If we set the coefficients for the $e_1$ or $e_2$ to $-1$ we project out the 
sector $S+b_1$. We have therefore obtained a constraint for the sector to 
survive. The number of families from this sector depends on the projector
\begin{equation}\label{eq:p1}
P^{(1)}_{0000} \propto \frac{1 - \cc{e_1}{S + b_1}}{2}\frac{1  -\cc{e_2}{S + b_1}}{2}.
\end{equation}
Imposing the $b_1$ projection reveals no results of interest for the number of 
generations. 
The GSO projection induced by the $b_1$ vector is
\begin{equation}
GSO_{b_1} = - \cc{S+b_1}{b_1} =  - \cc{b_1}{S + b_1} = -1.
\end{equation}
This result induces a constraint not relevant for this calculation. The GSO 
projection induced by the $b_2$ vector is
\begin{equation}\label{eq:GSOb1b2}
GSO_{b_2} = - \cc{S+b_1}{b_2} = - \cc{b_2}{S+b_1}.
\end{equation}
This induces a constraint on the chirality of the $\bar{\psi}^{1,\ldots,5}$
\begin{equation}
\chir(\chi^{12})\chir(y^5\bar{y}^5)\chir(y^6\bar{y}^6)\chir(\bar{\psi}^{1,
\ldots,5}) = - \cc{b_2}{S+b_1}.
\end{equation}
The chirality of the internal fermions is set by the coefficients $\cc{e_i}{S+b_1}
$. We only consider the states where the spacetime spinor is in the \emph{up} state by convention.
Due to the GSO projection of the $S$ vector we can solve the chirality of the 
$\chi^{1,2}$ state. The chirality of the $\bar{\psi}^{1,\ldots,5}$ is 
\begin{equation}
\chir(\bar{\psi}^{1,\ldots,5}) = \cc{b_2+e_5+e_6}{S + b_1}.
\end{equation}
The GSO projection of the last vector $z$ completes the projector $P^{(1)}_{0000}$
\begin{equation}\label{eq:p2}
P^{(1)}_{0000} \propto \frac{1 - \cc{z_1}{S+b_1}}{2}\frac{1 - \cc{z_2}{S+b_1}}{2}.
\end{equation}
The chirality of the $\bar{\psi}^{1,\ldots,5}$ contributes to the total number 
of generations per plane. In the $S+b_1$ sector the number of generations therefore is
\begin{multline}
\chir(\bar{\psi}^{1,\ldots,5}) = \cc{b_2+e_5 +e_6}{S+b_2} \times 
\frac{1 - \cc{e_1}{S+b_1}}{2}\frac{1 - \cc{e_2}{S+b_1}}{2} \\
\times \frac{1 - \cc{z_1}{S+b_1}}{2}\frac{1 - \cc{z_2}{S+b_1}}{2} \equiv X_{0000}^{(1)}P_{0000}^{(1)},
\end{multline}
where we have identified a chirality operator $X_{pqrs}^{(1)}$ and a projector operator $P_{pqrs}^{(1)}$. Similarly we can derive these formulae for the other planes and sectors as 
well. 

The number of
surviving spinorials per sector (\ref{ss}) is given by
\begin{subequations}\label{ps3}
\begin{eqnarray}
P_{pqrs}^{(1)}&=&
\frac{1}{16}\,\prod_{i=1,2}\left(1-\cc{e_i}{B_{pqrs}^{(1)}}\right)\,
\prod_{m=1,2}\left(1-\cc{z_m}{B_{pqrs}^{(1)}}\right) ,\label{pa}\\
P_{pqrs}^{(2)}&=&
\frac{1}{16}\,\prod_{i=3,4}\left(1-\cc{e_i}{B_{pqrs}^{(2)}}\right)\,
\prod_{m=1,2}\left(1-\cc{z_m}{B_{pqrs}^{(2)}}\right) ,\label{pb}
\\
P_{pqrs}^{(3)}&=&
\frac{1}{16}\,\prod_{i=5,6}\left(1-\cc{e_i}{B_{pqrs}^{(3)}}\right)\,
\prod_{m=1,2}\,\left(1-\cc{z_m}{B_{pqrs}^{(3)}}\right) ,\label{pc}
\end{eqnarray}
\end{subequations}
where $P_{pqrs}^i$ is a projector that takes values $\{0,1\}$.
The chirality of the surviving spinorials is given by
\begin{subequations}\label{cs3}
\begin{equation}
X_{pqrs}^{(1)}=\cc{b_2+(1-r) e_5+(1-s) e_6}{B_{pqrs}^{(1)}},\label{ca}
\end{equation}
\begin{equation}
X_{pqrs}^{(2)}=\cc{b_1+(1-r) e_5+(1-s) e_6}{B_{pqrs}^{(2)}},\label{cb}
\end{equation}
\begin{equation}
X_{pqrs}^{(3)}=\cc{b_1+(1-r) e_3+(1-s) e_4}{B_{pqrs}^{(3)}},\label{cc}
\end{equation}
\end{subequations}
where $X_{pqrs}^i=+$ corresponds to a ${\bf16}$ of $SO(10)$(or ${\bf27}$
in the case of $\mathbbm{E}_6$) and $X_{pqrs}^i=-$ corresponds
to a ${\bf\overline{16}}$ (or ${\bf\overline{27}}$) and we have chosen
the space-time chirality $C(\psi^\mu)=+1$.
The net number of spinorials and thus the net number of families is given by
\begin{equation}
N_F=\sum_{i=1}^3\sum_{p,q,r,s=0}^1X^{(i)}_{pqrs}
P_{pqrs}^{(i)}\label{nf}.
\end{equation}
Similar formulas can be easily derived for the number of vectorials
and the number of singlets and can be extended
to the $U(1)$ charges but in this work we restrict to the
spinorial calculation.

Formulas \eqref{ps3} allow us to identify the mechanism
of spinorial reduction, or in other words
the fixed point reduction, in the fermionic language. For a particular
sector ($B_{pqrs}^{(i)}$) of the orbifold plane $i$
there exist two shift vectors ($e_{2i-1}, e_{2i}$)
and the two $z$ vectors ($z_1,z_2$) that have no common elements
with $B_{pqrs}^{(i)}$. Setting the relative projection coefficients \eqref{ps3}
to $-1$ each of the above four vectors acts as a projector that cuts
the number of fixed points in the
associated sector by a factor of two. Since four such projectors are
available for each sector the number of fixed points
can be reduced from $16$ to $1$ per plane.

The projector action \eqref{ps3} can be expanded and
written in a simpler form
\begin{equation}
\Delta^{(i)} W^{(i)}=Y^{(i)}, \label{proj}
\end{equation}
where
\begin{subequations}
\begin{eqnarray}
\Delta^{(1)}&=&\left[
\begin{array}{cccc}
\oo{e_1}{e_3}&\oo{e_1}{e_4}&\oo{e_1}{e_5}&\oo{e_1}{e_6}\\
\oo{e_2}{e_3}&\oo{e_2}{e_4}&\oo{e_2}{e_5}&\oo{e_2}{e_6}\\
\oo{z_1}{e_3}&\oo{z_1}{e_4}&\oo{z_1}{e_5}&\oo{z_1}{e_6}\\
\oo{z_2}{e_3}&\oo{z_2}{e_4}&\oo{z_2}{e_5}&\oo{z_2}{e_6}
\end{array}
\right],
\nn
\\
\Delta^{(2)}&=&\left[
\begin{array}{cccc}
\oo{e_3}{e_1}&\oo{e_3}{e_2}&\oo{e_3}{e_5}&\oo{e_3}{e_6}\\
\oo{e_4}{e_1}&\oo{e_4}{e_2}&\oo{e_4}{e_5}&\oo{e_4}{e_6}\\
\oo{z_1}{e_1}&\oo{z_1}{e_2}&\oo{z_1}{e_5}&\oo{z_1}{e_6}\\
\oo{z_2}{e_1}&\oo{z_2}{e_2}&\oo{z_2}{e_5}&\oo{z_2}{e_6}
\end{array}
\right],
\\
\Delta^{(3)}&=&\left[
\begin{array}{cccc}
\oo{e_5}{e_1}&\oo{e_5}{e_2}&\oo{e_5}{e_3}&\oo{e_5}{e_4}\\
\oo{e_6}{e_1}&\oo{e_6}{e_2}&\oo{e_6}{e_3}&\oo{e_6}{e_4}\\
\oo{z_1}{e_1}&\oo{z_1}{e_2}&\oo{z_1}{e_3}&\oo{z_1}{e_4}\\
\oo{z_2}{e_1}&\oo{z_2}{e_2}&\oo{z_2}{e_3}&\oo{z_2}{e_4}
\end{array}
\right],
\nn
\end{eqnarray}
and
\begin{align}
Y^{(1)}&=
\left[
\begin{array}{c}
\oo{e_1}{b_1}\\
\oo{e_2}{b_1}\\
\oo{z_1}{b_1}\\
\oo{z_2}{b_1}
\end{array}
\right], &
Y^{(2)}&=
\left[
\begin{array}{c}
\oo{e_3}{b_2}\\
\oo{e_4}{b_2}\\
\oo{z_1}{b_2}\\
\oo{z_2}{b_2}
\end{array}
\right], &
Y^{(3)}=
\left[
\begin{array}{c}
\oo{e_5}{b_3}\\
\oo{e_6}{b_3}\\
\oo{z_1}{b_3}\\
\oo{z_2}{b_3}
\end{array}
\right],
\end{align}
and
\begin{equation} 
W^i= \left[ \begin{array}{c}
p_i\\
q_i\\
r_i\\
s_i
\end{array}\right],
\end{equation}
\end{subequations}
where
\begin{equation}\label{eq:exponent}
\cc{\alpha}{\beta} = e^{i\pi \oo{\alpha}{\beta}}.
\end{equation}
Note that we have fixed $\oo{\alpha}{1} = 1 = \oo{\alpha}{S}$. They form three systems of equations of the form $\Delta^i\,W^i=Y^i$
(one for each orbifold plane).
Each system contains  4 unknowns $p_i, q_i, r_i, s_i$ which correspond
to the
labels of surviving spinorials in the plane $i$. We call the set of
solutions of each system
$\Xi_i$. The net number of families (\ref{nf}) can be written as
\ba
N_F=\sum_{i=1}^3\sum_{p,q,r,s\in\Xi_i}X^{(i)}_{pqrs}.
\ea
The chiralities \eqref{cs3} can be further expanded in the
exponential form defined in equation \eqref{eq:exponent}
\begin{subequations}\label{chis3}
\ba
\chi_{pqrs}^{(1)} &=& 1+\oo{b_1}{b_2} + (1-r)\oo{e_5}{b_1}
+ (1-s) \oo{e_6}{b_1} + p \oo{e_3}{b_2}
\nn\\
&&+ q \oo{e_4}{b_2}
+ r \oo{e_5}{b_2} + s \oo{e_6}{b_2}
+ p (1-r) \oo{e_3}{e_5}
\nn\\
&& + p (1-s) \oo{e_3}{e_6}
+ q (1-r) \oo{e_4}{e_5}
+ q (1-s) \oo{e_4}{e_6}\nn\\
&&+ (r+s) \oo{e_5}{e_6} \mod 2,
\label{chia}
\ea
\ba
\chi_{pqrs}^{(2)} &=& 1+\oo{b_1}{b_2} + (1-r)\oo{e_5}{b_2}
+ (1-s) \oo{e_6}{b_2} + p \oo{e_1}{b_1}
\nn\\
&&+ q \oo{e_2}{b_1}
+ r \oo{e_5}{b_1} + s \oo{e_6}{b_1}
+ p (1-r) \oo{e_1}{e_5}
\nn\\
&& + q (1-r) \oo{e_2}{e_5}
+ p (1-s) \oo{e_1}{e_6}
+ q (1-s) \oo{e_2}{e_6}\nn\\
&&+ (r+s) \oo{e_5}{e_6} \mod 2,
\label{chib}
\ea
\ba
\chi_{pqrs}^{(3)} &=& 1+\oo{b_1}{b_2} + (1-p)\oo{e_1}{b_1}
+ (1-q) \oo{e_2}{b_1} + \oo{e_5+e_6}{b_1}
\nn\\
&& + (1-r) \oo{e_3}{b_2} + (1-s) \oo{e_4}{b_2}\nn\\
&& + (1-r)(1-p) \oo{e_3}{e_1} + (1-r)(1-q) \oo{e_3}{e_2}\nn\\
&&+ (1-r) \oo{e_3}{e_5} + (1-r) \oo{e_3}{e_6} + (1-s) \oo{e_4}{e_6}
\nn\\
&&+ (1-r) \oo{e_3}{z_1+z_2} + (1-s) \oo{e_4}{z_1+z_2}\nn\\
&&+\oo{b_1}{z_1+z_2} \mod 2.
\label{chic}
\ea
\end{subequations}
We remark here that the projection coefficient $\cc{b_1}{b_2}$ simply
fixes the overall chirality
and that our equations depend only on
\ba\label{allps3}
\oo{e_i}{e_j},\ \oo{e_i}{b_k},
\oo{e_i}{z_n},\ \oo{z_n}{b_k},
i=1,\dots,6, k=1,2, n=1,2.
\ea
However, the following six parameters do not appear in the expressions
$\oo{e_1}{e_2}$, $\oo{e_3}{e_4}$, $\oo{e_3}{b_1}$, $\oo{e_4}{b_1}$, $\oo{e_1}{b_2}$, $\oo{e_2}{b_2}$
and thus in a generic model the chiral content depends on 37 discrete parameters.

\section{The Simple $S^2V$ Free Fermionic Models}\label{sec:s2v1}

The subclass of models under consideration is generated by a set of $10$ basis
vectors.
\begin{equation}
V_1 = \{ v_1, \ldots, v_{10} \},
\end{equation}
where
\begin{eqnarray}
v_1=1		&=&\{\psi^\mu,\
		\chi^{1,\dots,6},y^{1,\dots,6},\omega^{1,\dots,6}|\bar{y}^{1,\dots,6},
		\bar{\omega}^{1,\dots,6},\bar{\eta}^{1,2,3},
		\bar{\psi}^{1,\dots,5},\bar{\phi}^{1,\dots,8}\},\nn\\
v_2=S		&=&\{\psi^\mu,\chi^{1,\dots,6}\},\nn\\
v_{2+i}=e_i	&=&\{y^{i},\omega^{i}|\bar{y}^i,\bar{\omega}^i\}, \
i=1,\dots,6,\label{vectors}\\
v_{9}=b_1	&=&\{\chi^{34},\chi^{56},y^{34},y^{56}|\bar{y}^{34},\bar{y}^{56},
\bar{\eta}^1,\bar{\psi}^{1,\dots,5}\},\nn\\
v_{10}=b_2	&=&\{\chi^{12},\chi^{56},y^{12},y^{56}|\bar{y}^{12},\bar{y}^{56},
\bar{\eta}^2,\bar{\psi}^{1,\dots,5}\}.\nn
\end{eqnarray}
The vectors $1,S$ generate a $N=4$ super symmetric model. The vectors
$e_i$, $i=1,\dots,6$ give rise to
all possible symmetric shifts of internal fermions
($y^i,\omega^i,\bar{y}^i,\bar{\omega}^i$) while $b_1$ and $b_2$
stand for  the $\mathbbm{Z}_2\times \mathbbm{Z}_2$ orbifold twists. This is the most general basis,
with symmetric shifts for the internal
fermions, that is compatible with
Kac--Moody level one $SO(10)$ gauge group.
The associated projection coefficients are denoted by
$\cc{v_i}{v_j}, i,j=1,\dots,10$ and can take the values $\pm1$.
They
are related by modular invariance
$\cc{v_i}{v_j}=\exp\{i\frac{\pi}{2}v_i\cdot v_j\} \cc{v_j}{v_i}$
and $\cc{v_i}{v_i}=\exp\{i\frac{\pi}{4}v_i\cdot v_i\} \cc{v_j}{1}$
leaving $2^{45}$ independent coefficients. Out of them, the requirement
of $N=1$ super symmetric spectrum fixes
(up to  a phase convention) all $\cc{S}{v_i},i=1,\dots,10$. Moreover,
without loss of generality we can set $\cc{1}{1}=-1$,
and leave the remaining 36 coefficients free. Therefore, a simple counting
gives $2^{36}$ (that is approximately $10^{10.8}$) distinct models in
the subclass under consideration.

\subsection{The gauge group}\label{sec:n4gauge}

The gauge group in this subclass of models at the $N=4$ level is not enhanced. The group induced by the
vectors $\{1, S, e\}$, where $e$ is defined as in equation \eqref{eq:internalseparation} is
\begin{equation}
G = SO(12) \times SO(32).
\end{equation}
We find that no other sector induces gauge bosons and the gauge group cannot 
be
enhanced. The internal gauge group $SO(12)$ can be broken to its Cartan
generators by means of a suitable choice of the GSO coefficients induced by 
the
$e_i$ i.e. the vectors $e_i$ can break $SO(12) \to U(1)^6$. The $SO(32)$ is
broken to $SO(10) \times U(1)^2 \times SO(18)$ by means of the vectors $b_i$ at the $N=1$ level.
We do not get find enhancements in this subclass at the $N=1$ level.

\subsection{Observable matter spectrum}\label{sec:s2v1matter}

The untwisted sector is common to all models in this subclass. The differences
between the models become apparent in the twisted sector. Therefore, we focus
 on the twisted sector.

The sectors that give rise to massless chiral matter are
\begin{subequations}
\begin{eqnarray}
B^{(1)}_{pqrs} &=& S + b_1 + p\ e_3 + q\ e_4 + r\ e_5 + s\ e_6 ,\\
B^{(2)}_{pqrs} &=& S + b_2 + p\ e_1 + q\ e_2 + r\ e_5 + s\ e_6.
\end{eqnarray}
\end{subequations}
These give 16 different sectors on each of the first two orbifold planes. The
upper index $i$ represents the plane and the lower index
$p,q,r,s=\{0,1\}$ represents the sector on the plane $i$.

Due to the GSO projections we can derive the projectors for the different
sectors as a function of the GSO coefficients. They are
\begin{subequations}\label{ps2v1}
\begin{eqnarray}
P^{(1)}_{pqrs} &=& \frac{1}{4} \prod_{i=1,2} \left( 1 -
\cc{e_i}{B^{(1)}_{pqrs}} \right),\label{pa1}\\
P^{(2)}_{pqrs} &=& \frac{1}{4} \prod_{i=3,4} \left( 1 -
\cc{e_i}{B^{(1)}_{pqrs}} \right),\label{pb1}
\end{eqnarray}
\end{subequations}
where $P_{pqrs}^{(i)}$ is a projector that takes values $\{0,1\}$.
The chirality of the surviving spinorials is given by
\begin{subequations}\label{cs2v1}
\begin{eqnarray}
X^{(1)}_{pqrs} &=& \cc{b_2 + (1-r)e_5 +
(1-s)e_6}{B^{(1)}_{pqrs}},\label{chia1}\\
X^{(2)}_{pqrs} &=& \cc{b_1 + (1-r)e_3 +
(1-s)e_4}{B^{(2)}_{pqrs}},\label{chib1}
\end{eqnarray}
\end{subequations}
where $X_{pqrs}^i=+$ corresponds to a ${\bf16}$ of $SO(10)$
and $X_{pqrs}^i=-$ corresponds
to a ${\bf\overline{16}}$ and we have chosen
the space-time chirality $C(\psi^\mu)=+1$.
The net number of spinorials and thus the net number of families is given by
\begin{equation}\label{eq:nfs2v1}
N_F=\sum_{i=1}^2\sum_{p,q,r,s=0}^1X^{(i)}_{pqrs}
P_{pqrs}^{(i)}.
\end{equation}

Similar formulas can be easily derived for the number of vectorials
and the number of singlets and can be extended
to the $U(1)$ charges but in this work we restrict to the
spinorial calculation.

Formulas \eqref{ps2v1} allow us to identify the mechanism
of spinorial reduction, or in other words
the fixed point reduction, in the fermionic language. For a particular
sector ($B_{pqrs}^{(i)}$) of the orbifold plane $i$
there exist two shift vectors ($e_{2i-1}, e_{2i}$).
Setting the relative projection coefficients \eqref{ps2v1}
to $-1$ each of the above two vectors acts as a projector that cuts
the number of fixed points in the
associated sector by a factor of two. Since two such projectors are
available for each sector the number of fixed points
can be reduced from $16$ to $4$ per plane.

Note that contrary to section \ref{sec:ferms3}, it is of no use to put
these formulas in a simpler form using the matrix formulation developed
therein. The reason being that in these cases we have two projectors for the
simple $SO(8)$ models, while we still have $2^4$ different sectors on each plane. We are
therefore always left with two free coefficients.
This is the exact origin of the fact that in these models we cannot reduce the
number of generations down to one per plane.

We remark here that the projection coefficient $\cc{b_1}{b_2}$ simply
fixes the overall chirality and that our equations depend only on
\ba\label{allps2v1}
\oo{e_i}{e_j},\ \oo{e_i}{b_A}.
\ea

\section{The Extended $S^2V$ Free Fermionic Models}\label{sec:s2v2}

The $S^2V$ $SO(8)$ subclass of models is generated by a set of $11$ basis
vectors.
\begin{equation}
V_1 = \{ v_1, \ldots, v_{11} \},
\end{equation}
where
\begin{eqnarray}
v_1=1		&=&\{\psi^\mu,\
		\chi^{1,\dots,6},y^{1,\dots,6},\omega^{1,\dots,6}|\bar{y}^{1,\dots,6},
		\bar{\omega}^{1,\dots,6},\bar{\eta}^{1,2,3},
		\bar{\psi}^{1,\dots,5},\bar{\phi}^{1,\dots,8}\},\nn\\
v_2=S		&=&\{\psi^\mu,\chi^{1,\dots,6}\},\nn\\
v_{2+i}=e_i	&=&\{y^{i},\omega^{i}|\bar{y}^i,\bar{\omega}^i\}, \
i=1,\dots,6,\nn\\
v_{9}=b_1	&=&\{\chi^{34},\chi^{56},y^{34},y^{56}|\bar{y}^{34},\bar{y}^{56},
		\bar{\eta}^1,\bar{\psi}^{1,\dots,5}\},\label{eq:second.set}\\
v_{10}=b_2	&=&\{\chi^{12},\chi^{56},y^{12},y^{56}|\bar{y}^{12},\bar{y}^{56},
		\bar{\eta}^2,\bar{\psi}^{1,\dots,5}\},\nn\\
v_{11}=z_2	&=&\{\bar{\phi}^{5,\dots,8}\}.\nn
\end{eqnarray}
The remaining fermions not affected by the action
of the previous vectors are $\bar{\eta}^3, \bar{\phi}^i,i=1,\dots,8$ which
normally give
rise to the hidden sector gauge group.
 The vector $z_2$ divide these nine fermions in two sets, one giving rise to
the additional $SO(10)$ and the other giving rise to a hidden $SO(8)$.

The associated projection coefficients are denoted by
$\cc{v_i}{v_j}, i,j=1,\dots,11$ and can take the values $\pm1$.
Due to modular invariance they give rise to $2^{55}$ independent coefficients.
Out of them, the requirement
of $N=1$ super symmetric spectrum fixes
(up to  a phase convention) all $\cc{S}{v_i},i=1,\dots,11$. Moreover,
without loss of generality we can set $\cc{1}{1}=-1$,
and leave the remaining 45 coefficients free. Therefore, a simple counting
gives  $2^{45}$ (that is approximately $10^{13.5}$) distinct models in
the class under consideration.

In the following we study this class of models by deriving analytic
formulas for the gauge group and the
spectrum and then using these formulas for the classification. It is easy to
see that the third twisted plane has massless states since $b_3=b_1 + b_2$
gives rise to vectorial representations of the observable $SO(10)$.

\subsection{The gauge group}\label{sec:s2v2gauge}

The gauge group in this subclass of models at the $N=4$ level, induced by the
vectors $\{1,S, e, z_2\}$ without enhancements is
\begin{equation}
G = SO(12) \times SO(24) \times SO(8).
\end{equation}
Enhancements can occur due to the $z_2$ sector. Both the SO(12) and the SO(24)
can be enhanced by a suitable choice of the GSO coefficient $\cc{e}{z_2}$. The
enhancements are listed in table \ref{tab:enhancements}

\begin{table}
\begin{center}
\begin{tabular}{r | r}
$\cc{e}{z_2}$ & Group $G$\\\hline
$-1$ & $SO(20) \times SO(24)$\\
$1$  & $SO(12) \times SO(32)$
\end{tabular}
\end{center}
\caption{The configuration of the gauge group of the $N=4$ theory. We have
separated a
 priori the internal gauge group using the vectors $e$ and $z_2$. Introducing
the other vectors $e_i$ and $b_i$ only induce breaking of these
groups.}\label{tab:enhancements}
\end{table}
The internal gauge group $SO(12)$ can be reduced to its Cartan generators
$U(1)^6$. The $SO(24)$ is broken to $SO(10)^2 \times U(1)^2$. 
We therefore find the gauge group at the $N=1$ level
\begin{equation}\label{eq:gauge.second}
SO(10) \times U(1)^2 \times SO(10) \times SO(8).
\end{equation}
In this subclass we always get enhancement. We have
argued that the hidden sector gets enhanced when $\cc{e}{z_2}=1$. This is the
configuration where the internal sector is completely separated from the gauge
sector. At the $N=1$ level we can get enhancement of $SO(8) \times U(1) \to SO(10)$. We discuss this case in more detail below. In a similar fashion we can get enhancement of $SO(8) \times SO(10) \to SO(18)$. Matter states under this enhanced gauge group are not realised. The enhancement in the adjoint representation is only found when $-\cc{b_1}{z_2}=-\cc{b_1}{z_2}=\cc{e_3}{z_2}=\cc{e_4}{z_2}=1$. This enhancement is only realised at the $N=1$ level as is explained in section \ref{sec:liftability}.

\subsection{Observable matter spectrum}\label{sec:s2v2matter}

The untwisted sector is common to all models in this subclass. The differences
between the models become apparent in the twisted sector. Therefore, we focus
 on the twisted sector.

The sectors that give rise to massless chiral states are split up into two
types. The first type gives spinorial representations of the first $SO(10)$ and
the second type gives spinorial representations of the second $SO(10)$. Due to
the sector $z_2$ we can get enhancement of the $SO(8) \times U(1) \to SO(10)$.
This sector can give rise to spinorial representations as well and we label
this as the type $3$ spinorials. The sectors that give chiral matter of the
first type are
\begin{subequations}
\begin{eqnarray}
B^{(1)}_{pqrs} &=& S + b_1 + p\ e_3 + q\ e_4 + r\ e_5 + s\ e_6, \\
B^{(2)}_{pqrs} &=& S + b_2 + p\ e_1 + q\ e_2 + r\ e_5 + s\ e_6,
\end{eqnarray}
\end{subequations}
and the sectors giving chiral matter of the second type are
\begin{subequations}
\begin{eqnarray}
\bar{B}^{(1)}_{pqrs} &=& S + b_1 + x + {p}\ e_3 + {q}\ e_4 +
{r}\ e_5 + {s}\ e_6 ,\\
\bar{B}^{(2)}_{pqrs} &=& S + b_2 + x + {p}\ e_1 + {q}\ e_2 +
{r}\ e_5 + {s}\ e_6,
\end{eqnarray}
\end{subequations}
where
\begin{equation}\label{x}
x = 1 + S + \sum_{i=1}^{6}e_i + z_2.
\end{equation}
These give 16 different sectors on the first two orbifold planes for each type.
The upper index $i$ represents the plane and the lower index
$p,\-q,\-r,\-s=\{0,1\}$ represents the sector on the plane $i$. The sectors
giving rise to chiral states of the second type are represented by $\bar{B}$.

The spinorial $SO(8) \times U(1) \to SO(10)$ enhancement can occur in this
subclass of models due to the sectors
\begin{equation}
B^{(3)}_{pqrs} = S + b_1 + b_2 + z_2 + p\ e_1 + q\ e_2 + r\ e_3 + s\ e_4
\end{equation}
and the sector 
\begin{equation}
V^{(3)}_{pqrs} = S + b_1 + b_2 + p\ e_1 + q\ e_2 + r\ e_3 + s\ e_4.
\end{equation}
Vectorials of the $SO(8)$ are only realised
when $\cc{b_3}{z_2}=-1$. The sector $V^{(3)}_{pqrs}$ then gives rise to the $8_v$ and the sector $B^{(3)}_{pqrs}$ gives $8_s$ to complete the $16$ or $\bar{16}$, depending on the
GSO coefficients. In order to find the chirality of the spinorial
representation of $SO(10)$ we look at which $U(1) \subset SO(10) \times U(1)^2
\times SO(10) \times SO(8)$ is used. We therefore focus on the vectorial
bilinear realisation of the enhancement. This enhancement is realised by the
$NS$ sector and the $z_2$ sector giving $45 = 28 + 1 + 8^+ + 8^-$. When
$-\cc{b_1}{z_2}=\cc{b_2}{z_2}=\cc{e_5}{z_2}=\cc{e_6}{z_2}=1$ the state
\begin{equation}
\psi^\mu_{1,2}\ \bar{\eta}^{1} \left[ \bar{\phi}^{5,\ldots,8} \right] | 0 >
\end{equation}
survives and the enhancement is realised with the $U(1)$ induced by
$\bar{\eta}^{1}$. When $\cc{b_1}{z_2}=-\cc{b_2}{z_2}=\cc{e_1}{z_2}=\cc{e_2}{z_2}=1$ the state
\begin{equation}
\psi^\mu_{1,2}\ \bar{\eta}^{2} \left[ \bar{\phi}^{5,\ldots,8} \right] | 0 >
\end{equation}
survives and the enhancement is realised with the $U(1)$ induced by
$\bar{\eta}^{2}$. Depending now on the charge of state under the correct $U(1)$
we get either a $16$ or $\bar{16}$ of the $SO(10)$ by $16=8_s(1) + 8_v(-1)$ or
$\bar{16}=8_s(-1) + 8_v(1)$. We have therefore listed the sign of the $U(1)$
charge of the correct state to describe the chirality of the type three
$SO(10)$. Note that this $SO(10)$ is realised on the third plane.

Since the enhancement is only present when $\cc{b_3}{z_2}=-1$, we can conclude
that the enhancement only takes place at the $N=1$ level of the model. Models
that are liftable to an $N=4$ theory therefore do not exhibit this type of
enhancement. For completeness we do give our results for the spinorial
representations of the third $SO(10)$, that is found on the third plane. When
enhancement does not occur we do find spinorial representations of a $SO(8)$
gauge group. The chirality of these spinors is determined by
$\cc{z_2}{B^{(3)}_{pqrs}}$. Again only two projectors can be realised leading
to a reduction of the number of generations under the $SO(8)$ of $16 \to 4$.
Since we focus in this classification on the generations of an observable
$SO(10)$ we do not consider these states here.

Due to the GSO projections we can derive the projectors for the different
sectors as a function of the GSO coefficients. They are for the first type
\begin{subequations}\label{ps2v21}
\begin{eqnarray}
P^{(1)}_{pqrs} &=& \frac{1}{8} \prod_{i=1,2} \left( 1 -
\cc{e_i}{B^{(1)}_{pqrs}} \right) \left( 1 - \cc{z_2}{B^{(1)}_{pqrs}}
\right),\label{pa2}\\
P^{(2)}_{pqrs} &=& \frac{1}{8} \prod_{i=3,4} \left( 1 -
\cc{e_i}{B^{(1)}_{pqrs}} \right) \left( 1 - \cc{z_2}{B^{(2)}_{pqrs}}
\right),\label{pb2}
\end{eqnarray}
\end{subequations}
and for the second type
\begin{subequations}\label{ps2v22}
\begin{eqnarray}
\bar{P}^{(1)}_{pqrs} &=& \frac{1}{8} \prod_{i=1,2} \left( 1 -
\cc{e_i}{\bar{B}^{(1)}_{pqrs}} \right) \left( 1 -
\cc{z_2}{\bar{B}^{(1)}_{pqrs}} \right),\label{pc2}\\
\bar{P}^{(2)}_{pqrs} &=& \frac{1}{8} \prod_{i=3,4} \left( 1 -
\cc{e_i}{\bar{B}^{(2)}_{pqrs}} \right) \left( 1 -
\cc{z_2}{\bar{B}^{(2)}_{pqrs}} \right),\label{pd2}
\end{eqnarray}
\end{subequations}
and for the third type
\begin{equation}
P^{(3)}_{pqrs} = \frac{1}{4} \prod_{i=5,6} \left(1 - \cc{e_i}{B^{(3)}_{pqrs}}
\right).
\end{equation}

Again we have represented the projectors for the second type as $\bar{P}$.
$P_{pqrs}^{(i)}$, or $\bar{P}_{pqrs}^{(i)}$ for the second type chiral states,
is a projector that takes values $\{0,1\}$.
The chirality of the surviving spinorials of the first type is given by
\begin{subequations}\label{cs2v21}
\begin{eqnarray}
X^{(1)}_{pqrs} &=& \cc{b_2 + (1-r)e_5 +
(1-s)e_6}{B^{(1)}_{pqrs}},\label{chia2}\\
X^{(2)}_{pqrs} &=& \cc{b_1 + (1-r)e_3 +
(1-s)e_4}{B^{(2)}_{pqrs}},\label{chib2}
\end{eqnarray}
\end{subequations}
while for the second type they are
\begin{subequations}\label{cs2v22}
\begin{eqnarray}
\bar{X}^{(1)}_{pqrs} &=& \cc{b_2 + e_3 + e_4 + {r}\ e_5 +
{s}\ e_6}{\bar{B}^{(1)}_{pqrs}},\label{chic2}\\
\bar{X}^{(2)}_{pqrs} &=& \cc{b_1 + e_1 + e_2 + {r}\ e_5 +
{s}\ e_6}{\bar{B}^{(2)}_{pqrs}},\label{chid2}
\end{eqnarray}
\end{subequations}
and for the third type they are when $-\cc{b_1}{z_2}=\cc{b_2}{z_2}=1$
\begin{subequations}\label{cs2v23}
\begin{equation}
X^{(3)}_{pqrs} = \cc{(1-r)e_3 + (1-s)e_4 + b_1}{B^{(3)}_{pqrs}},
\end{equation}
and when $\cc{b_1}{z_2}=-\cc{b_2}{z_2}=1$
\begin{equation}
X^{(3)}_{pqrs} = \cc{(1-p)e_1 + (1-q)e_2 + b_2}{B^{(3)}_{pqrs}},
\end{equation}
\end{subequations}
where $X_{pqrs}^i=+1$ ($\bar{X}_{pqrs}^i=+1$) corresponds to a ${\bf16}$ of the
first (second) $SO(10)$
and $X_{pqrs}^i=-1$ ($\bar{X}_{pqrs}^i=-1$) corresponds
to a ${\bf\overline{16}}$ and we have chosen
the space-time chirality $C(\psi^\mu)=+1$.
The net number of spinorials of the first type and therefore the net number of
families is given by
\begin{equation}\label{eq:nfs2v21}
N_F=\sum_{i=1}^2\sum_{p,q,r,s=0}^1X^{(i)}_{pqrs} P_{pqrs}^{(i)},
\end{equation}
while for the second type
\begin{equation}\label{eq:nfs2v22}
\bar{N}_F = \sum_{i=1}^2\sum_{p,q,r,s=0}^1 \bar{X}^{(i)}_{pqrs}
\bar{P}^{(i)}_{pqrs},
\end{equation}
and for the third type
\begin{equation}
N_F= \sum_{p,q,r,s=0}^1X^{(i)}_{pqrs} P_{pqrs}^{(i)}.
\end{equation}

Similar formulas can be easily derived for the number of vectorials
and the number of singlets and can be extended
to the $U(1)$ charges but in this work we restrict to the
spinorial calculation.

Formulas \eqref{ps2v21} -- \eqref{ps2v22} allow us to identify the mechanism
of spinorial reduction, or in other words
the fixed point reduction, in the fermionic language. For a particular
sector ($B_{pqrs}^{(i)}$) of the orbifold plane $i$
there exist two shift vectors ($e_{2i-1}, e_{2i}$) and one $z$ vector $z_2$.
Setting the relative projection coefficients \eqref{ps2v21} -- \eqref{ps2v22}
to $-1$ each of the above three vectors acts as a projector that cuts
the number of fixed points in the
associated sector by a factor of two. Since three such projectors are
available for each sector the number of fixed points
can be reduced from $16$ to $2$ per plane. This argumentation holds for the
spinorial representations of both the $SO(10)$'s.

Note that contrary to section \ref{sec:ferms3}, it is of no use to put
these formulas in a simpler form using the matrix formulation developed
therein. The reason being that in these cases we have three projectors for the
extended $SO(8)$ models, while we still have $2^4$ different sectors on each plane. We are
therefore always left with one free coefficients.
This is the exact origin of the fact that in these models we cannot reduce the
number of generations down to one per plane.

We remark here that the projection coefficient $\cc{b_1}{b_2}$ simply
fixes the overall chirality and that our equations depend only on
\ba\label{allps2v2}
\oo{e_i}{e_j},\ \oo{e_i}{b_A},
\oo{e_i}{z_2},\ \oo{z_2}{b_A},\ i=1,\dots,6,\  A=1,2.
\ea

\section{The $SV^2$ and $V^3$ Free Fermionic Models}\label{remnants}

In this section we discuss the remaining models. We start by
describing the $SV^2$ models and continue with
the $V^3$ models. We show that these models do not give rise to phenomenologically interesting models. As a result we do not derive the projectors and chirality operators of these models as we have done in sections \ref{sec:ferms3}, \ref{sec:s2v1} and \ref{sec:s2v2}.

Again we start from the $N=4$ partition function written down
in equation \eqref{eq:fermionpt44}. Reducing the number of super symmetries 
from $N=4$ to $N=1$ is realised using a $\mathbbm{Z}_2 \times \mathbbm{Z}_2$ twist. This
is the origin of the four different subclasses of models.

Below we show that the $SV^2$ and $V^3$ subclasses do not contain realistic
models.

\subsection{The $SV^2$ free fermionic models}\label{sec:sv2}

Intuitively one may think that the $SV^2$ models are realised using the two
twisting vectors
$b_1$ and $b_2$ in the fermionic construction.
\begin{eqnarray}
b_1 &=& \{ \chi^{3,4}, \chi^{5,6}, y^{3,4}, y^{5,6}\
| \bar{y}^{3,4}, \bar{y}^{5,6},\
\bar{\eta}^{1}, \bar{\psi}^{1,\ldots,5} \} ,\label{eq:sv2.1}\\
b_2 &=& \{ \chi^{1,2}, \chi^{5,6}, y^{1,2}, y^{5,6}\
| \bar{y}^{1,2}, \bar{y}^{5,6},\
\bar{\eta}^{1}, \bar{\eta}^{3} \} .
\label{eq:sv2.2}
\end{eqnarray}
Indeed we find that this gives spinorial representations on the first
twisted plane and vectorial representations on the second plane. The third
plane, however, gives rise to spinorial representations as well. In defining
the second
twisting vector as equation \eqref{eq:sv2.2}, we have merely redefined the
$S^2V$ models.

Let us therefore consider all possible options. We fix the basis vector $b_1$ 
and consider different configurations of the vector $b_2$. We can relax the 
constraint of
vectorial representations on the second and third plane. We require only that 
there are no spinorials on the second and third plane of the observable 
$SO(10)$.
This leads to four possible $b_2$ vectors. Due to modular
invariance we can either choose $8$ real fermions or $16$ real fermions on the
right moving side. Choosing the latter gives us three options. In the latter
case we can either choose, due to modular invariance, that the overlap of the
fermions of the $b_1$ and $b_2$ vector on the right moving side are $4$, $8$ or
$12$ real fermions. We have labeled the different options in table
\ref{tab:options}.

\begin{table}
\begin{center}
\begin{tabular}{l | r | r}
Subclass 	& \# right-moving 	& \# overlapping 	\\
 		& real fermions 	& real fermions 	\\\hline
A 		& 16			& 4	\\
B		& 16			& 8	\\
C		& 16			& 12	\\
D		& 8			& 4
\end{tabular}
\caption{The a priori different possible options for the realisation of the
$SV^2$ models.}\label{tab:options}
\end{center}
\end{table}

We now discuss the different options in more detail.
\begin{itemize}
\item[\bf A] One possible vector to describe this subclass of models defined in table \ref{tab:options} is given
by
\begin{equation}
b_2 = \{ \chi^{1,2}\ \chi^{5,6}\ y^{1,2}\ y^{5,6}\ |\ \bar{y}^{1,2}\
\bar{y}^{5,6}\ \bar{\eta}^{1}\ \bar{\eta}^{3} \bar{\phi}^{1,\ldots,4} \}.
\end{equation}
At the level of the $10$ basis vectors $\{1,S, b_1, b_2, e_i\}$ it is easy to
show that the gauge group induced by the gauge fermions is
\begin{equation}
G = SO(10)^3 \times U(1).
\end{equation}
One $SO(10)$ gauge group is realised on each of the three planes. On the first
two planes there are spinorial representations of their respective $SO(10)$
gauge group. At this level we can only realise two projector operators which
leads to a total reduction of the number of generations to $4$ on each of the
two planes. Adding $z_2$ as defined in equation \eqref{eq:second.set} we obtain
the gauge group described in equation \eqref{eq:gauge.second}. It is easy to
see that this addition of the vector $z_2$ realises an equivalent description
of the extended $S^2V$ models. We therefore conclude that adding $z_1$ results
in the $S^3$ models. The only models relevant for this subclass are those that
do not have a separated $SO(8)$ gauge group and we have seen that this model
can only realise $4,8,16$ generations on each plane of their respective
$SO(10)$ gauge groups.\\
\item[\bf B] One possible vector to describe this subclass of models defined in table \ref{tab:options} is given
by
\begin{equation}
b_2 = \{ \chi^{1,2}\ \chi^{5,6}\ y^{1,2}\ y^{5,6}\ |\ \bar{y}^{1,2}\
\bar{y}^{5,6}\ \bar{\psi}^{1,2,3} \bar{\phi}^{1,2,3} \}.
\end{equation}
At the level of the $10$ vectors $\{1,S, b_1, b_2, e_i\}$ it is easy to show
that the gauge group induced by the gauge fermions is
\begin{equation}
G = SO(6)^3 \times SO(14).
\end{equation}
We are not able to generate an $SO(10)$ unification group in this subclass of
models. Adding $z_2$ breaks $SO(14)\to SO(8) \times SO(6)$ but allows for
enhancements to appear. However enhancements to an $SO(10)$ gauge group are
never realised. We can now add the alternative vector $\tilde{z}_1$
\begin{equation}
\tilde{z}_1 = \{ \bar{\eta}^{2,3}\bar{\phi}^{1,2} \}.
\end{equation}
This breaks $SO(6) \times SO(6) \to SU(2) \times SU(2) \times U(1) \times SU(2)
\times SU(2) \times U(1)$. The addition of this last vector does not realise an
enhancement to an observable $SO(10)$. \\
\item[\bf C] One possible vector to describe this subclass of models defined in table \ref{tab:options} is given
by
\begin{equation}
b_2 = \{ \chi^{1,2}\ \chi^{5,6}\ y^{1,2}\ y^{5,6}\ |\ \bar{y}^{1,2}\
\bar{y}^{5,6}\ \bar{\eta}^{2}\ \bar{\psi}^{1,\ldots,5} \}.
\end{equation}
This vector is identical to equation \eqref{eq:twistvectors2s3}. We have
therefore found either the $S^3$ or the $S^2V$ subclass of models.
\item[\bf D] One possible vector to describe this subclass of models defined in table \ref{tab:options} is given
by
\begin{equation}
b_2 = \{ \chi^{1,2}\ \chi^{5,6}\ y^{1,2}\ y^{5,6}\ |\ \bar{y}^{1,2}\
\bar{y}^{5,6}\ \bar{\eta}^{1}\ \bar{\eta}^{3}  \}.
\end{equation}
This vector is identical to equation \eqref{eq:sv2.2}. We have therefore found
the $S^2V$ subclass of models.
\end{itemize}

We have shown that in this subclass of models we are not able to generate
phenomenologically interesting models. The reasons are that in case {\bf A} we are
either not able to reduce the number of generations of one $SO(10)$ gauge group
both to $3$ or $6$ or we have realised a redefinition of the extended $S^2V$ models. In case {\bf B} a $SO(10)$ unification gauge group is not realised. In case {\bf C} and {\bf D} the $S^2V$ models were again realised.

\subsection{The $V^3$ free fermionic models}

In this section we describe the $V^3$ models. We show that when an observable $SO(10)$ is realised the $V^3$ subclass reduces to the $S^3$ subclass of models.
 
The $V^3$ models are initially realised using the two twisting vectors
$b_1$ and $b_2$ defined as
\begin{eqnarray}
b_1 &=& \{ \chi^{3,4}, \chi^{5,6}, y^{3,4}, y^{5,6}\
| \bar{y}^{3,4}, \bar{y}^{5,6},\
\bar{\eta}^{2}, \bar{\eta}^{3} \} ,\label{eq:v3.1}\\
b_2 &=& \{ \chi^{1,2}, \chi^{5,6}, y^{1,2}, y^{5,6}\
| \bar{y}^{1,2}, \bar{y}^{5,6},\
\bar{\eta}^{1}, \bar{\eta}^{3} \}.
\label{eq:v3.2}
\end{eqnarray}
In this notation vectorial representations are realised on all the three
planes where the vectorials of the third plane are realised by means of
the vector $b_3 = b_1 + b_2$. However as soon as we separate out $SO(8)$ gauge
groups from the hidden sector we automatically generate spinorial
representations
of a $SO(8)$, or even a $SO(10)$, gauge group. To see this we isolate one block
of theta
functions by means of
\begin{equation}
z_2 = \{ \bar{\phi}^{5,\ldots, 8} \}.
\end{equation}
Adding this basis vector gives rise to new massless states of the form
\begin{equation}
b_1 + z_2 = \{ \chi^{3,4}, \chi^{5,6}, y^{3,4}, y^{5,6}\
| \bar{y}^{3,4}, \bar{y}^{5,6},\
\bar{\eta}^{2}, \bar{\eta}^{3}, \bar{\phi}^{5,\ldots, 8} \}.
\end{equation}
giving spinorial representations of the $SO(8)$ formed by
$\bar{\phi}^{5,\ldots, 8}$. The gauge group of this set of basis vectors
is a priori $SU(2) \times SU(2) \times SO(28) \to U(1)^3 \times SO(26)$.
Separating the theta block we  break a priori the group further to
$U(1)^3 \times SO(26) \to U(1)^3 \times SO(18) \times SO(8)$. At this
level the $SO(10)$ unification group can not be realised.

However, the additional sectors induced by the $z_2$ vector can enhance $SO(8)
\times U(1) \to SO(10)$. This depends on the GSO coefficients of the specific
model. If the enhancement is realised the GSO coefficients are
$\cc{z_1}{b_1}=\cc{z_1}{b_2}=-1$ necessarily. This however removes the
vectorial states induced by the sectors $b_1$ and $b_2$. We have then
effectively realised the $S^2V$ models which are discussed in detail in sections \ref{sec:s2v1} and \ref{sec:s2v2}.

To break $SO(18) \to SO(10) \times SO(8)$ we can separate another $SO(8)$ gauge
group. This separation gives us a redefinition of the $S^3$ models. To show
that we define
\begin{equation}
z_1 = \{ \bar{\phi}^{1,\ldots, 4} \}.
\end{equation}
The spinors on the three planes are then realised as
\begin{eqnarray}
B_{spinor}^1 &=& b_1 + x,\\
B_{spinor}^2 &=& b_2 + x,\\
B_{spinor}^3 &=& b_1 + b_2 + x,
\end{eqnarray}
where
\begin{equation}
x = 1 + S + \sum_{i=1}^6 e_i + \sum_{k=1}^2 z_k.
\end{equation}
These models are analysed in detail in \cite{Faraggi:2004rq, Faraggi:2003yd}. The only difference is that
the generalised GSO coefficients are redefined in this subclass of models.

We have therefore shown that the $SV^2$ and $V^3$ subclasses either do not
contain realistic models or reduce to the $S^3$ or $S^2V$ models. Because of
this we do not give the partition functions for the $SV^2$ and $V^3$ models and
do not give the chiral and projector operators. These models are also not
included in chapter \ref{chap:models}.

	\part{Results and Conclusions}
\chapter{Some Sample $\mathbbm{Z}_2 \times \mathbbm{Z}_2$ Orbifold Models}\label{chap:models}

In this chapter we give some examples of models in the subclasses $S^3$ and $S^2V$ that exhibit some interesting features. We focus on these subclasses of models because only these are interesting for phenomenology as explained in chapter \ref{chap:fermform}. We describe their chiral content and their gauge group.

\section{Three $S^3$ Models}\label{sec:modelss3}

We apply here the formalism developed in chapter \ref{chap:fermform} in order to derive sample models in the free
fermionic formulation.

\noindent{\it The $\mathbbm{Z}_2\times \mathbbm{Z}_2$ symmetric orbifold}\\[5pt]
The simplest example is the symmetric $\mathbbm{Z}_2\times \mathbbm{Z}_2$ orbifold.  Here we set
all the free  GSO  phases \eqref{allps3}  to zero. The full GSO phase matrix takes
the form ($\cc{v_i}{v_j}=\exp[i\pi(v_i|v_j)]$) with 
{\small
\begin{equation}
(v_i|v_j)\ \ =\ \
\bordermatrix{
&1 & S & e_1 & e_2 & e_3 & e_4 & e_5 & e_6 & b_1 & b_2 & z_1 & z_2\cr
   1   & 1 & 1  &  1 &  1 &  1 &  1 &  1 &  1 &  1 &  1 &  1 & 1 \cr
   S   & 1 & 1  &  1 &  1 &  1 &  1 &  1 &  1 &  1 &  1 &  1 & 1\cr
  e_1  & 1 &  1 &  0 &  0 &  0 &  0 &  0 &  0 &  0 &  0 &  0 & 0\cr
  e_2  & 1 &  1 &  0 &  0 &  0 &  0 &  0 &  0 &  0 &  0 &  0 & 0\cr
  e_3  & 1 &  1 &  0 &  0 &  0 &  0 &  0 &  0 &  0 &  0 &  0 & 0\cr
  e_4  & 1 &  1 &  0 &  0 &  0 &  0 &  0 &  0 &  0 &  0 &  0 & 0\cr
  e_5  & 1 &  1 &  0 &  0 &  0 &  0 &  0 &  0 &  0 &  0 &  0 & 0\cr
  e_6  & 1 &  1 &  0 &  0 &  0 &  0 &  0 &  0 &  0 &  0 &  0 & 0\cr
  b_1  & 1 &  0 &  0 &  0 &  0 &  0 &  0 &  0 &  1 &  1 &  0 & 0\cr
  b_2  & 1 &  0 &  0 &  0 &  0 &  0 &  0 &  0 &  1 &  1 &  0 & 0\cr
  z_1  & 1 &  1 &  0 &  0 &  0 &  0 &  0 &  0 &  0 &  0 &  1 & 1\cr
  z_2  & 1 &  1 &  0 &  0 &  0 &  0 &  0 &  0 &  0 &  0 &  1 & 1\cr
  }.
\end{equation}}\null
This model is equivalent to the model where $\Delta^{(i)}=W^{(i)}=0$ in  equation \eqref{proj}.
All projectors become inactive and thus the number of surviving twisted
sector spinorials takes its maximum
value which is 48 with all chiralities positive according to equations
\eqref{chis3}.
 Moreover three spinorials and three anti-spinorials arise from the
untwisted sector.
Following (\ref{gb1}), (\ref{gb2}) the gauge group enhances to
$\mathbbm{E}_6\times{U(1)}^2\times \mathbbm{E}_8$ and the spinorials of $SO(10)$
combine with vectorials and singlets to produce 48+3=51 families ($\bf 27$) and 3 anti-families ($\bf \overline{27}$) of $\mathbbm{E}_6$.

\noindent{\it A three generation $\mathbbm{E}_6$ model}\\[5pt]
We can obtain a three family $\mathbbm{E}_6$ model by choosing the following set of projection coefficients
{\small
\begin{equation}
(v_i|v_j)\ \ =\ \
\bordermatrix{
       &1  &S   & e_1& e_2& e_3& e_4&e_5 & e_6& b_1& b_2&z_1 &z_2\cr
   1   &1  &1  &1  &1  &1  &1  &1  &    1  &    1  &    1  &    1  &  1\cr
   S   &1  &1  &1  &1  &1  &1  &1  &    1  &    1  &    1  &    1  &  1\cr
  e_1  &1  &1  &0  &0  &1  &0  &0  &    1  &    0  &    0  &    0  &  0\cr
  e_2  &1  &1  &0  &0  &0  &0  &1  &    0  &    0  &    0  &    0  &  1\cr
  e_3  &1  &1  &1  &0  &0  &0  &1  &    0  &    0  &    0  &    0  &  0\cr
  e_4  &1  &1  &0  &0  &0  &0  &1  &    0  &    0  &    0  &    1  &  0\cr
  e_5  &1  &1  &0  &1  &1  &1  &0  &    1  &    0  &    0  &    0  &  0\cr
  e_6  &1  &1  &1  &0  &0  &0  &1  &    0  &    0  &    0  &    1  &  1\cr
  b_1  &1  &0  &0  &0  &0  &0  &0  &    0  &    1  &    0  &    0  &  0\cr
  b_2  &1  &0  &0  &0  &0  &0  &0  &    0  &    0  &    1  &    0  &  0\cr
  z_1  &1  &1  &0  &0  &0  &1  &0  &    1  &    0  &    0  &    1  &  1\cr
  z_2  &1  &1  &0  &1  &0  &0  &0  &    1  &    0  &    0  &    1  &  1\cr
  }.
\end{equation}}\null
The full gauge group is $\mathbbm{E}_6\times {U(1)}^2\times {SO(8)}^2$. Three families
$({\bf 27})$, one
from each plane,  arise from the sectors $S+b_i+(x), i=1,2,3$. Another set of
three families and three
anti-families arise from the untwisted sector.
The hidden sector consists of  nine 8-plets under each $SO(8)$. In addition
there exist a number of non-Abelian
gauge group singlets. The model could be phenomenologically acceptable
provided one finds a way
of breaking $\mathbbm{E}_6$. Since it is not possible to generate the $\mathbbm{E}_6$ adjoint
(not in Kac-Moody level one),
we need to realize the breaking by an additional Wilson-line like vector.
However,  a detailed
investigation of acceptable basis vectors, shows that the $\mathbbm{E}_6$ breaking
is accompanied by
truncation of the fermion families. Thus this kind of perturbative $\mathbbm{E}_6$
breaking is not compatible
with the presence of three generations. 

\noindent{\it A six generation $\mathbbm{E}_6$ model}\\[5pt]
Similarly a six family $\mathbbm{E}_6\times{U(1)}^2\times \mathbbm{E}_8$ model can be obtained using the following
projection coefficients
{\small
\begin{equation}
(v_i|v_j)\ \ =\ \
\bordermatrix{
      &1  &S   & e_1& e_2& e_3& e_4&e_5 & e_6& b_1& b_2&z_1 &z_2\cr
   1  &1  &   1  &   1  &   1  &   1  &   1  &   1  &   1  &   1  &   1  &   1&   1\cr
   S  &1  &   1  &   1  &   1  &   1  &   1  &   1  &   1  &   1  &   1  &   1&   1\cr
  e_1 &1  &   1  &   0  &   0  &   0  &   0  &   1  &   1  &   0  &   0  &   0&   0\cr
  e_2 &1  &   1  &   0  &   0  &   1  &   0  &   0  &   1  &   0  &   0  &   0&   0\cr
  e_3 &1  &   1  &   0  &   1  &   0  &   0  &   0  &   1  &   0  &   0  &   0&   0\cr
  e_4 &1  &   1  &   0  &   0  &   0  &   0  &   1  &   0  &   0  &   0  &   0&   1\cr
  e_5 &1  &   1  &   1  &   0  &   0  &   1  &   0  &   0  &   0  &   0  &   0&   0\cr
  e_6 &1  &   1  &   1  &   1  &   1  &   0  &   0  &   0  &   0  &   0  &   0&   1\cr
  b_1 &1  &   0  &   0  &   0  &   0  &   0  &   0  &   0  &   1  &   0  &   0&   0\cr
  b_2 &1  &   0  &   0  &   0  &   0  &   0  &   0  &   0  &   0  &   1  &   0&   0\cr
  z_1 &1  &   1  &   0  &   0  &   0  &   0  &   0  &   0  &   0  &   0  &   1&   1\cr
  z_2 &1  &   1  &   0  &   0  &   0  &   1  &   0  &   1  &   0  &   0  &   1&   1\cr
 }.
\end{equation}}\null
In this model we have six families from the twisted sector, two from each plane together with
three families and three anti-families from the untwisted sector, accompanied by a number of singlets and
8-plets of both hidden $SO(8)$'s.


\section{Two Simple $S^2V$ Models}\label{sec:modelss2v1}


In this subclass of models we present an example of a model with $32$
generations coming from the twisted sector, $16$ coming from the first two
planes. The second example is a model with $8$ generations coming from the
twisted sector, $4$ coming from each of the first two planes.\\

\noindent{\it The $\mathbbm{Z}_2\times \mathbbm{Z}_2$ symmetric orbifold}\\[5pt]
The simplest example is the symmetric $\mathbbm{Z}_2\times \mathbbm{Z}_2$ orbifold.  Here we set
all the free  GSO  phases \eqref{allps2v1} to zero. This gives a model of $32$
generations where $16$ are coming from each of the first two planes. The full
GSO phase matrix takes
the form ($\cc{v_i}{v_j}=\exp[i\pi(v_i|v_j)]$)
{\small\begin{equation}
(v_i|v_j)\ \ =\ \
\bordermatrix{
&1 & S & e_1 & e_2 & e_3 & e_4 & e_5 & e_6 & b_1 & b_2 \cr
   1   & 1 & 1  &  1 &  1 &  1 &  1 &  1 &  1 &  1 &  1 \cr
   S   & 1 & 1  &  1 &  1 &  1 &  1 &  1 &  1 &  1 &  1 \cr
  e_1  & 1 &  1 &  0 &  0 &  0 &  0 &  0 &  0 &  0 &  0 \cr
  e_2  & 1 &  1 &  0 &  0 &  0 &  0 &  0 &  0 &  0 &  0 \cr
  e_3  & 1 &  1 &  0 &  0 &  0 &  0 &  0 &  0 &  0 &  0 \cr
  e_4  & 1 &  1 &  0 &  0 &  0 &  0 &  0 &  0 &  0 &  0 \cr
  e_5  & 1 &  1 &  0 &  0 &  0 &  0 &  0 &  0 &  0 &  0 \cr
  e_6  & 1 &  1 &  0 &  0 &  0 &  0 &  0 &  0 &  0 &  0 \cr
  b_1  & 1 &  0 &  0 &  0 &  0 &  0 &  0 &  0 &  1 &  0 \cr
  b_2  & 1 &  0 &  0 &  0 &  0 &  0 &  0 &  0 &  0 &  1 \cr
  }.
\end{equation}
}

\noindent{\it An eight generation $SO(10)$ model}\\[5pt]
We can obtain an eight generation $SO(10)$ model, four coming from each of the
first two planes, by choosing the following set of projection coefficients
{\small\begin{equation}\label{model1}
(v_i|v_j)\ \ =\ \
\bordermatrix{
       &1  &S   & e_1& e_2& e_3& e_4&e_5 & e_6& b_1& b_2 \cr
   1   &1  &1   & 1  & 1  & 1  & 1  & 1  & 1  & 1  &  1  \cr
   S   &1  &1   & 1  & 1  & 1  & 1  & 1  & 1  & 1  &  1  \cr
  e_1  &1  &1   & 0  & 0  & 0  & 0  & 1  & 1  & 0  &  0  \cr
  e_2  &1  &1   & 0  & 0  & 1  & 0  & 0  & 1  & 0  &  0  \cr
  e_3  &1  &1   & 0  & 1  & 0  & 0  & 0  & 1  & 0  &  0  \cr
  e_4  &1  &1   & 0  & 0  & 0  & 0  & 1  & 1  & 0  &  0  \cr
  e_5  &1  &1   & 1  & 0  & 0  & 1  & 0  & 0  & 0  &  0  \cr
  e_6  &1  &1   & 1  & 1  & 1  & 1  & 0  & 0  & 0  &  0  \cr
  b_1  &1  &0   & 0  & 0  & 0  & 0  & 0  & 0  & 1  &  0  \cr
  b_2  &1  &0   & 0  & 0  & 0  & 0  & 0  & 0  & 0  &  1  \cr
  }.
\end{equation}
}


\section{Three Extended $S^2V$ Models}\label{sec:modelss2v2}

In this subclass of models we present a model with $32$ positively chiral
spinors of the first SO(10), $16$ coming from the first two planes and $32$
negatively chiral spinors of the second $SO(10)$, again $16$ coming from the
first two planes. Secondly we present a model that has four generations only of
the first $SO(10)$, $2$ coming from each of the first two planes and zero net
generations of the second $SO(10)$, $2$ positive chiral from the first and $2$
negative chiral from the second plane.

\noindent{\it The $\mathbbm{Z}_2\times \mathbbm{Z}_2$ symmetric orbifold}\\[5pt]
The simplest example is the symmetric $\mathbbm{Z}_2\times \mathbbm{Z}_2$ orbifold.  Here we set
all the free  GSO  phases \eqref{allps2v2} to zero. The full GSO phase matrix takes
the form ($\cc{v_i}{v_j}=\exp[i\pi(v_i|v_j)]$)
{\small\begin{equation}
(v_i|v_j)\ \ =\ \
\bordermatrix{
       & 1 & S  & e_1& e_2& e_3& e_4& e_5& e_6& b_1& b_2&  z_2\cr
   1   & 1 & 1  &  1 &  1 &  1 &  1 &  1 &  1 &  1 &  1 & 1\cr
   S   & 1 & 1  &  1 &  1 &  1 &  1 &  1 &  1 &  1 &  1 & 1\cr
  e_1  & 1 &  1 &  0 &  0 &  0 &  0 &  0 &  0 &  0 &  0 & 0\cr
  e_2  & 1 &  1 &  0 &  0 &  0 &  0 &  0 &  0 &  0 &  0 & 0\cr
  e_3  & 1 &  1 &  0 &  0 &  0 &  0 &  0 &  0 &  0 &  0 & 0\cr
  e_4  & 1 &  1 &  0 &  0 &  0 &  0 &  0 &  0 &  0 &  0 & 0\cr
  e_5  & 1 &  1 &  0 &  0 &  0 &  0 &  0 &  0 &  0 &  0 & 0\cr
  e_6  & 1 &  1 &  0 &  0 &  0 &  0 &  0 &  0 &  0 &  0 & 0\cr
  b_1  & 1 &  0 &  0 &  0 &  0 &  0 &  0 &  0 &  1 &  0 & 0\cr
  b_2  & 1 &  0 &  0 &  0 &  0 &  0 &  0 &  0 &  0 &  1 & 0\cr
  z_2  & 1 &  1 &  0 &  0 &  0 &  0 &  0 &  0 &  0 &  0 & 1\cr
}.
\end{equation}}\null
In this model we find $32$ generations of the observable $SO(10)$, $16$ coming
from each of the first two planes, and $32$ anti-generations of the hidden
$SO(10)$, $16$ coming from each of the first two planes.

\noindent{\it A four generation $SO(10)$ model}\\[5pt]
We can obtain a four generation $SO(10)$ model by choosing the following set of
projection coefficients
{\small
\begin{equation}\label{model2}
(v_i|v_j)\ \ =\ \
\bordermatrix{
       & 1 & S  &e_1 & e_2& e_3& e_4& e_5& e_6& b_1& b_2&z_2\cr
   1   & 1 & 1  &  1 &  1 &  1 &  1 &  1 &  1 &  1 &  1 & 1 \cr
   S   & 1 & 1  &  1 &  1 &  1 &  1 &  1 &  1 &  1 &  1 & 1 \cr
  e_1  & 1 &  1 &  0 &  1 &  0 &  0 &  0 &  1 &  0 &  0 & 1 \cr
  e_2  & 1 &  1 &  1 &  0 &  1 &  0 &  0 &  0 &  0 &  0 & 0 \cr
  e_3  & 1 &  1 &  0 &  1 &  0 &  0 &  0 &  1 &  0 &  0 & 0 \cr
  e_4  & 1 &  1 &  0 &  0 &  0 &  0 &  1 &  1 &  0 &  0 & 1 \cr
  e_5  & 1 &  1 &  0 &  0 &  0 &  1 &  0 &  1 &  0 &  0 & 1 \cr
  e_6  & 1 &  1 &  1 &  0 &  1 &  1 &  1 &  0 &  0 &  0 & 0 \cr
  b_1  & 1 &  0 &  0 &  0 &  0 &  0 &  0 &  0 &  1 &  0 & 0 \cr
  b_2  & 1 &  0 &  0 &  0 &  0 &  0 &  0 &  0 &  0 &  1 & 0 \cr
  z_2  & 1 &  1 &  0 &  0 &  0 &  0 &  0 &  0 &  0 &  0 & 1 \cr
  }.
\end{equation}
}


In this model we find two generations of the observable $SO(10)$ coming from
each of the first two planes and two generations of the hidden $SO(10)$ coming
from the first plane and two anti-generations of the hidden $SO(10)$ coming
from the second plane. This model is completely determined at the $N=4$ level. For more details on this property we refer to section \ref{sec:liftability}.
Therefore we do not find an enhancement of the $SO(8) \times U(1)\to SO(10)$
and the $SO(8)$ remains completely separate as explained in section \ref{sec:s2v2gauge}.

\noindent{\it A six generation $SO(10)$ model}\\[5pt]
We can obtain a six generation $SO(10)$ model by choosing the following set of
projection coefficients
{\small
\begin{equation}\label{model3}
(v_i|v_j)\ \ =\ \
\bordermatrix{
       & 1 & S  &e_1 & e_2& e_3& e_4& e_5& e_6& b_1& b_2&z_2\cr
   1   & 1 &  1 &  1 &  1 &  1 &  1 &  1 &  1 &  1 &  1 & 1 \cr
   S   & 1 &  1 &  1 &  1 &  1 &  1 &  1 &  1 &  1 &  1 & 1 \cr
  e_1  & 1 &  1 &  0 &  0 &  0 &  0 &  0 &  0 &  0 &  0 & 1 \cr
  e_2  & 1 &  1 &  0 &  0 &  1 &  0 &  0 &  1 &  0 &  0 & 0 \cr
  e_3  & 1 &  1 &  0 &  1 &  0 &  0 &  0 &  1 &  0 &  0 & 1 \cr
  e_4  & 1 &  1 &  0 &  0 &  0 &  0 &  1 &  1 &  0 &  0 & 0 \cr
  e_5  & 1 &  1 &  0 &  0 &  0 &  1 &  0 &  1 &  0 &  0 & 1 \cr
  e_6  & 1 &  1 &  0 &  1 &  1 &  1 &  1 &  0 &  0 &  0 & 0 \cr
  b_1  & 1 &  0 &  0 &  0 &  0 &  0 &  0 &  0 &  0 &  0 & 0 \cr
  b_2  & 1 &  0 &  0 &  0 &  0 &  0 &  0 &  0 &  0 &  1 & 0 \cr
  z_2  & 1 &  1 &  0 &  0 &  0 &  0 &  0 &  0 &  0 &  0 & 1 \cr
  }.
\end{equation}
}


In this model we find four generations of the observable $SO(10)$ on the first
plane and two generations of the observable $SO(10)$ on the second plane. We
find one generation of the hidden $SO(10)$ and one anti-generation of the
hidden $SO(10)$ on the first plane. This model is completely determined at the $N=4$ level. For more details on this property we refer to section \ref{sec:liftability}.
Therefore, we do not find spinorials of the third hidden $SO(10)$ on the third
plane as explained in section \ref{sec:s2v2gauge}.


\chapter{General Results}\label{chap:results}

In this chapter we give the general results of the classification. We show how this description uncovers a subset of models that have a well defined $N=4$ origin. We also highlight the mechanism that realises three generations. We discuss the method we have employed in the classification. We end this section with an overview of the characteristics of realistic models or semi-realistic models in the chiral $\mathbbm{Z}_2 \times \mathbbm{Z}_2$ free fermionic models with symmetric shifts.

\section{$N=4$ Liftable Vacua}\label{sec:liftability}

In the models considered in sections \ref{sec:modelss3}, \ref{sec:modelss2v1} and \ref{sec:modelss2v2} we have managed to separate the  orbifold
twist action (represented here by $b_1$, $b_2$) from
the shifts (represented by $e_i$) and the Wilson lines $(z_1,z_2)$. However,
these actions are further correlated
through the GSO projection coefficients $\cc{v_i}{v_j}$.  Nevertheless, we
remark that the twist action can be
 decoupled from the other two in the case
\begin{equation}
\cc{b_n}{z_k}=\cc{b_m}{e_i}=+1\ , i=1,\dots,6,\ k=1,2,\ m,n=1,2,3\label{lifte}
\end{equation}
for the $S^3$ models. The twist action can be decoupled from the other two actions in the case 
\begin{subequations}
\begin{equation}
\cc{b_m}{e_i}=+1\ , i=1,\dots,6,\ k=1,2,\ m,n=1,2,3
\end{equation}
for the simple $S^2V$ models or
\begin{equation}
\cc{b_n}{z_2}=\cc{b_m}{e_i}=+1\ , i=1,\dots,6,\  m,n=1,2,3
\end{equation}
\end{subequations}
for the extended $S^2V$ models.
The above relation defines a subclass of $N=1$ four dimensional vacua with
interesting phenomenological properties
and includes three generation models in the $S^3$ subclass of models. Due to the decoupling of the orbifold
twist action these vacua are direct descendants
of $N=4$ vacua so we refer to these models as $N=4$ lift-able models.
In this subclass of models  some important phenomenological properties of
the vacuum, as the number of generations, are
predetermined at the $N=4$ level as it is related to the
$\oo{e_i}{e_j}$ and $\oo{z_i}{e_j}$ phases in the $S^3$ subclass of models. In the $S^2V$ subclass of models they are related to the $\oo{e_i}{e_j}$ and also, in the case where we have introduced the $z_2$
vector, the $\oo{z_2}{e_j}$ phases.
The orbifold action reduces the super symmetries and the gauge group and
makes chirality apparent, however the number of generations is selected by
the $N=4$ vacuum structure. At the
$N=1$ level this is understood as follows: the $\mathbbm{Z}_2\times \mathbbm{Z}_2$ orbifold
has $48$ fix points in the $S^3$ subclass and $32$ fix points in the $S^2V$ subclass of models.
Switching on some of the above phase correspond to a free action that
removes some of the fixed points
and thus reduces the number of spinorials. Moreover, in the $S^3$ subclass of models,
the chirality of the surviving spinorials
is again related as seen by equations \eqref{chis3} to the $\oo{e_i}{e_j}$
and $\oo{e_i}{z_k}$ coefficients, which are
all fixed at the $N=4$ level.
The observable gauge group of liftable models is always $\mathbbm{E}_6$ and this can
be easily seen by applying \eqref{lifte} to \eqref{gb1} and \eqref{gb2}.
In the $S^2V$ subclass of models
the chirality of the surviving spinorials
is related as seen by \eqref{cs2v1} and \eqref{cs2v21} --
\eqref{cs2v22} to
the $\oo{e_i}{e_j}$ and $\oo{e_i}{z_k}$ coefficients, which are
all fixed at the $N=4$ level.

Typical examples of liftable vacua in the $S^3$ subclass of models are the three and six generation
$\mathbbm{E}_6\times {U(1)}^2\times {SO(8)}^2$ models presented
in section \ref{sec:modelss3}.
A careful counting, taking into account some symmetries among the
coefficients, shows that this class of models consists of $2^{20}$
models, or $2^{21}$ if we include $\oo{b_1}{b_2}$.
Typical examples of liftable vacua in the simple and extended $S^2V$ models
are presented in equation
\eqref{model1} and \eqref{model2} respectively.
A careful counting, taking into account some symmetries among the
coefficients, shows that the simple $S^2V$ class consists of $2^{10}$
models, or $2^{11}$ if we include $\oo{b_1}{b_2}$. The extended $S^2V$ class consists of $2^{15}$
models, or $2^{16}$ if we include $\oo{b_1}{b_2}$.
All these vacua are interesting because they have a clear geometrical
interpretation.

From the orbifold description we learn that all breakings of the
hidden and observable gauge group are induced using Wilson lines. From the
$4D$  point of view the internal gauge group is broken in a similar fashion
using Wilson lines. The twisted planes in equation \eqref{eq:twisted}
describe the
removal of the free moduli using twists. When a group is broken using Wilson
lines the field corresponding to this Wilson line obtains a nonzero VEV.
The fixing of
the moduli using twists can be interpreted as the removal of the quantum
fluctuations of the fields identified with the Wilson lines.
These Wilson lines
become discrete Wilson lines and the VEV becomes a fixed value.

In the extended $S^2V$ models we find that enhancement of $SO(8) \times U(1) \to
SO(10)$ is possible only at the $N=1$ level as is explained in section \ref{sec:s2v2gauge}. The models that exhibit this type
of enhancement are therefore not liftable to a $N=4$ theory.

\section{The Method of Classification}

As we discussed in chapter \ref{chap:fermform} and \ref{chap:models}, the free GSO phases of the $N=1$ partition
function control the number of chiral generations in a given
model. In chapter \ref{chap:fermform} we have given analytic formulas that enable
the calculation of the number of generations for any given set of
phases. To gain an insight into the structure of this class of vacua
we can proceed with a computer evaluation of these formulas and
thus classify the space of these vacua with respect to the number
of generations. This also allows detailed examination of the
structure of these vacua and in particular how the generations are
distributed among the three orbifold planes. The main obstacle to
this approach is the huge number of vacua under consideration. As
a first step in this direction, we restrict to the class
of liftable  vacua that is  physically appealing and contains
representative models with the right number of generations. 
As
stated in section \ref{sec:liftability} the liftable $S^3$ subclass consists in principle of $2^{21}$
models or less in the $S^2V$ subclass of models. 
Their complete classification  takes a few minutes on a personal computer
using an appropriate computer program.
The program is written in {\sc fortran} and goes over all GSO coefficients. 
It takes into account the constraints that are required for liftablility. For each configuration 
of GSO coefficients it uses equation \eqref{nf} for the $S^3$ models, equation \eqref{eq:nfs2v1} 
for the simple $S^2V$ models and equations \eqref{eq:nfs2v21} and  \eqref{eq:nfs2v22} for the 
extended $S^2V$ models to calculate the number of generations. The program gives for each plane 
the total number of generations and the 
total number of anti-generations. If this model has been found already by permuting the 
three twisted planes the model is dropped from the end result while if it is a new model, 
this model is recorded. At the end we have acquired a complete list of all models that are 
physically inequivalent. We have required in the program that the models exhibit $N=1$ supersymmetry.

We have first analysed the $S^3$ subclass of models.
The different configurations are  used to calculate the number of
generations
using formulae \eqref{ps3} -- \eqref{nf}. For the analysis
of the gauge group we use formulae \eqref{gb1} -- \eqref{gs3}.
The results are presented in Tables \ref{tab:real11} -- \ref{tab:real3}.
In these tables we list the number of generations coming from the twisted
sectors. They are listed per plane. The number of positive chiral
generations is separated from the number of negative chiral generations
on each plane. The total number is then listed before listing the total
net number of generations. As the sign of the chirality is determined by
the coefficient $\oo{b_1}{b_2}$, as is clear from equations \eqref{chis3}, we
have listed models that have a positive net number of generations.
In order to maintain a complete separation of the hidden gauge group we
have set $\oo{z_1}{z_2}=1$. The tables are ordered by the total net number
of chiral states.

We find that there are no liftable models with a $SO(10)$
observable gauge group, which is always extended to $\mathbbm{E}_6$,
and the states from the vector $x$ are not projected out.
Since the models admit a geometrical interpretation, it means that they
must descend from the ten dimensional  $\mathbbm{E}_8\times \mathbbm{E}_8$
heterotic--string on $\mathbbm{Z}_2\times \mathbbm{Z}_2$ Calabi--Yau threefold.

In a small number 
of all the models the hidden gauge group
is enhanced to $SO(8)\times SO(8) \to SO(16)$. We find that a fraction of the 
liftable models are enhanced to $SO(8)\times SO(8) \to \mathbbm{E}_8$. We find that
the 24 generations \nahe\ model as explained in section \ref{sec:general} is present
in table \ref{tab:real11}.

Since the total number
of different models is relatively small in the $S^2V$ subclass of models we can do a complete
classification of these models. The results of the classification of
the simple $S^2V$ models are listed in table \ref{tab:results}. The
models listed in chapter
\ref{chap:models} are liftable. We have listed these models due
to their clear $N=4$ origin. Since the extended $S^2V$ subclass of models do
not give phenomenologically interesting models we have not listed
the physically different models we find in this subclass. We do find
however that we cannot reduce the number of families down to $3$,
i.e. $2$ on one plane and $1$ on the second plane. We are able to
reduce the number of generations down to $4$, or two on each plane,
but at the cost of introducing another $SO(10)$ gauge group with
spinorials on the first two planes. In the subclass of liftable extended $S^2V$  models we 
find that it is not possible to have only spinorial
representations of the first $SO(10)$ in the massless spectrum. An
example of the models which have net zero generations from the
second $SO(10)$ is given in equation \eqref{model2}.

\section{Reduction of Generations and the Internal Manifold}

In chapter \ref{chap:orbifold} we discussed a direct translation between the
bosonic
formulation and the fermionic formulation of the heterotic string
compactifications. $\mathbbm{Z}_2\times \mathbbm{Z}_2$ orbifold
compactifications are relevant for this thesis.
These orbifolds contain three twisted
sectors, or three twisted planes. A priori we may have the
possibility that all three twisted planes produce spinorial
$SO(10)$ representations. We refer to this subclass of models
as $S^3$ models. The alternatives are models in which spinorial
representations may be obtained from only two, one, or none of the  twisted
planes, and the others produce vectorial representations. We refer
to these cases as $S^2V$, $SV^2$ and $V^3$ models, respectively.
The \nahe\ set is an element of the $S^3$ subclass of the chiral $\mathbbm{Z}_2 \times \mathbbm{Z}_2$ models. 
The $S^3$ subclass allows, depending
on the one--loop GSO projection coefficients, the possibility
of spinorials on each plane. In specific models in this subclass
each Standard Model family is obtained from a distinct orbifold
plane. Such models therefore produce three generation models
and may be phenomenologically interesting.
The only other phenomenologically viable option can intuitively come from the subclass
$S^2V$ models as this class of models may intuitively contain a model with for example $2$
generations coming from the first plane and $1$ generation coming from the
second plane and none from the third. However, we find that the $S^2V$ subclass of models do not give 
phenomenologically interesting models as is seen in chapter \ref{chap:fermform} and \ref{chap:models}. 
The $SV^2$ class of models cannot
produce a
physical model because it is not possible to reduce the number of families to
$3$ as they would have to be coming from one plane and $3$ cannot be written
as a power of $2$. Similarly the $V^3$ subclass of models will not contain
phenomenologically interesting models.

Since the projectors are constructed using the complete separation of the
internal manifold we see that three generation models are only possible when
\begin{equation}
\Gamma_{6,6}\  =\  \Gamma_{2,2}^3\ \  \longrightarrow\ \  \Gamma_{1,1}^6.
\end{equation}
These $\Gamma_{1,1}$ internal parts do not describe a complex manifold. They
describe internal real circles. If we use solely complex manifolds,
of the type $\Gamma_{6,6}=\Gamma_{2,2}^3$, and using only
symmetric shifts, we find that
there are no $3$ generation models.
We therefore conclude that the net number of generations can
never be equal to three
in the framework of $\mathbbm{Z}_2\times \mathbbm{Z}_2$ Calabi--Yau compactifications.
This implies the necessity of non zero torsion in CY $\mathbbm{Z}_2\times \mathbbm{Z}_2$
compactifications in order to obtain semi--realistic three
generation models. This is one of the main results of the analysis and it reveals, at least in the context of the three generation free fermionic models, that the geometrical structures that underly these models may not be simple Calabi--Yau manifolds, but it corresponds to geometries that are yet to be defined.

In the realistic free fermionic
models the reduction of the number of families together with
the breaking of the observable $SO(10)$ is realized by isolating full
multiplets at two fixed points on the internal manifold.
In reducing the number of families
down to one, different component of each family are attached
to the two distinct fixed points.
We remove one full multiplet and simultaneously break the
$SO(10)$ symmetry. We
therefore keep a full multiplet on each twisted plane.
In the $SO(10)$ models described here a whole $16$ or $\overline{16}$
of $SO(10)$ is attached to one fixed point.
We are therefore not able to break the $SO(10)$, and simultaneously
preserve the full Standard Model multiplets.
For this reason we find that the
observable $SO(10)$ cannot be broken perturbatively
in this class of three generation models, and may only
be broken non perturbatively.
It is therefore not possible to reduce both the number of
families down to $3$ and break the observable gauge group $SO(10)$ down
to its subgroups perturbatively.

We conclude that there is a method to reduce the number of generations from
$48$ to $3$. Since we need $4$ projectors we need to separate the hidden gauge
group using $SO(8)$ characters
\begin{equation}
\Gamma_{0,8} \to \Gamma_{0,4}\ \Gamma_{0,4}
\end{equation}
and we need to break the internal complex manifold to an internal
real manifold
\begin{equation}
\Gamma_{6,6} \to \left[ \Gamma_{1,1}\ \Gamma_{1,1} \right]^3.
\end{equation}
If we reduce the number of generations to $3$ we cannot break the $SO(10)$
observable group to its subgroups, while maintaining a full multiplet. The
$SO(10)$
observable gauge group cannot therefore be broken perturbatively.
We can reduce the number of generations from $48$ to $6$ using $3$ projectors.
This entails that we can choose either to separate the hidden gauge group
using $SO(16)$ characters, or to leave the internal manifold complex.

We argued above that we cannot break $SO(10)$ down to
a subgroup perturbatively, while reducing the number of generations
to $3$.
If we want to break the $SO(10)$ symmetry perturbatively,
and keep a full $SO(10)$ multiplet from a given twisted plane,
we can only reduce the number of generations to $6$.
This can be achieved if we define three different
projectors like the ones defined in equations \eqref{ps3}.
We are therefore left with two options.
First, we can use $SO(16)$ characters for the separation of the hidden gauge
group. We have then constructed only one projector which leaves us no other
 option
 than to break the complex structure using symmetric shifts
\begin{equation}
\Gamma_{6,6} \to \Gamma_{1,1}^6.
\end{equation}
Second, we can use $SO(8)$ characters for the separation of the hidden gauge
group. In doing so we have constructed two projectors. The third can be
realized
by the symmetric shifts that leave the complex structure of the internal
manifold
 intact
\begin{equation}
\Gamma_{6,6} \to \Gamma_{2,2}^3.
\end{equation}

We have therefore concluded the classification of the chiral 
$\mathbbm{Z}_2 \times \mathbbm{Z}_2$ free fermionic models with symmetric 
shifts. In this chapter we have shown how liftable models appear in the 
models discussed.  We have elaborated on the method we have used to 
classify all the models. Although we have not listed all models of the 
$S^3$ subclass we are able to derive some general properties of these 
additional models. We have shown that in this class of models we do not 
find realistic three generation models with a Calabi-Yau compactification.


\chapter{Discussion}\label{chap:conclusions}

In this chapter we give a short summary of the issues discussed in this thesis. We end with a discussion on the realistic free fermionic models and give some suggestions for further research.

In this thesis we have classified the chiral content of the heterotic $\mathbbm{Z}_2 \times \mathbbm{Z}_2$ orbifolds. This classification was possible since we can choose \emph{any} point in the moduli space to describe the chiral content as explained in section \ref{sec:orbifoldsandshifts}. This choice gave us the opportunity to describe the chiral content of the heterotic $\mathbbm{Z}_2 \times \mathbbm{Z}_2$ orbifolds in the free fermionic construction. In this thesis we presented a direct translation between symmetric shifts in the orbifold construction to symmetric shifts in the free fermionic construction. As the hidden $\mathbbm{E}_8$ gauge group is broken to $SO(8) \times SO(8)$ in the realistic free fermionic models we have included this in our classification. The classification of the chiral content therefore includes all models where symmetric shifts are realised on the internal space and where the hidden gauge group is broken to $SO(8) \times SO(8)$ at \emph{any} point in the moduli space. 

We have started by giving an overview of the general construction of string models in chapter \ref{chap:heterotic.setup}. We gave an introduction to the orbifold description of string theory in chapter \ref{chap:orbifold.setup}. We set the string on a circle after which we expanded to the general manifold. We then introduced shifts and ended this chapter with a discussion of $\mathbbm{Z}_2$ twists. In this chapter we have shown that the twisted sector \emph{does not} depend on the moduli of the internal space. This is an essential ingredient that has allowed this classification. In chapter \ref{chap:fermionic} we constructed the partition function of the heterotic string. We isolated some constraints on the partition function due to modular invariance of the theory. This lead to the formulation of the free fermionic string. Since bosons can only be interchanged with free fermions at the maximal symmetry point, the target space in the free fermionic construction is inherently at the maximal symmetry point in the moduli space. In chapter \ref{chap:realistic} we discuss the practicalities of constructing a free fermionic model and its spectrum. We isolate some sectors that are relevant for the remainder of the thesis. Two semi-realistic models are discussed in detail. 

In chapter \ref{chap:orbifold} we use the orbifold description developed in chapter \ref{chap:orbifold.setup} to isolate the free parameters of the $\mathbbm{Z}_2 \times \mathbbm{Z}_2$ orbifold with symmetric shifts. We discuss the translation of the orbifold description to the free fermionic description. We have specified the basis vectors for the free fermionic formulation that describe a shift on the internal space. We have shown that the basis vectors that induce the twist on the internal space are completely fixed by requiring spinorial representations. In chapter \ref{chap:fermform} we use the free fermionic language to derive formulas that describe the chiral content of the $\mathbbm{Z}_2 \times \mathbbm{Z}_2$ orbifold with symmetric shifts using the methods we explained in chapter \ref{chap:realistic}. Using these formulas we have written a computer program in {\sc fortran} to analyse all heterotic $\mathbbm{Z}_2 \times \mathbbm{Z}_2$ orbifold models. Using this computer program we gave some sample models in chapter \ref{chap:models}. In chapter \ref{chap:results} we discussed the general results we obtained from the classification.

In chapter \ref{chap:realistic} we have discussed two semi-realistic models. Similarly realistic models can be constructed using the \nahe\ set basis vectors. All these models have one feature in common, which is their $\mathbbm{Z}_2 \times \mathbbm{Z}_2$ orbifold origin. In chapter \ref{chap:results} we came to the conclusion that there are no three generation models with a $\mathbbm{Z}_2 \times \mathbbm{Z}_2$ Calabi-Yau space. The realistic free fermionic models do however exhibit three generations. We have shown in section \ref{sec:general} that the last vector $\gamma$ that realises the reduction of the number of generations to three, cannot be written as a linear combination of symmetric shifts. This vector necessarily induces a different type of reduction than the one we have described in this thesis.

In the free fermionic formulation the moduli of the internal manifold are all fixed to the maximal symmetry point in the moduli space. In the orbifold description we have left all the moduli untouched. It may therefore be the case that the vector that breaks the observable $SO(10)$ and simultaneously reduces the number of generations to three requires the moduli to be fixed at the maximal symmetry point. Research in this direction may prove very interesting and may shed some light on the dynamical mechanism that singles out the vacuum we observe in nature. 

In this thesis we have focused on the chiral content of the $\mathbbm{Z}_2 \times \mathbbm{Z}_2$ fermionic models of the heterotic string. It is fairly easy to extend this analysis to the vectorial sector. Similarly the total $U(1)$ charge can be calculated as a function of the free GSO coefficients. In chapter \ref{chap:fermform} we have isolated the GSO coefficients that completely determine the chiral content. It is therefore believed that the remaining coefficients determined the other sectors of the model. Furthermore, as we have explained, some very interesting results are to be expected from analysing the dual descriptions of the free fermionic formulation of the heterotic string. This should be possible as the complete structure of partition function is known at the level of the $\mathbbm{Z}_2 \times \mathbbm{Z}_2$ orbifold with symmetric shifts. Likewise we can analyse threshold corrections, helicity supertraces and other properties of the semi-realistic models.

Having given an overview of the thesis together with some suggestions for future work, we have come to a closure.

	\part{Appendix}
		\appendix
			\renewcommand{\chaptername}{Appendix}
			\renewcommand{\sectionname}{Appendix}
\refstepcounter{chapter}\label{app:spectrum}
\addcontentsline{toc}{chapter}%
	{\protect\numberline{\thechapter}The Spectrum of the Two Advanced Models}
\refstepcounter{section}\label{sec:anomalyfree}
\addcontentsline{toc}{section}%
	{\protect\numberline{\thesection}The Spectrum of the Model Without Enhanced Symmetry}
\lhead{}\chead{}\rhead{{\sc \sectionname\ \thesection.\ The Spectrum of the Model Without Enhanced Symmetry}}
\noindent\null\clearpage\addtocounter{page}{-2}
\begin{table}[H]
\begin{center}
\begin{sideways}
\begin{tabular}{|c|c|c|rrrr|c||c|rr|}\hline
 SEC 				  &   $SU(4) \times SU(2)$	  
& $Q_{R}$ & $Q_1$ & $Q_2$ & $Q_4$ & $Q_5$ &
$SU(2)_{3} \times SU(2)_{6}$ & $SU(4)_H \times SU(2)_{H_3} \times$ & 
$Q_{7}$ & $Q_{8}$ \\
   				  &   		  &   	  &   	  
&   	  &   	  &   	  &    &
$SU(2)_{H_1} \times SU(2)_{H_2} $	  &   	  &   	\\\hline
  $b_1$ 			  &   $(\bar{4},1)$	  &   4	  &   -2  
&   0	  &   -2  &   0	  &
$(1,1)$	  &   $(1,1,1,1)$	  &   0	  &   0	\\
   				  &   $(\bar{4},1)$	  &   4	  &   2	  
&   0	  &   2	  &   0	  &
$(1,1)$	  &   $(1,1,1,1)$	  &   0	  &   0	\\
   				  &   $(\bar{4},1)$	  &   -4  &   -2  
&   0	  &   -2  &   0	  &
$(1,1)$	  &   $(1,1,1,1)$	  &   0	  &   0	\\
   				  &   $(\bar{4},1)$	  &   -4  &   2	  
&   0	  &   2	  &   0	  &
$(1,1)$	  &   $(1,1,1,1)$	  &   0	  &   0	\\\hline

  $b_2$ 			  &   $(\bar{4},1)$	  &   4	  &   0	  
&   -2  &   0	  &   2	  &
$(1,1)$	  &   $(1,1,1,1)$	  &   0	  &   0	\\
   				  &   $(\bar{4},1)$ 	  &   4	  &   0	  
&   2	  &   0	  &   -2  &
$(1,1)$	  &   $(1,1,1,1)$	  &   0	  &   0	\\
   				  &   $(\bar{4},1)$	  &   -4  &   0	  
&   -2  &   0	  &   2	  &
$(1,1)$	  &   $(1,1,1,1)$	  &   0	  &   0	\\
   				  &   $(\bar{4},1)$ 	  &   -4  &   0	  
&   2	  &   0	  &   -2  &
$(1,1)$	  &   $(1,1,1,1)$	  &   0	  &   0	\\\hline

  $b_3$ 			  &   $(4,2)$	  &   0	  &   0	  &   0	  
&   0	  &   0	  &
$(1,2)$	  &   $(1,1,1,1)$	  &   0	  &   0	\\\hline

  $S + b_1 + b_2 + $		  &   $(1,1)$	  &   0	  &   2  
&   -2	  &   0	  &   0
  &   $(1,1)$	  &   $(1,1,1,2)$	  &   0	  &   0	\\
  $\alpha + \beta$		  &   $(1,1)$	  &   0	  &   -2 &   2	  
&   0	  &   0	  &
 $(1,1)$	  &   $(1,1,2,1)$	  &   0	  &   0	\\
			 	  &   $(1,1)$	  &   0	  &   -2 &   2	  
&   0	  &   0	  &   $(1,1)$	  &
$(1,1,1,2)$	  &   0	  &   0	\\
				  &   $(1,1)$	  &   0	  &   2  &   -2   
&   0	  &   0	  &   $(1,1)$	  &
$(1,1,2,1)$	  &   0	  &   0	\\\hline

  $b_3 + \beta + 2\gamma$	  &   $(1,1)$	  &   0	  &   0	  &   0	  
&   -2  &
  2	  &   $(2,1)$	  &   $(1,1,1,1)$	  &   4	  &   0	\\
   				  &   $(1,1)$	  &   0	  &   0	  &   0	  
&   2	  &   2	  &   $(1,2)$	  &
 $(1,1,1,1)$	  &   4	  &   0	\\
   				  &   $(1,1)$	  &   0	  &   0	  &   0	  
&   -2  &   -2  &   $(1,2)$	  &
 $(1,1,1,1)$	  &   4	  &   0	\\
   				  &   $(1,1)$	  &   0	  &   0	  &   0	  
&   2	  &   -2  &   $(2,1)$	  &
 $(1,1,1,1)$	  &   4	  &   0	\\
  				  &   $(1,1)$	  &   0	  &   0	  &   0	  
&   -2  &   2	  &   $(2,1)$	  &
$(1,1,1,1)$	  &   -4  &   0	\\
   				  &   $(1,1)$	  &   0	  &   0	  &   0	  
&   2	  &   2	  &   $(1,2)$	  &
 $(1,1,1,1)$	  &   -4  &   0	\\
   				  &   $(1,1)$	  &   0	  &   0	  &   0	  
&   -2  &   -2  &   $(1,2)$	  &
 $(1,1,1,1)$	  &   -4  &   0	\\
   				  &   $(1,1)$	  &   0	  &   0	  &   0	  
&   2	  &   -2  &   $(2,1)$	  &
 $(1,1,1,1)$	  &   -4  &   0	\\\hline

  $S + 2\gamma$			  &   $(1,1)$	  &   4	  &   0	  
&   0	  &   0	  &   0	  &
$(1,1)$	  &   $(4,2,1,1)$	  &   0	  &   4	\\
  				  &   $(1,1)$	  &   -4  &   0	  &   0	  
&   0	  &   0	  &   $(1,1)$	  &
$(\bar{4},2,1,1)$	  &   0	  &   -4\\\hline

  $1 + S + b_3 +$	  &   $(1,2)$	  &   0	  &   2	  &   2	  &   0	  
&   0	  &
$(1,1)$	  &   $(1,1,2,1)$	  &   0	  &   0	\\
  $ \alpha + \beta + 2\gamma$	  &   $(1,2)$	  &   0	  &   -2  
&   -2  &   0
 &   0	  &   $(1,1)$	  &   $(1,1,2,1)$	  &   0	  &   0	\\\hline

  $1 + b_1 + $	  &   $(4,1)$	  &   0	  &   0	  &   0	  &   0	  
&   0	  &
$(2,1)$	  &   $(1,1,1,2)$	  &   0	  &   0	\\
  $b_2 + 2\gamma$		  &   		  &   	  &   	  &   	  
&   	  &   	  &   		  &
	 	  &   	  &   	\\\hline

  $1 + S + b_1 +$	  &   $(1,2)$	  &   0	  &   0	  &   0	  &   4	  
&   0	  &
$(1,1)$	  &   $(1,1,1,2)$	  &   0	  &   0	\\
  $b_2 + b_3 + 2\gamma$		  &   $(1,2)$	  &   0	  &   0	  
&   0	  &   -4  &
  0	  &   $(1,1)$	  &   $(1,1,1,2)$	  &   0	  &   0	\\\hline
\end{tabular}
\end{sideways}
\caption{The spectrum of the model described in table \ref{tab:su421without} and matrix \eqref{phasesmodel1}, which does not exhibit an enhanced symmetry.}
\end{center}
\end{table}
\newpage

\refstepcounter{section}\label{sec:anomalous}
\addcontentsline{toc}{section}%
	{\protect\numberline{\thesection}The Spectrum of the Model With Enhanced Symmetry}
\lhead{}\chead{}\rhead{{\sc \sectionname\ \thesection.\ The Spectrum of the Model With Enhanced Symmetry}}
\noindent
\begin{table}[H]
\begin{center}
\begin{sideways}
\begin{tabular}{|c|c|r|rrrrrr||c|rr|}\hline
 SEC	\& 	 	 &  $SU(4)$  & $Q_{R}$ & $Q_1$ & $Q_2$ & $Q_3$ & 
$Q_4$ & $Q_5$ &
$Q_6$ &  $SU(4)_H \times SU(2)_{H_3} \times$ 	  &  $Q_{7}$	 &  $Q_{8}$
	\\
Field 				 &$\times SO(5)$  &   	  &   	  &   	  &
&   	  &   	  &   	  &  $SU(2)_{H_2}$ 	 &   	 & 	\\
\hline
  $\bf 0$:                 &  &  &  &  &  &  &  &  & & & \\
  $\phi_1$, ${\bar\phi}_1$ & (1,1) &  0  &  0   & $\mp 1$ & $\mp 1$  &  0 
&  0
&  0 & $(1,1,1)$ & 0 & 0\\
  $\phi_2$, ${\bar\phi}_2$ & (1,1) &  0  &  0   & $\pm 1$ & $\mp 1$  &  0 
&  0
&  0 & $(1,1,1)$ & 0 & 0\\
  $\phi_3$, ${\bar\phi}_3$ & (1,1) &  0  &$\mp 1$&  0     & $\mp 1$  &  0 
&  0
&  0 & $(1,1,1)$ & 0 & 0\\
  $\phi_4$, ${\bar\phi}_4$ & (1,1) &  0  &$\pm 1$&  0     & $\mp 1$  &  0 
&  0
&  0 & $(1,1,1)$ & 0 & 0\\
  $\phi_5$, ${\bar\phi}_3$ & (1,1) &  0  &$\mp 1$& $\mp 1$&  0       &  0 
&  0
&  0 & $(1,1,1)$ & 0 & 0\\
  $\phi_6$, ${\bar\phi}_6$ & (1,1) &  0  &$\mp 1$& $\pm 1$&  0       &  0 
&  0
&  0 & $(1,1,1)$ & 0 & 0\\
\hline
  $b_1$ & $(\bar{4},1)$ &  -4 	  &  2 	  &  0 	  &  0 	  &  2 	  
&  0
&  0 	  &  $(1,1,1)$ 	 	 &  0 	 &  0	\\
   				 & $(\bar{4},1)$ &  4 	  &  2 	  &  0 	  
&  0 	  &  -2	  &  0 	  &  0
&  $(1,1,1)$ 	 	 &  0 	 &  0	\\
   				 & $(\bar{4},1)$ &  4 	  &  -2	  &  0 	  
&  0 	  &  2 	  &  0 	  &  0
&  $(1,1,1)$ 	 	 &  0 	 &  0	\\
 & $(\bar{4},1)$ &  -4 	  &  -2	  &  0 	  &  0 	  &  -2	  &  0 	  &  0
 &  $(1,1,1)$ 	 	 &  0 	 &  0	\\ \hline

  $b_2$ 			 & $(\bar{4},1)$ &  4 	  &  0 	  &  -2	  
&  0 	  &  0 	  &  -2	  &   0   &  $(1,1,1)$ 	  &  0 	 &  0	\\
 & $(\bar{4},1)$ &  -4 	  &  0 	  &  -2	  &  0 	  &  0 	  &  2 	  &  0
 &  $(1,1,1)$ 	  &  0 	 &  0	\\
& $(\bar{4},1)$ &  -4 	  &  0 	  &  2 	  &  0 	  &  0 	  &  -2	  &  0
 &  $(1,1,1)$ 	  &  0 	 &  0	\\
  & $(\bar{4},1)$ &  4 	  &  0 	  &  2 	  &  0 	  &  0 	  &  2 	  &  0
&  $(1,1,1)$ 	  &  0 	 &  0	\\\hline

  $b_3 \oplus b_3 + $	 &  $(4,4)$  	  &  0 	  &  0 	  
   &  0 &  2 	  &  0 	  &  0 	  &  -2	  &  $(1,1,1)$ 	  &  0 	 &  0	\\
$\zeta+ 2\gamma $&  $(4,4)$ 	  &  0 	  &  0 	  &  0 	  &  2 	  &  0 	  &  0 	  &  2 	  &
 $(1,1,1)$ 	  &  0 	 &  0	\\\hline
$S + 2\gamma$:			 &  $(1,1)$ 	  &  -4	  &  0 	  &  0 
&  0 	  &  0 	  &  0 &  0 	  &  $(6,1,1)$ 	  &  4	 &  0	\\
&  $(1,1)$ 	  &  4	  &  0 	  &  0 	  &  0 	  &  0 	  &  0 	  &  0 	 
& $(6,1,1)$ 	  &  -4	 &  0	\\
$S_1$, ${\bar S}_1$		 &  $(1,1)$ 	  & $\mp 4$	  &  0 	  
&  0 	  &  0 	  &  0 &  0 	  &  0 	  &  $(1,1,1)$ 	  & $\mp 4$ & 
$\pm 8$	\\
$S_2$, ${\bar S}_2$              &  $(1,1)$ 	  & $\mp 4$	  &  0 	  
& 0 	  &  0
  &  0 	  &  0 	  &  0 	  &
   				 $(1,1,1)$ 	  & $\mp 4$	 & $\mp 8$\\
\hline
\end{tabular}
\end{sideways}
\caption{The spectrum of the model described in table \ref{tab:su421with} and matrix \eqref{phasesmodel2}, which does exhibit an enhanced symmetry.}\label{tab:spectrumsu421with1}
\end{center}
\end{table}

\noindent
\begin{table}[H]
\begin{center}
\begin{sideways}
\begin{tabular}{|c|c|r|rrrrrr||c|rr|}\hline
 SEC	\& 	 	 &  $SU(4)$  & $Q_{R}$ & $Q_1$ & $Q_2$ & $Q_3$ & 
$Q_4$ & $Q_5$ &
$Q_6$ &  $SU(4)_H \times SU(2)_{H_3} \times$ 	  &  $Q_{7}$	 &  $Q_{8}$
	\\
Field 				 &$\times SO(5)$  &   	  &   	  &   	  &
&   	  &   	  &   	  &  $SU(2)_{H_2}$ 	 &   	 & 	\\
\hline
  $b_2 + \alpha + 2\gamma$&  &  &  &  &  &  &  &  & & & \\
 $S_3$, ${\bar S}_3$  				 &  $(1,1)$	  &  0 	  
&0 	  & $\mp 2$& 0 	  &$\mp
2$&$\mp 2$&$\mp 2$&  $(1,1,1)$ 	  &$\mp 4$&  0	\\
 $S_4$, ${\bar S}_4$  				 &  $(1,1)$	  &  0 	  
&0 	  & $\mp 2$& 0 	  &$\mp
2$&$\pm 2$&$\mp 2$&  $(1,1,1)$ 	  &$\pm 4$&  0	\\
  $S_5$, ${\bar S}_5$  				 &  $(1,1)$	  &  0 	  
&0 	  & $\mp 2$& 0 	  &$\mp
2$&$\mp 2$&$\pm 2$&  $(1,1,1)$ 	  &$\mp 4$&  0	\\
  $S_6$, ${\bar S}_6$ 				 &  $(1,1)$	  &  0 	  
&0 	  & $\mp 2$& 0 	  &$\mp
2$&$\pm 2$&$\pm 2$&  $(1,1,1)$ 	  &$\pm 4$&  0	\\
  $S_7$, ${\bar S}_7$  				 &  $(1,1)$	  &  0 	  
&0 	  & $\mp 2$& 0 	  &$\pm
2$&$\mp 2$&$\mp 2$&  $(1,1,1)$ 	  &$\pm 4$&  0	\\
  $S_8$, ${\bar S}_8$   	                 &  $(1,1)$ 	  &  0 	  
&0 	  & $\mp
2$& 0 	  &$\pm 2$&$\pm 2$&$\mp 2$&  $(1,1,1)$ 	  &$\mp 4$&  0	\\
  $S_9$, ${\bar S}_9$  				 &  $(1,1)$	  &  0 	  
&0 	  & $\mp 2$& 0 	  &$\pm
2$&$\mp 2$&$\pm 2$&  $(1,1,1)$ 	  &$\pm 4$&  0	\\
 $S_{10}$, ${\bar S}_{10}$			 &  $(1,1)$	  &  0 	  
&0 	  & $\mp 2$& 0 	  &$\pm
2$&$\pm 2$&$\pm 2$&  $(1,1,1)$ 	  &$\mp 4$&  0	\\
\hline
  $S + b_2 + $		 &  $(1,1)$ 	  &  2 	  &  0 	  
&  2 	  &  2 	  &  0
 &  -2	  &  0 	  &  $(4,1,1)$ 	  &  0 	 &  0	\\
  $b_3 + \alpha + $	 &  $(1,1)$	  &  2 	  &  0 	  
&  -2	  &  2
&  0 	  &  2 	  &  0 	  &  $(4,1,1)$ 	  &  0 	 &  0	\\

 $\beta \pm \gamma$  				 &  $(1,1)$	  &  2 	  &  0 	  &  -2	  
&  2 	  &  0 	  &  2 	  &  0 	  &
$(1,2,1)$ 	  &  -2	 &  0	\\
   				 &  $(1,1)$	  &  2 	  &  0 	  &  2 	  
&  2 	  &  0 	  &  -2	  &  0 	  &
$(1,2,1)$ 	  &  -2	 &  0	\\

  				 &  $(1,1)$ 	  &  -2	  &  0 	  &  -2	  
&  2 	  &  0 	  &  -2	  &  0 	  &
$(\bar{4},1,1)$ 	  &  0 	 &  0	\\
   				 &  $(1,1)$	  &  -2	  &  0 	  &  2 	  
&  2 	  &  0 	  &  2 	  &  0 	  &
$(\bar{4},1,1)$ 	  &  0 	 &  0	\\

   				 &  $(1,1)$	  &  -2	  &  0 	  &  2 	  
&  2 	  &  0 	  &  2 	  &  0 	  &
$(1,2,1)$ 	  &  2 	 &  0	\\
   				 &  $(1,1)$	  &  -2	  &  0 	  &  -2	  
&  2 	  &  0 	  &  -2	  &  0 	  &
$(1,2,1)$ 	  &  2 	 &  0	\\ \hline

\end{tabular}
\end{sideways}
\caption{Table \ref{tab:spectrumsu421with1} continued}\label{tab:spectrumsu421with2}
\end{center}
\end{table}

\noindent
\begin{table}[H]
\begin{center}
\begin{sideways}
\begin{tabular}{|c|c|r|rrrrrr||c|rr|}\hline
 SEC				 &  $SU(4) \times $ 	  & $Q_{R}$ 
& $Q_1$ & $Q_2$ & $Q_3$ & $Q_4$ &
$Q_5$ & $Q_6$ &  $SU(4)_H \times SU(2)_{H_3}$ 	  &  $Q_{7}$	 
&  $Q_{8}$
\\
   				 &  $SO(5)$ 	  &   	  &   	  &   	  &   	  
&   	  &   	  &   	  &
$SU(2)_{H_2}$ 	  &   	 & 	\\
\hline
  $S + b_1 + b_3 + $		 &  $(1,1)$ 	  &  2 	  &  -2   
&  0 	  &  2 	  &  -2
 & 0	&  0 	  &  $(4,1,1)$ 	  &  0 	 &  0	\\
  $\alpha + \beta \pm \gamma$	 &  $(1,1)$	  &  2 	  &  2 	  
&  0 	  &  2
&  2 	  & 0	&  0 	  &  $(4,1,1)$ 	  &  0 	 &  0	\\

   				 &  $(1,1)$	  &  2 	  &  2    &  0	  
&  2 	  &  2 	  & 0	&  0 	  &
$(1,2,1)$ 	  &  -2	 &  0	\\
   				 &  $(1,1)$	  &  2 	  &  -2   &  0	  
&  2 	  &  -2	  & 0	&  0 	  &
$(1,2,1)$ 	  &  -2	 &  0	\\

  				 &  $(1,1)$ 	  &  -2	  &  2 	  &  0 	  
&  2 	  &  -2	  & 0	&  0 	  &
$(\bar{4},1,1)$ 	  &  0 	 &  0	\\
   				 &  $(1,1)$	  &  -2	  &  -2   &  0 	  
&  2 	  &  2 	  & 0	&  0 	  &
$(\bar{4},1,1)$ 	  &  0 	 &  0	\\

   				 &  $(1,1)$	  &  -2	  &  -2   &  0 	  
&  2 	  &  2 	  & 0	&  0 	  &
$(1,2,1)$ 	  &  2 	 &  0	\\
   				 &  $(1,1)$	  &  -2	  &  2    &  0	  
&  2 	  &  -2	  & 0	&  0 	  &
$(1,2,1)$ 	  &  2 	 &  0	\\\hline

  $S + b_1 + b_2 + $		 &  $(1,4)$ 	  &  0 	  &  -2   
&  -2	  &  0 	  &  0
 &  0 	  &  0 	  &  $(1,1,2)$ 	  &  0 	 &  0	\\
  $\alpha + \beta \ \oplus $	 &  	 	  &   	  &   	  &   	  
&   	  &   	  &
  &   	  &  	 	  &   	 &  	\\
  $S + b_1 +b_2 + $		 &  $(1,4)$	  &  0 	  &  2	  
&  2	  &  0 	  &
0 	  &  0 	  &  0	  &  $(1,1,2)$ 	  &  0 	 &  0	\\
  $\alpha + \beta+\zeta+2\gamma$	 &  	 	  &   	  &   	  &   	  
&   	  &   	  &
&   	  &  	 	  &   	 &  	\\ \hline

  $1 + b_1 + $	 &  $(1,1)$ 	  &  2 	  &  0 	  &  2	  & 2	  &  0 	  
&  -2	  &  0
	  &  $(1,1,2)$ 	  &  0 	 &  0	\\
  $\alpha + \beta \pm \gamma$	 &  $(1,1)$	  &  2 	  &  0 	  
&  -2	  & 2	  &
0 	  &  2 	  &  0 	  &  $(1,1,2)$ 	  &  0 	 &  0	\\
   				 &  $(1,1)$	  &  -2	  &  0 	  &  -2	  
& 2	  &  0 	  &  -2	  &  0 	  &
$(1,1,2)$ 	  &  0 	 &  0	\\
   				 &  $(1,1)$	  &  -2	  &  0 	  &  2	  
& 2	  &  0 	  &  2 	  &  0 	  &
$(1,1,2)$ 	  &  0 	 &  0	\\ \hline

  $1 + b_2 + $	 &  $(1,1)$ 	  &  2 	  &  2	  &  0 	  & 2	  &  2 	  
&  0 	  &  0
	  &  $(1,1,2)$ 	  &  0 	 &  0	\\
  $\alpha + \beta \pm \gamma$	 &  $(1,1)$	  &  2 	  &  -2	  
&  0 	  & 2	  &
-2	  &  0 	  &  0 	  &  $(1,1,2)$ 	  &  0 	 &  0	\\
   				 &  $(1,1)$	  &  -2	  &  -2	  &  0 	  
& 2	  &  2 	  &  0 	  &  0 	  &
$(1,1,2)$ 	  &  0 	 &  0	\\
   				 &  $(1,1)$	  &  -2	  &  2	  &  0 	  
& 2	  &  -2	  &  0 	  &  0 	  &
$(1,1,2)$ 	  &  0 	 &  0	\\\hline
\end{tabular}
\end{sideways}
\caption{Table \ref{tab:spectrumsu421with2} continued}\label{tab:spectrumsu421with3}
\end{center}
\end{table}
	


\refstepcounter{chapter}\label{app:tables}
\addcontentsline{toc}{chapter}%
	{\protect\numberline{\thechapter}Tables of the Classified Models}
\refstepcounter{section}
\addcontentsline{toc}{section}%
	{\protect\numberline{\thesection}Tables of the $S^3$ Models}
\lhead{}\chead{}\rhead{{\sc \sectionname\ \thesection.\ Tables of the $S^3$ Models}}
\noindent
\begin{table}[H]
\begin{center}
\begin{tabular}{r || r r | r r | r r | r r | r }
&\multicolumn{2}{c|}{$1$}&\multicolumn{2}{c|}{$2$}&\multicolumn{2}{c|}{$3$}&\multicolumn{2}{c|}{total}&\multicolumn{1}{c}{net}\\
No.  & $+$& $-$& $+$& $-$& $+$& $-$&  $+$& $-$ &  \\\hline\hline
1&16&0&8&0&8&0&32&0&32\\
2&8&0&8&0&8&0&24&0&24\\
3&8&0&8&0&4&0&20&0&20\\
4&8&0&6&2&4&0&18&2&16\\
5&8&0&4&0&4&0&16&0&16\\
6&12&4&4&0&4&0&20&4&16\\
7&8&0&8&0&4&4&20&4&16\\
8&6&2&4&0&4&0&14&2&12\\
9&4&0&4&0&4&0&12&0&12\\
10&8&0&2&0&2&0&12&0&12\\
11&4&0&4&0&2&0&10&0&10\\
12&4&0&4&0&3&1&11&1&10\\
13&6&2&4&0&2&0&12&2&10\\
14&4&4&4&0&4&0&12&4&8\\
15&4&0&4&0&2&2&10&2&8\\
16&4&0&3&1&2&0&9&1&8\\
17&4&0&2&0&2&0&8&0&8\\
18&6&2&3&1&2&0&11&3&8\\
19&6&2&2&0&2&0&10&2&8\\
20&10&6&2&0&2&0&14&6&8\\
21&6&2&4&0&2&2&12&4&8\\
22&3&1&3&1&2&0&8&2&6\\
23&3&1&2&0&2&0&7&1&6\\
24&2&0&2&0&2&0&6&0&6\\
25&4&0&2&2&2&0&8&2&6\\
26&4&0&2&0&1&1&7&1&6\\
27&4&0&1&0&1&0&6&0&6\\
28&6&2&1&0&1&0&8&2&6\\
29&3&1&3&1&1&0&7&2&5\\
30&2&0&2&0&1&0&5&0&5
\end{tabular}
\caption{\cption{$\mathbbm{E}_6 \times U(1)^2 \times SO(8) \times SO(8)$}}
\label{tab:real11}
\end{center}
\end{table}

\begin{table}[H]
\begin{center}
\begin{tabular}{r || r r | r r | r r | r r | r }
&\multicolumn{2}{c|}{$1$}&\multicolumn{2}{c|}{$2$}&\multicolumn{2}{c|}{$3$}&\multicolumn{2}{c|}{total}&\multicolumn{1}{c}{net}\\
No.  & $+$& $-$& $+$& $-$& $+$& $-$&  $+$& $-$ &  \\\hline\hline
31&3&1&2&0&1&0&6&1&5\\
32&3&1&2&0&2&2&7&3&4\\
33&2&2&2&0&2&0&6&2&4\\
34&4&4&2&0&2&0&8&4&4\\
35&4&0&2&2&2&2&8&4&4\\
36&3&1&2&0&1&1&6&2&4\\
37&2&0&2&0&1&1&5&1&4\\
38&2&0&1&0&1&0&4&0&4\\
39&3&1&1&0&1&0&5&1&4\\
40&1&1&3&1&3&1&7&3&4\\
41&2&0&1&0&1&1&4&1&3\\
42&3&1&1&1&1&0&5&2&3\\
43&1&0&1&0&1&0&3&0&3\\
44&2&0&1&1&1&1&4&2&2\\
45&2&0&2&0&1&3&5&3&2\\
46&2&2&2&0&1&1&5&3&2\\
47&1&1&1&0&1&0&3&1&2\\
48&2&2&1&0&1&0&4&2&2\\
49&4&4&1&0&1&0&6&4&2\\
50&1&1&1&1&3&1&5&3&2\\
51&1&1&1&0&1&1&3&2&1\\
52&1&1&0&1&3&1&4&3&1\\
53&2&2&2&2&2&2&6&6&0\\
54&2&0&2&2&1&3&5&5&0\\
55&2&2&1&1&1&1&4&4&0\\
56&4&4&2&2&2&2&8&8&0\\
57&4&4&1&1&1&1&6&6&0\\
58&1&1&1&1&1&1&3&3&0\\
59&2&2&2&2&1&1&5&5&0\\
60&1&3&1&0&1&0&3&3&0\\
61&4&4&4&4&4&4&12&12&0
\end{tabular}
\caption{Table \ref{tab:real11} continued.}
\label{tab:real12}
\end{center}
\end{table}


\begin{table}[H]
\begin{center}
\begin{tabular}{r || r r | r r | r r | r r | r }
&\multicolumn{2}{c|}{$1$}&\multicolumn{2}{c|}{$2$}&\multicolumn{2}{c|}{$3$}&\multicolumn{2}{c|}{total}&\multicolumn{1}{c}{net}\\
No.  & $+$& $-$& $+$& $-$& $+$& $-$&  $+$& $-$ &  \\\hline\hline
1&16&0&8&0&8&0&32&0&32\\
2&8&0&8&0&8&0&24&0&24\\
3&8&0&6&2&4&0&18&2&16\\
4&8&0&4&0&4&0&16&0&16\\
5&12&4&4&0&4&0&20&4&16\\
6&8&0&8&0&4&4&20&4&16\\
7&6&2&4&0&4&0&14&2&12\\
8&4&0&4&0&4&0&12&0&12\\
9&4&4&4&0&4&0&12&4&8\\
10&4&0&4&0&2&2&10&2&8\\
11&4&0&3&1&2&0&9&1&8\\
12&4&0&2&0&2&0&8&0&8\\
13&6&2&3&1&2&0&11&3&8\\
14&6&2&2&0&2&0&10&2&8\\
15&10&6&2&0&2&0&14&6&8\\
16&6&2&4&0&2&2&12&4&8\\
17&3&1&3&1&2&0&8&2&6\\
18&3&1&2&0&2&0&7&1&6\\
19&2&0&2&0&2&0&6&0&6\\
20&3&1&2&0&2&2&7&3&4\\
21&2&2&2&0&2&0&6&2&4\\
22&4&4&2&0&2&0&8&4&4\\
23&4&0&2&2&2&2&8&4&4\\
24&3&1&2&0&1&1&6&2&4\\
25&2&0&2&0&1&1&5&1&4\\
26&1&1&3&1&3&1&7&3&4\\
27&2&0&1&1&1&1&4&2&2\\
28&1&1&1&1&3&1&5&3&2\\
29&2&2&2&2&2&2&6&6&0\\
30&2&0&2&2&1&3&5&5&0\\
31&2&2&1&1&1&1&4&4&0\\
32&4&4&2&2&2&2&8&8&0\\
33&4&4&1&1&1&1&6&6&0\\
34&1&1&1&1&1&1&3&3&0\\
35&4&4&4&4&4&4&12&12&0
\end{tabular}
\caption{\cption{$\mathbbm{E}_6  \times U(1)^2 \times SO(16)$}}
\label{tab:real21}
\end{center}
\end{table}


\begin{table}[H]
\begin{center}
\begin{tabular}{r || r r | r r | r r | r r | r }
&\multicolumn{2}{c|}{$1$}&\multicolumn{2}{c|}{$2$}&\multicolumn{2}{c|}{$3$}&\multicolumn{2}{c|}{total}&\multicolumn{1}{c}{net}\\
No.  & $+$& $-$& $+$& $-$& $+$& $-$&  $+$& $-$ &  \\\hline\hline
1&16&0&16&0&16&0&48&0&48\\
2&12&4&8&0&8&0&28&4&24\\
3&8&0&8&0&8&0&24&0&24\\
4&10&6&4&0&4&0&18&6&12\\
5&6&2&6&2&4&0&16&4&12\\
6&6&2&4&0&4&0&14&2&12\\
7&4&0&4&0&4&0&12&0&12\\
8&3&1&3&1&3&1&9&3&6\\
9&4&4&2&2&2&2&8&8&0\\
10&4&4&4&4&4&4&12&12&0\\
11&2&2&2&2&2&2&6&6&0
\end{tabular}
\caption{\cption{$\mathbbm{E}_6 \times U(1)^2  \times \mathbbm{E}_8$}}
\label{tab:real3}
\end{center}
\end{table}

\refstepcounter{section}
\addcontentsline{toc}{section}%
	{\protect\numberline{\thesection}Tables of the Simple $S^2V$ Models}
\lhead{}\chead{}\rhead{{\sc \sectionname\ \thesection.\ Tables of the Simple $S^2V$ Models}}
\noindent

\begin{table}[H]
\begin{center}
\begin{tabular}{r || r r | r r | r r | r }
&\multicolumn{2}{c|}{$1$}&\multicolumn{2}{c|}{$2$}&\multicolumn{2}{c|}{total}&\multicolumn{1}{c}{net}\\
No.  & $+$& $-$& $+$& $-$& $+$& $-$ &  \\\hline\hline
    1&    16&     0&    16&     0&    32&     0&    32\\
    2&    12&     4&     8&     0&    20&     4&    16\\
    3&     8&     0&     8&     0&    16&     0&    16\\
    4&    10&     6&     4&     0&    14&     6&     8\\
    5&     6&     2&     6&     2&    12&     4&     8\\
    6&     4&     0&     4&     0&     8&     0&     8\\
    7&     3&     1&     3&     1&     6&     2&     4\\
    8&     8&     8&     8&     8&    16&    16&     0\\
    9&     4&     4&     4&     4&     8&     8&     0\\
   10&     2&     2&     2&     2&     4&     4&     0\\
   11&     0&     0&     0&     0&     0&     0&     0
\end{tabular}
\caption{Inequivalent simple $S^2V$ models. The chiral content of
each model is listed per plane and
numbered, $+$ lists all the positive chiral states per plane
while $-$ lists all the negative states per plane. The total
sum of all the planes is then listed and subsequently the net
total number of chiral states. The list is ordered by the total
net number of chiral states.}
\label{tab:results}
\end{center}
\end{table}

\backmatter

			\lhead{}\chead{}\rhead{{\sc Bibliography}}
	\bibliography{library}
	\bibliographystyle{siam}

\end{document}